\documentclass[twocolumn]{IEEEtran}
\usepackage[final]{graphicx}
\usepackage{amsmath,epsfig,amssymb,verbatim,amsopn,cite,subfigure,multirow}
\usepackage{balance}
\usepackage[usenames,dvipsnames]{color}
\usepackage[all]{xy}  
\usepackage{url}
\usepackage{amsfonts}
\usepackage{amssymb}
\usepackage{epsfig}
\usepackage{bm}
\usepackage{epsfig}
\usepackage{amsmath,amssymb, amstext}
\usepackage{graphicx}
\usepackage{epstopdf}
\usepackage{algorithm}
\usepackage{algorithmic}
\usepackage{tabularx,booktabs}
\usepackage{url}
\usepackage{multirow}
\usepackage{array}
\usepackage{makecell}
\usepackage{tikz}
\usepackage{enumerate}
\usepackage{amsmath,epsfig,amssymb,verbatim,amsopn,cite,subfigure,multirow,amsthm}
\usepackage{balance}
\usepackage[usenames,dvipsnames]{color}
\usepackage[all]{xy}  
\usepackage{url}
\usepackage{amsfonts}
\usepackage{amssymb}
\usepackage{epsfig}
\usepackage{bm}
\usepackage{epsfig}
\usepackage{amsmath,amssymb, amstext}
\usepackage{graphicx}
\usepackage{epstopdf}
\usepackage{algorithm}
\usepackage{algorithmic}
\usepackage{tabularx,booktabs}
\usepackage{url}
\usepackage{multirow}
\usepackage{array}
\usepackage{makecell}
\usepackage{balance}
\usepackage{bbm}
\usepackage{dsfont}
\usepackage{multirow}
\usepackage{booktabs}
\newtheorem{remark}{Remark}
\usepackage{longtable}

\usetikzlibrary{shapes.geometric, arrows}
\usetikzlibrary{positioning}

\DeclareMathOperator*{\argmax}{arg\,max}

\tikzstyle{rect} = [rectangle, rounded corners, minimum width=1cm, minimum height=0.8cm,text centered, draw=black, fill=white!30]
\tikzstyle{arrow} = [thick,->,>=stealth]

\begin{document}

\title{A Tutorial on Chirp Spread Spectrum for LoRaWAN: Basics and Key Advances

\thanks{Alireza Maleki and Ebrahim Bedeer are with the Department of Electrical and Computer Engineering, University of Saskatchewan, Saskatoon, Canada S7N5A9. Emails: \{alireza.maleki and e.bedeer\}@usask.ca.}
\thanks{Ha H. Nguyen, deceased, was with the Department of Electrical and Computer Engineering, University of Saskatchewan, Saskatoon, Canada S7N5A9. Email: \{ha.nguyen\}@usask.ca.}
\thanks{R. Barton is with Cisco Systems Inc. Email: robbarto@cisco.com.}
\thanks{This work was supported by the NSERC/Cisco Industrial Research Chair program.}
\thanks{The co-authors dedicate this paper to Prof. Ha H. Nguyen, who passed away before the submission of the paper.}
}
\author{Alireza Maleki, Ha H. Nguyen, Ebrahim Bedeer, and Robert Barton}
\maketitle

\begin{abstract}
Chirps spread spectrum (CSS) modulation is the heart of long-range (LoRa) modulation used in the context of long-range wide area network (LoRaWAN) in internet of things (IoT) scenarios. Despite being a proprietary technology owned by Semtech Corp., LoRa modulation has drawn much attention from the research and industry communities in recent years. However, to the best of our knowledge, a comprehensive tutorial, investigating the CSS modulation in the LoRaWAN application, is missing in the literature. Therefore, in the first part of this paper, we provide a thorough analysis and tutorial of CSS modulation modified by LoRa specifications, discussing various aspects such as signal generation, detection, error performance, and spectral characteristics. Moreover, a summary of key recent advances in the context of CSS modulation applications in IoT networks is presented in the second part of this paper under four main categories of transceiver configuration and design, data rate improvement, interference modeling, and synchronization algorithms. 
\end{abstract}

\IEEEpeerreviewmaketitle

\section{Introduction}
\IEEEPARstart{T}{he} internet of things (IoT) has emerged as a groundbreaking technology that enables intelligent sensing and actuation for a wide range of objects. By facilitating information exchange with a core network, IoT empowers individuals to remotely manage and monitor device behavior from systems located hundreds of kilometers away. Its applications include smart homes, intelligent transportation, smart hospitals, and smart cities \cite{Vaezi_2022,Jouhari_2023,Milarokostas_2023}. These emerging applications lead to the rapid growth of IoT network sizes which results in massive IoT networks. The IoT causes a significant transformation in business and consumer culture and ignites a new industrial revolution, with over $20$ billion connected devices in 2020 \cite{Vaezi_2022}. More devices or objects were connected to the Internet in 2009 than there were people. Despite the wide variations in the IoT market volume predictions made by various analysts and consulting firms, all of them concur that the market is enormous and expanding quickly. By 2025, there will be more than $30$ billion active IoT devices, excluding smartphones, according to IoT Analytics \cite{Lueth_2020}. By 2025, the number of IoT devices is expected to reach more than $64$ billion, according to Business Insider \cite{5G}. With a compound annual growth rate (CAGR) of $25.4\%$, the IoT industry is predicted to increase from $381.3$ billion USD in 2021 to $1,854.76$ billion USD in 2028 \cite{market}. 
\begin{table*}[t]
\caption{Summary of recent works focusing on CSS application in LoRaWAN-based IoT scenarios}
\label{comp}
\centering
\begin{tabularx}{\textwidth}{cccccc}
\toprule[0.5pt]
\textbf{Article} & \textbf{Modulation} & \textbf{Error performance} & \textbf{Interference and orthogonality} & \textbf{Spectral characteristics} & \textbf{Synchronization effects}\\ 
\midrule[0.5pt] 
Vangelista \cite{Vangelista_2017} & \checkmark & \checkmark & -- & -- & -- \\
Passolini \cite{Pasolini_2022} & \checkmark & \checkmark & -- & -- & -- \\
Chiani \textit{et al.} \cite{Chiani_2019} & \checkmark & -- & -- & \checkmark & -- \\
Elshabrawy \textit{et al.} \cite{Elshabrawy_2018_IntBER} & \checkmark & \checkmark & \checkmark & -- & -- \\
Benkhelifa \textit{et al.} \cite{Benkhelifa_2022} & \checkmark & \checkmark & \checkmark & -- & -- \\
Nguyen \textit{et al.} \cite{TTNguyen_2019} & \checkmark & \checkmark & -- & -- & -- \\
Presented paper & \checkmark & \checkmark & \checkmark & \checkmark & \checkmark \\
\bottomrule[0.5pt]
\end{tabularx}
\end{table*}
The proliferation of IoT devices and the increasing demands for connectivity necessitate addressing several challenges in developing efficient IoT systems. As the number of devices continues to surge, the ability to manage and maintain seamless connectivity becomes crucial. Additionally, meeting the diverse requirements of different applications within the IoT ecosystem poses a significant challenge. The design and implementation of intelligent, adaptable, energy-efficient, and cost-effective systems that can handle massive data transmissions, which are not delay sensitive, and ensure reliable communication in the face of scalability constraints is a complex task. To answer these needs, low-power wide area network (LPWAN) has emerged as the favored connectivity option for IoT networks due to its extensive communication range, energy efficiency, and cost-effectiveness. It is particularly suitable for delay-tolerant applications with limited device throughput, making it an ideal choice for low-power battery-operated devices. LPWAN protocols use modulations such as differential phase-shift keying (DPSK) in SigFox, $\frac{\pi}{2}$-binary PSK ($\frac{\pi}{2}$-BPSK) and $\frac{\pi}{2}$-quadrature PSK ($\frac{\pi}{2}$-QPSK) in narrowband-IoT (NB-IoT), BPSK, QPSK, $16$ and $64$ quadrature amplitude modulation (QAM) in long-term evolution for machines (LTE-M), and chirp spread spectrum (CSS) in LoRaWAN to ensure robust transmission over long distances by tolerating interference and propagation noise \cite{Jouhari_2023}.

In all LPWAN technologies, the main characteristics of long-range communication, low-power consumption, and resistance to interference or simultaneous transmission collisions of LoRaWAN communication systems have made them highly regarded for creating a wide range of IoT applications such as space-to-ground communications, localization systems, smart buildings, and environmental monitoring \cite{Jouhari_2023}. At the physical layer, LoRaWAN exploits a proprietary modulation scheme developed by Semtech Corporation designed to enable the mentioned features of LoRaWAN networks. LoRa modulation was introduced by Semtech in 2012 as a key component of their LoRaWAN protocol \cite{LoRa_White} with the goal of providing a cost-effective and energy-efficient solution for long-range IoT connectivity. As stated in \cite{LoRa_White}, the heart of LoRa modulation is CSS based on which LoRa offers a trade-off between sensitivity and data rate. The mathematical definition of LoRa modulation was first presented by Vangelista in \cite{Vangelista_2017}. Ever since its publication, this paper has served as the foundation for numerous research works conducted in academia and industry, focusing on LoRa modulation. These research endeavors aim to enhance the performance of the LoRaWAN network across various domains, leveraging the insights and findings presented in the paper. 

The scope of this paper is to present a comprehensive mathematical foundation for CSS modulation in the context of the LoRaWAN network. It is worth noting that throughout this paper, ``LoRa modulation'' and ``CSS modulation'' terms are used interchangeably. There are several works in the literature that focus on one aspect of CSS modulation in the context of LoRaWAN. For example, the work of \cite{Vangelista_2017} discusses the basic CSS modulation and its error performance, and the work of \cite{Chiani_2019} deals with the spectral characteristics of CSS modulation. However, the current paper can be considered the first tutorial on CSS modulation along with a discussion on the recent key advances of the subject in LoRaWAN-based IoT scenarios. A comparison between the existing works and the presented paper is provided in Table \ref{comp}. As can be seen, there is a lack of work in the existing literature that covers all important aspects of CSS modulation along with an investigation of key recent works. To address this issue, the followings are presented as the main contributions of this paper:
\begin{enumerate}
    \item Presenting a comprehensive mathematical foundation of CSS modulation with the application in LoRaWAN-based IoT scenarios. We shed light on the basic foundations of CSS modulation and its signal generation procedures. Also, we discuss the analytical process of CSS signal detection at the receiver exploring dechirping operation. Additionally, we evaluate the analytical expressions provided in the literature for error performance of CSS modulation in different LoRaWAN specifications in the presence of noise and fading. Moreover, detailed power spectral characteristic of CSS waveforms is presented. Also, a mathematical-based orthogonality analysis of CSS-modulated signals is provided.
    \item We provide a detailed analysis of the state-of-the-art research works of LoRa modulation by dividing them into four main categories, i.e., transceiver configuration and design, data rate improvements, interference modeling, and synchronization algorithms.    
\end{enumerate}
To the best of our knowledge, this work is the first one focusing on providing a tutorial on CSS modulation and discussing the recent key advances of the subject in IoT networks. We hope the tutorial serves as a solid foundation for future researchers in the context of LoRa modulation providing an integrated CSS signal introduction discussing the most relevant research topics in LoRa modulation.

For the convenience of the readers, Table \ref{not} is provided to contain all acronyms used in this paper. In the remainder of this paper, we present an analytical representation of CSS modulation basics in Section \ref{section:Base}. In Section \ref{section:Tax}, we provide a categorization of the key advances related to CSS modulation in LoRaWAN-based IoT scenarios. Recent key advances in transceiver configuration and design of CSS transmission systems are investigated in Section \ref{section:TxRx}. Key recent research works with the aim of data rate improvement of CSS
modulation are discussed in Section \ref{section:DR}. In Section \ref{section:Int}, the key works related to interference modeling of a CSS-based communication system are analyzed. Section \ref{section:Sync} contains the discussion on LoRa systems synchronization and the operations performed on the received CSS signals to recover time and frequency offsets. Future research directions are provided in Section \ref{section:Fut} and the paper is concluded in Section \ref{section:Con}.

\begin{table}[t]
\caption{Acronyms used in this paper.}
\label{not}
\centering
\begin{tabularx}{0.48\textwidth}{ll}
\hline
\textbf{Acronym}            & \textbf{Description}   \\               \hline
        AWGN & Additive white Gaussian noise \\ 
        AF & Ambiguity function \\ 
        ADC & Analog-to-digital converter \\ 
        ACS & Asymmetry chirp signal \\ 
        BPSK & Binray phase-shift keying \\ 
        BER & Bit error rate \\ 
        CFO & Carrier frequency offset \\ 
        CSI & Channel state information \\ 
        CSS & Chirp spread spectrum \\ 
        CAGR & Compound annual growth rate \\ 
        CDF & Cumulative distribution function \\ 
        DCSS & Differential CSS \\ 
        DDS & Direct digital synthesis \\ 
        DCRK-CSS & Discrete chirp rate keying CSS \\ 
        DFT & Discrete Fourier transform \\ 
        DFS & Doppler frequency shift \\ 
        EICS-LoRa & Enhanced ICS-LoRa \\ 
        ePSK-LoRa & Enhanced PSK-LoRa \\ 
        FFT & Fast Fourier transform \\ 
        FCW & Folded chirp waveform \\ 
        FCrSK & Folded chirp-rate shift keying \\ 
        FPAR & Frequency peak average rate \\ 
        FSCSS & frequency shift chirp spread spectrum \\ 
        FBI-LoRa & Frequency-bin-index LoRa \\ 
        FSK & Frequency-shift keying \\ 
        FF  & Fundamental frequencies \\ 
        IM & Index modulation \\ 
        IQCIM & In-phase and quadrature chirp index modulation \\ 
        IQCSS & In-phase and quadrature CSS \\ 
        ICS-LoRa & Interleaved chirp spreading LoRa \\ 
        IoT & Internet of things \\ 
        LoRaWAN & Long-range wide area network \\ 
        LUT & Look-up table \\ 
        LEO & Low Earth orbit \\ 
        LPWAN & Low power wide area network \\ 
        MF & Matched filter \\ 
        ML & Maximum likelihood \\ 
        MPC & Multi-path channel \\ 
        MAI & Multiple access interference \\ 
        MIMO & Multiple-input multiple-output \\ 
        OCG & Orthogonal chirp generator \\ 
        OFDM & Orthogonal frequency-division multiplexing \\ 
        PLL & Phase-locked loop \\ 
        PSK-LoRa & Phase-shift keying LoRa \\ 
        SFO & Sampling frequency offset \\ 
        STO & Sampling time offset \\ 
        SIR & Signal-to-interference ratio \\ 
        SSK-LoRa & Slope-shift keying LoRa \\ 
        SDR & Software-defined radio \\ 
        STBC & Space-time block coding \\ 
        SE & Spectral efficiency \\ 
        SF & Spreading factor  \\ 
        SFD & Start frame delimiter \\ 
        SIC & Successive interference cancellation \\ 
        SER & Symbol error rate \\ 
        SCS & Symmetry chirp signal \\ 
        SC-MCR & Symmetry chirp with multiple chirp rates \\ 
        TDM-LoRa & Time domain multiplexed LoRa \\ 
        TDM SC-MCR & Time domain multiplexed SC-MCR \\ 
        WSS & Wide-sense stationary \\ 
\hline
\end{tabularx}
\end{table}
\section{Chirp Spread Spectrum: Basic Mathematical Concepts}
\label{section:Base}
\subsection{CSS Modulation}
The CSS modulation transforms each data symbol into a chirp in which the signal frequency increases (up-chirp) or decreases (down-chirp) with time over the bandwidth (BW). The main feature of this up-chirp is that when the symbol frequency reaches its maximum (minimum in down-chirp), it will wrap over and start from the minimum frequency (maximum in down-chirp) and keep on sweeping until it sweeps the BW once. Note that hereinafter, by chirp, LoRa, and CSS signals, we refer to up-chirp unless otherwise stated. The continuous-time form of the chirp can be formulated as \cite{Pasolini_2022}:
\begin{equation}
\label{1}
c\left(t\right)=A_0\cos{\left(2\pi\int_{t_0}^{t}f\left(\rho\right){\rm d}\rho+\phi_0\right),\ \ }t_0\le t\le t_0+T_{\rm{sym}},
\end{equation}
where $A_0$, $t_0$, $f\left(t\right)$, $\phi_0$, and $T_{\rm{sym}}$ represent the chirp amplitude, starting time of the chirp, the instantaneous frequency at time $t$, initial phase, and chirp duration (symbol duration), respectively.

\begin{figure}[t]
  \centering
\includegraphics[width=0.5\textwidth]{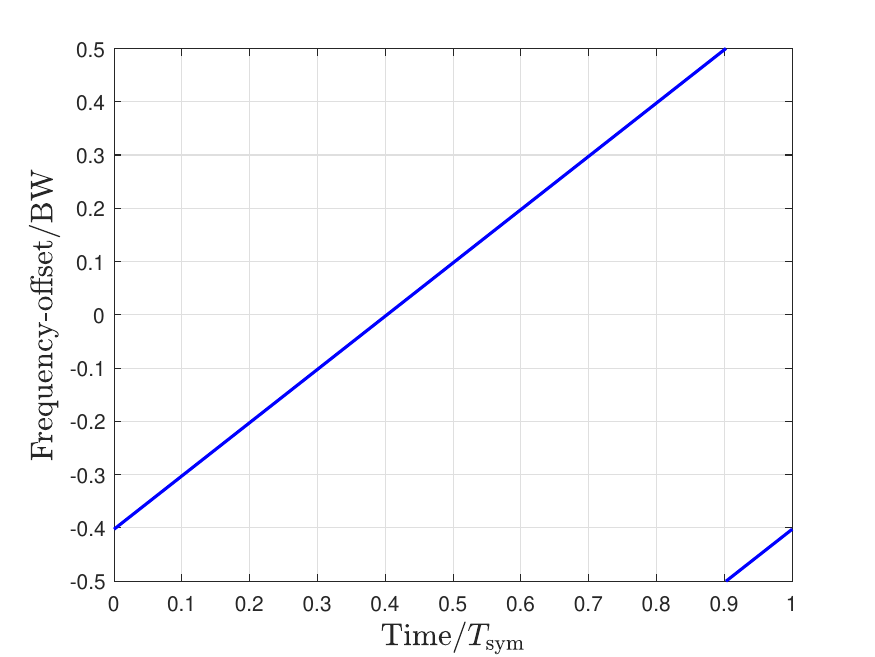}
  \caption{Instantaneous frequency-offset for $s=50$, $\rm{SF}=9$, $\rm{BW}=250$ kHz, and $T_{\rm sym}=2.048$ ms.}
  \label{Chir}
\end{figure}

Without loss of generality, we assume that $f_0$ is the center of available BW, i.e., $\left[f_0-{\rm{BW}}/2,f_0+{\rm{BW}}/2\right]$, and the starting time and the initial signal phase are set to zero. Then we have:
\begin{equation}
\label{2}
c\left(t\right)=A_0\cos{\left(2\pi f_0t+2\pi\int_{0}^{t}\Delta f\left(\rho\right){\rm d}\rho\right),\ }0\le t\le T_{\rm{sym}},
\end{equation}
where $\Delta f\left(t\right)$ is the instantaneous frequency-offset at time $t$. This chirp is denoted as the basic up-chirp which carries the zero symbol in the LoRa symbol set and it is the basis of LoRa modulation for generating $M$ symbols for a given data bit stream. With LoRa, the frequency shift at the beginning of the symbol serves as the information-carrying component, and the chirp resembles a certain type of carrier, according to \cite{Vangelista_2017}. An important parameter of LoRa modulation is the spreading factor (SF) which indicates the number of bits that each symbol can carry. i.e., ${\rm{SF}}=\log_2{M}$. Now, defining a symbol set  $\mathcal{S}=\left\{0,1,\ldots,M-1\right\}$, the starting frequency-offset for the symbol $s\in \mathcal{S}$ is equal to $-{\rm{BW}}/2+{\rm{BW}}\times s/M$ \cite{Pasolini_2022}. The CSS configuration presented by Semtech has a set of specific values of BW and SF for being exploited in IoT applications, i.e., $\rm{BW}\in\left\{125,250,500\right\}$ kHz and $\rm{SF}\in\left\{7,8,9,10,11,12\right\}$ \cite{LoRa_base}. Fig. \ref{Chir} shows the chirp frequency-offset for $s=50$, $\rm{SF}=9$, and $\rm{BW}=250$ kHz.

The time instant $T_w$ is introduced as the wrap-around time and is related to the symbol duration as $T_w=T_{\rm{sym}}-T_{\rm{sym}}\times s/M$. Based on this, the instantaneous frequency offset of the CSS signal modulating the symbol $s$ can be formulated as \cite{Pasolini_2022}:
\begin{equation}
\label{3}
\Delta f_s\left(t\right) =
              \begin{cases}
-\frac{\rm{BW}}{2}+\frac{\rm{BW}}{M}s+\frac{\rm{BW}}{T_{\rm{sym}}}t, & 0\le t< T_w, \\
                -\frac{3\rm{BW}}{2}+\frac{\rm{BW}}{M}s+\frac{\rm{BW}}{T_{\rm{sym}}}t, & T_w\le t\le T_{\rm{sym}}.
              \end{cases}
\end{equation}
The chirp rate is defined as the rate at which the chirp frequency varies over time and is formulated as:
\begin{equation}
\label{4}
R_{\rm{chirp}}=\frac{\rm{BW}}{T_{\rm{sym}}}=\frac{\rm{BW}^2}{M},\ \left(\rm{Hz}/\rm{s}\right),
\end{equation}
where $T_{\rm sym}=M/{\rm BW}$. As proved in \cite{Pasolini_2022}, the following equation holds for all values of $s$:
\begin{equation}
\label{5}
\int_{0}^{T_{\rm{sym}}}{\Delta f_{s}\left(\alpha\right){\rm d}\alpha=0},
\end{equation}
and considering the phase term in (\ref{2}) as:
\begin{equation}
\label{6}
\phi_{s}\left(t\right)=2\pi\int_{0}^{t}\Delta f_{s}\left(\alpha\right)d\alpha ,
\end{equation}
we obtain $\phi_s\left(T_{\rm{sym}}\right)=\phi_s\left(0\right)=0$ which means that the starting and final phases of each symbol are zero.

From the aforementioned discussions, we can introduce CSS as a continuous phase memoryless modulation scheme. This feature is particularly useful in practical implementations, as it will simplify the signal generation procedure at the transmitter end.

Considering the memoryless property of CSS modulation and also noting that the starting and final phases of each symbol are zero, the complex envelope of the CSS signal for $n$th symbol interval, i.e., $nT_{\rm sym}\leq t<(n+1)T_{\rm sym}$, can be written as:
\begin{equation}
\label{7}
x_{s_n}\left(t\right)=A_0{\exp{\left[j\phi_{s_n}\left(t\right)\right]}g_{T_{\rm{sym}}}\left(t-nT_{\rm{sym}}\right)},
\end{equation}
where $s_{n}$ is the symbol in the $n$th time interval and $g_{T_{\rm sym}}\left(t\right)=1$ for $0\le t\le {T_{\rm sym}}$ and $g_{T_{\rm sym}}(t)=0$ elsewhere.

As shown in \cite{LoRa_Tx_Patent}, the CSS transmitter exploits an IQ approach as depicted in Fig. \ref{Tx} where $x_{s_n}(t)=i_{s_n}(t)+jq_{s_n}(t)$. Therefore, the $i_{s_n}\left(t\right)$ and $q_{s_n}\left(t\right)$ signals can be formulated as:
\begin{figure}[t!]
\centering
\includegraphics[scale=0.45]{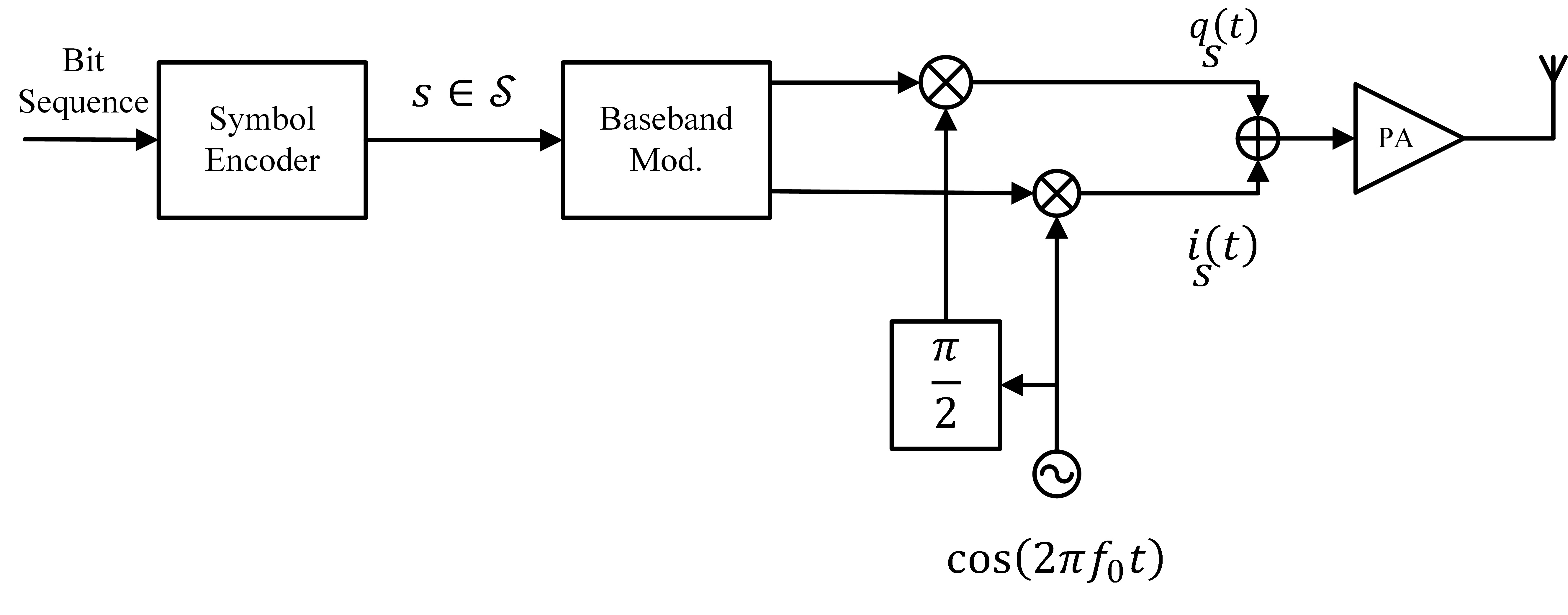}
  \caption{CSS modulation IQ-based transmitter structure.}
  \label{Tx}
\end{figure}
\begin{equation}
\label{8}
i_{s_n}\left(t\right)=A_0{\cos{\left[\phi_{s_n}\left(t\right)\right]}g_{T_{\rm{sym}}}\left(t-nT_{\rm{sym}}\right)},
\end{equation}
and
\begin{equation}
\label{9}
q_{s_n}\left(t\right)=A_0{\sin{\left[\phi_{s_n}\left(t\right)\right]}g_{T_{\rm{sym}}}\left(t-nT_{\rm{sym}}\right)}.
\end{equation}
From the standpoint of digital implementation, the transmitter's baseband step produces discrete-time signals $i_{s_n}\left(kT_{s}\right)$ and $q_{s_n}\left(kT_{s}\right)$ where $k\in\mathbb{Z}$ and $T_{s}$ is the proper sampling time at the transmitter. The baseband modulator's last stage, which contains a pair of digital-to-analog converters (DACs) to obtain (\ref{8}) and (\ref{9}), then converts these signals from digital to analog. Note that in Fig. \ref{Tx}, DACs are included inside the Baseband Mod. block.
\subsection{CSS Signal Generation}
The signal generation process in CSS modulation is of paramount importance as it forms the foundation of the modulation scheme and its suitability for application in long-range communication systems. The phase term of the basic up-chirp for a starting time of $t_0=0$ can be analytically derived from (\ref{3}) and (\ref{6}) as:
\begin{equation}
\label{10}
\phi_0\left(t\right)=2\pi\left[-\frac{\rm{BW}}{2}t+\frac{\rm{BW}}{2T_{\rm{sym}}}t^2\right],\ 0\le t\le T_{\rm{sym}},
\end{equation}
which we denote as the basic up-chirp phase. After discretizing $\phi_{0}(t)$ by obtaining samples using the sampling frequency of $f_s=1/T_s={\rm{BW}}={2^{\rm SF}}/{T_{\rm sym}}$, we have:
\begin{equation}
\label{11}
\left\{\phi_0^{\left(\rm{SF}\right)}\left[k\right]\right\}_{k=0}^{M-1}=\left\{k{\pi}\left[-1+\frac{k}{2^{\rm{SF}}}\right]\right\}_{k=0}^{M-1}.
\end{equation}
As proved in \cite{Pasolini_2022}, with the help of the reference up-chirp phase denoted as $\left\{\phi_0^{\left(12\right)}\left[k\right]\right\}_{k=0}^{2^{12}-1}$, it is possible to obtain the phase samples for any SF and any symbol $s$ using the following equation:

\begin{equation}
\label{12}
\begin{split}
\left\{\phi^{\left(\rm{SF}\right)}_{s}\left[k\right]\right\}=&\frac{1}{C}\left\{\phi_0^{\left(12\right)}\left[C\left(\left(k+s\right)\mod{2^{\rm{SF}}}\right)\right]\right\}\\
&-\frac{1}{C}\phi_0^{\left(12\right)}\left[sC\right]
\end{split}
\end{equation}
where $C=2^{12-\rm{SF}}$. Also, the term $-\frac{1}{C}\phi^{\left(12\right)}\left[sC\right]$ is due to phase continuity as mentioned before. By applying (\ref{11}) into (\ref{12}), the discrete-time CSS waveform in (\ref{7}) for the $n$th symbol interval can be formulated as follows:
\begin{equation}
\label{X_n_k}
\begin{split}
x_{s_n}[k]=A_0\exp\left(2\pi j\frac{k^2+2ks_n-kM}{2M}\right).
\end{split}
\end{equation}
\begin{remark}
\label{remark2} 
\textbf{\textit{Continuous-phase vs non-continuous phase versions of CSS.}} The equation of CSS waveform in (\ref{X_n_k}) is obtained for continuous-phase CSS in which the phase of each CSS signal, regardless of the symbol $s_n$, starts and ends at zero. However, in literature, several works, e.g., \cite{Almeida_2020,Baruffa_2021,An_2022}, exploit the non-continuous phase form of CSS modulation which can be obtained as:
\begin{equation}
\label{X_non}
x_{s_n}^{{\rm (NC)}}[k]=A_0\exp\left[2\pi j\frac{(k+s_n)^2-M(k+s_n)}{2M}\right].
\end{equation}
\end{remark}

Hereinafter, we use the following definitions for simplicity of notation:
\begin{enumerate}
    \item Down-chirp continuous-phase CSS signal as $x_{s_n,{\rm D}}[k]=A_0\exp\left(-2\pi j\frac{k^2+2ks_n-kM}{2M}\right)$.
    \item Down-chirp non-continuous phase CSS signal as $x_{s_n,{\rm D}}^{\rm (NC)}[k]=A_0\exp\left[-2\pi j\frac{(k+s_n)^2-M(k+s_n)}{2M}\right]$.
    \item Basic up-chirp CSS signal (in which $s_n=0$) as $x_{0}[k]$.
    \item Basic down-chirp CSS signal (in which $s_n=0$) as $x_{0,{\rm D}}[k]$
\end{enumerate}
\subsection{Dechirping and CSS Demodulation}
\begin{figure}[t!]
  \centering
\includegraphics[width=0.5\textwidth]{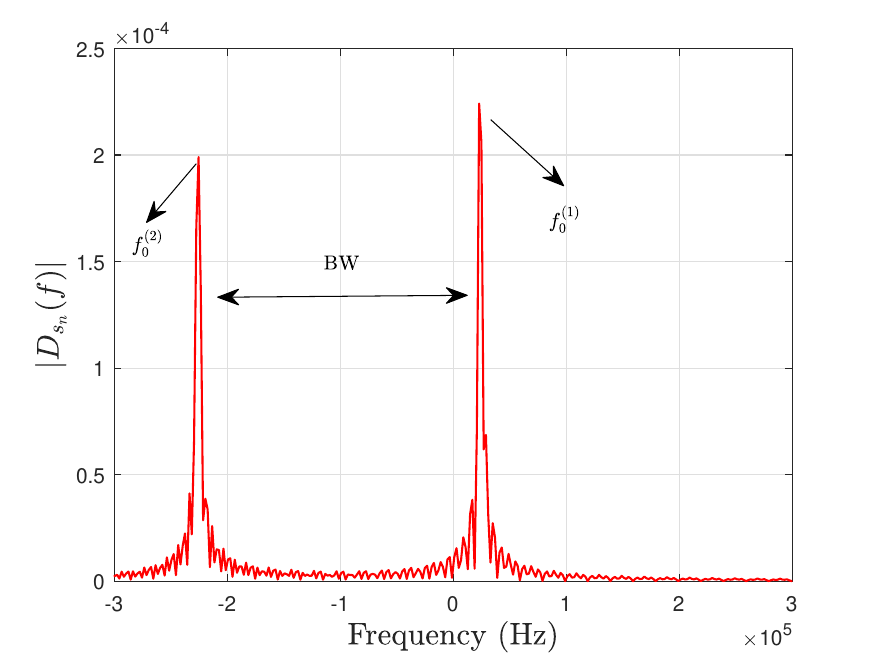}
  \caption{Amplitude spectrum for $s=50$, ${\rm{SF}=10}$ and ${\rm{BW}}=250$ kHz}
  \label{FFT}
\end{figure}

The dechirping process in CSS modulation holds significant importance as it facilitates the recovery of transmitted information. By analyzing the received chirp signal, the dechirping process enables the extraction of the original data which is a crucial step for accurate data retrieval, noise reduction, and enhancing the resilience of CSS modulation in various real-world scenarios. The dechirping includes the multiplication of a CSS signal by the complex conjugate of the basic up-chirp to obtain the dechirped CSS signal as follows:
\begin{equation}
\label{13}
d_{s_n}\left(t\right)=x_{s_n}\left(t\right)\times x^*_{0}(t),
\end{equation}
which yields (\ref{14}) at the top of next page where $T_{\hat{w}}$ is the residual chirp duration after folding, i.e., $T_{\hat{w}}=T_{\rm sym}-T_w$. Note that here we do not consider any noise and fading for the ease of explanation of the demodulation process and their effects of them will be discussed in detail in the Subsection \ref{BER}.
\begin{figure*}
\normalsize
\begin{equation}
\label{14}
\begin{split}
 d_{s_n}\left(t\right)=&
            \begin{cases}
A_0\exp{\left[j2\pi\frac{\rm{BW}}{M}{s_n}\left(t-nT_{\rm{sym}}\right)\right],} & nT_{\rm{sym}}\le t<nT_{\rm{sym}}+T_w\ , \\
                A_0\exp{\left[j2\pi\left(-{\rm{BW}}+\frac{\rm{BW}}{M}{s_n}\right)\left(t-nT_{\rm{sym}}\right)\right],} & nT_{\rm{sym}}+T_w\le t<\left(n+1\right)T_{\rm{sym}}\
              \end{cases}\\
 =& A_0\exp{\left[j2\pi\frac{\rm{BW}}{M}s_n\left(t-nT_{\rm{sym}}\right)\right]}\times g_{T_w}(t-nT_{\rm sym})\\
  &+A_0\exp{\left[j2\pi\left(-{\rm{BW}}+\frac{{\rm{BW}}}{M}s_n\right)\left(t-nT_{\rm{sym}}\right)\right]}\times g_{T_{\hat{w}}}(t-nT_{\rm sym}-T_w).
\end{split}
\end{equation}
\hrulefill
\vspace
*
{4pt}
\end{figure*}
By taking the continuous-time Fourier transform (CTFT) of (\ref{14}) and computing its modulus \cite{Pasolini_2022}, we obtain (\ref{15}) as indicated at top of the next page where $f_0^{\left(1\right)}=s_n{\rm{BW}}/M$ and $f_0^{\left(2\right)}=-{\rm{BW}}+s_n{\rm{BW}}/M$.
\begin{figure*}
\normalsize
\begin{equation}
\label{15}
\begin{split}
D_{s_n}\left(f\right)=&A_0 T_w {\rm{sinc}}\left[\left(f-f_0^{\left(1\right)}\right)T_w\right]\exp{\left[-j2\pi\left(f-f_0^{\left(1\right)}\right)\left(nT_{\rm{sym}}+\frac{T_w}{2}\right)\right]}\exp{\left(-j2\pi f_0^{\left(1\right)}nT_{\rm{sym}}\right)}\\
&+A_0 T_{\hat{w}}{\rm{sinc}}\left[\left(f-f_0^{\left(2\right)}\right)T_{\hat{w}}\right]\exp{\left[-j2\pi\left(f-f_0^{\left(2\right)}\right)\left(nT_{\rm{sym}}
+T_w+\frac{T_{\hat{w}}}{2}\right)\right]}\exp{\left(-j2\pi f_0^{\left(2\right)}nT_{\rm{sym}}\right)}.
\end{split}
\end{equation}
\hrulefill
\vspace
*
{4pt}
\end{figure*}
As depicted in Fig. \ref{FFT}, the result peaks at two frequency bins (which is obtained by dividing the swept BW by the value of $M$) of $f_0^{\left(1\right)}$ and $f_0^{\left(2\right)}$. Therefore, one can find the transmitted symbol $s_n$ by finding the peak value of the dechirped signal and the corresponding frequency bins. 
Based on this concept, the CSS demodulation procedure developed in \cite{LoRa_Tx_Patent} and \cite{CSS_Patent} is as follows:
\begin{itemize}
  \item Step 1: Obtaining the dechirped signal with a proper sampling frequency $f_s$.
  \item Step 2: Performing the FFT of the output from the previous step (note that FFT operation is exploited due to the sampling at the receiver in the practice).
  \item Step 3: Detection of the frequency bin containing the peak value.
\end{itemize}
The two peaks at the frequency domain of the dechirped CSS signal are non-symmetric in general. However, in \cite{LoRa_Tx_Patent} and \cite{CSS_Patent}, the detection is performed based on a single peak in the frequency domain.

\begin{figure}[t!]
  \centering
\includegraphics[width=0.47\textwidth]{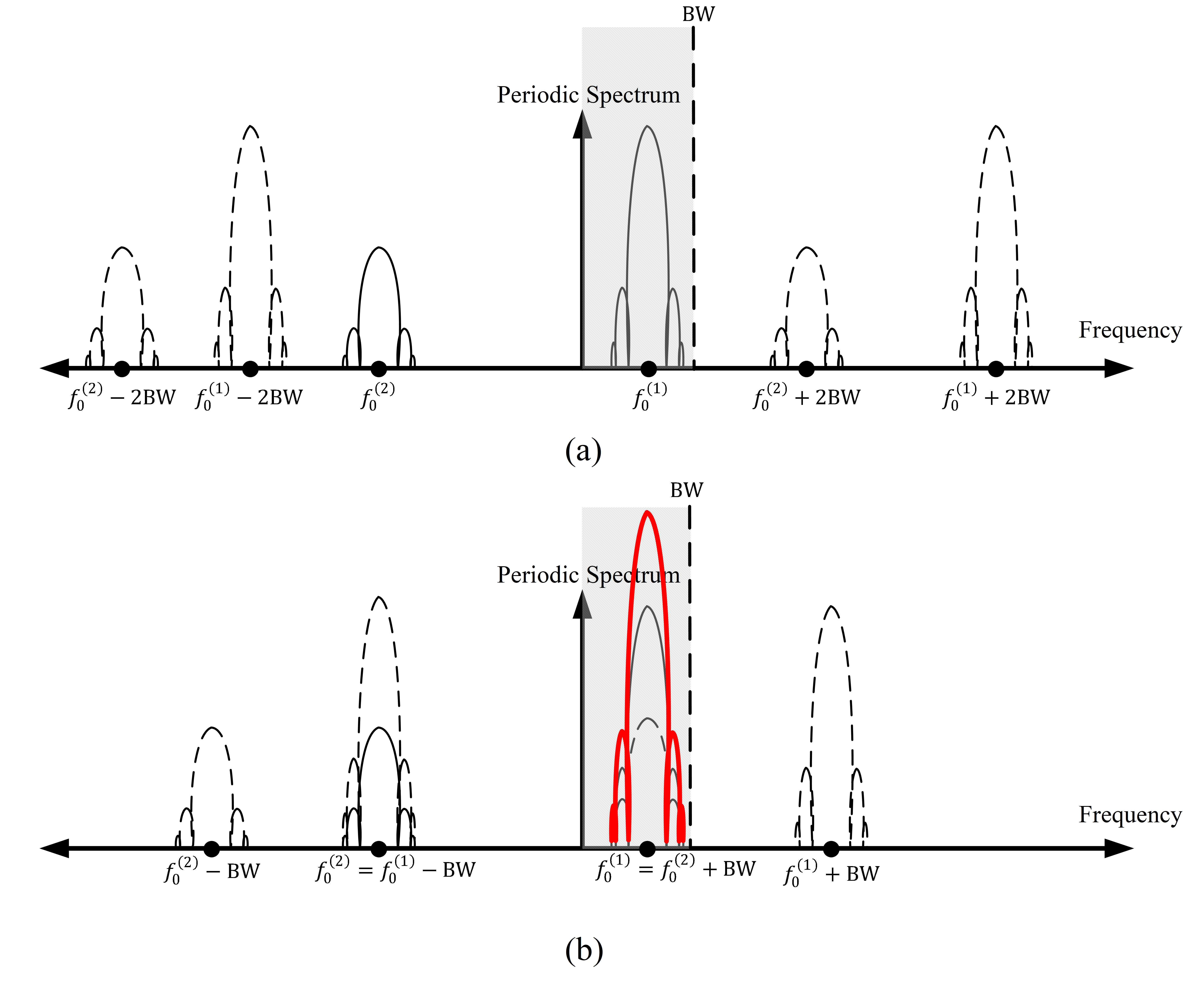}
  \caption{Periodic spectrum of discrete-time dechirped CSS signal for (a) $f_s\approx2\rm{BW}$, and (b) $f_s=\rm{BW}$.}
  \label{sampling}
\end{figure}

\begin{remark}
\label{rem1}
\textbf{\textit{Sampling frequency selection.}} Here, we intend to briefly discuss the resulting single peak that was investigated thoroughly in \cite{Pasolini_2022}. Considering that the CTFT of the dechirped signal $d_{s_n}(t)$ is $D_{s_n}(f)$, then the discrete-time signal with the sampling frequency of $f_s$ has a periodic spectrum of \cite{Proakis_2007}: 
\begin{equation}
\label{per}
D^{\rm (disc.)}(f)=f_s\sum^{\infty}_{i=-\infty}D_{s_n}(f-if_s).
\end{equation}
The key point here is to explain the proper selection of $f_s$ as $f_s={\rm{BW}}$ instead of complying with the sampling theorem requirement which suggests $f_s\geq 2{\rm{BW}}$. It is obvious that this under-sampling results in aliasing in the frequency domain. If we sample the signal $d_{s_n}(t)$, the resulting spectrum would be periodic with the period of the selected $f_s$. If we comply with the sampling theorem and select our sampling frequency as $f_s\approx2\rm{BW}$, we obtained the period spectrum as in Fig. \ref{sampling} (a). However, in the case of $f_s={\rm BW}$, and as can be seen in Fig. \ref{sampling} (b), due to the fact that the difference between $f_0^{(1)}$ and $f_0^{(2)}$ is equal to BW, we will have constructive aliasing (red diagram in Fig. \ref{sampling} (b)) at the processing BW. This results in a higher peak at the desired frequency bin and improves the detection performance. Also, it has been shown in \cite{Pasolini_2022} that this useful peak is an increasing function of SF and therefore, we can achieve better performances for higher SFs in terms of communication reliability.
\end{remark} 

\subsection{Orthogonality of CSS Waveforms}
Here in this subsection, we provide a discussion on the orthogonality of CSS waveforms. The discussion is carried out for four types of CSS signals, i.e., the continuous-time CSS, the continuous-time dechirped CSS, the discrete-time CSS, and the discrete-time dechirped CSS. We focus on calculating the cross-correlation between two CSS waveforms that transmit symbols $s_1$ and $s_2$. Without the loss of generality, symbol $s_1$ is transmitted using ${\rm{SF}}_1$, while symbol $s_2$ is transmitted using ${\rm{SF}}_2$. The respective bandwidths for the transmissions are ${\rm{BW}}_1$ and ${\rm{BW}}_2$. The symbol transmission times are denoted as $T_{{\rm{sym}},1}=2^{{\rm SF}_1}/{\rm{BW}}_1$ and $T_{{\rm{sym}},2}=2^{{\rm SF}_2}/{\rm{BW}}_2$, corresponding to $s_1$ and $s_2$, respectively. The starting frequencies of symbols $s_1$ and $s_2$ are $f_1$ and $f_2$, respectively. Moreover, without the loss of generality, we assume that ${\rm{SF}}_1$ is less than ${\rm{SF}}_2$, i.e., $\xi={\rm{SF}}_2-{\rm{SF}}_1$, where $\xi$ can be any integer from $1$ to $5$. 

The cross-correlation function between two continuous-time waveforms $x_{s_1}(t)$ and $x_{s_2}(t)$ can be defined as:
\begin{equation}
\label{orth_1}
\mathcal{C}^{\rm{(cont.)}}_{s_1,s_2}(\tau,f_1,f_2)=\int^{t_0 +\tau}_{t_0+T_{{\rm{sym}},1}}x_{s_1}(t)x^*_{s_2}(t-\tau) {\rm{d}}t,
\end{equation}
where $\tau$ is the time delay. The closed-form expression for this cross-correlation is calculated as (\ref{orth_2}) at the top of the next page where we have \cite{Benkhelifa_2022}:
\begin{figure*}
\normalsize
\begin{equation}
\label{orth_2}
\begin{split}
&\mathcal{C}^{\rm{(cont.)}}_{s_1,s_2}(\tau,f_1,f_2)=\\
&\begin{cases}
\mathcal{K}_{a_{12}}(b_{s_1,s_2}-{\rm{BW}}_1,\tau,T_{{\rm sym},1}), & {\rm if}\; t_{s_1}\leq \tau < T_{{\rm sym},1} \leq \tau+t_{s_2}, \\   
                \mathcal{K}_{a_{12}}(b_{s_1,s_2}-{\rm{BW}}_1,\tau,\tau+t_{s_2})
                +e^{-2j\pi{\rm{BW}_2}\tau}\mathcal{K}_{a_{12}}(b_{s_1,s_2}-{\rm{BW}}_1+{\rm{BW}}_2,\tau+t_{s_2},T_{{\rm sym},1}), & {\rm if}\; t_{s_1}\leq \tau < \tau+t_{s_2} \leq T_{{\rm sym},1}, \\
                \mathcal{K}_{a_{12}}(b_{s_1,s_2},\tau,t_{s_1})+\mathcal{K}_{a_{12}}(b_{s_1,s_2}-{\rm{BW}}_1,t_{s_1},T_{{\rm sym},1}), & {\rm if}\; \tau< t_{s_1} < T_{{\rm sym},1} \leq \tau+t_{s_2}, \\
                \mathcal{K}_{a_{12}}(b_{s_1,s_2},\tau,m_{s_1,s_2}(\tau))\\
                \;+e^{-2j\pi{\rm{BW}_2}\tau U(t_{s_1}-\tau-t_{s_2})}\mathcal{K}_{a_{12}}(b_{s_1,s_2}-{\rm{BW}}_c,m_{s_1,s_2}(\tau),M_{s_1,s_2}(\tau))\\
                \times \mathds{1}(t_{s_1}-\tau \neq t_{s_2}) +e^{-2j\pi{\rm{BW}_2}\tau }
                \times\mathcal{K}_{a_{12}}(b_{s_1,s_2}-{\rm{BW}_1}+{\rm{BW}_2},M_{s_1,s_2}(\tau),T_{{\rm sym},1}), & {\rm if}\; \tau< t_{s_1}\; \&\; \tau+t_{s_2}< T_{{\rm sym},1}.
              \end{cases}
\end{split}
\end{equation}
\hrulefill
\vspace
*
{4pt}
\end{figure*}
\begin{equation}
\label{orth_3}
\begin{split}
&\mathcal{K}_{a_{12}}(b_{s_1,s_2},t_1,t_2)=\\
&\begin{cases}
\frac{e^{2j\pi c_{s_2}}e^{-j\pi \frac{b^2_{s_1,s_2}}{a_{12}}}}{\sqrt{T_{{\rm sym},1}T_{{\rm sym},2}}}\\
\;\times\Bigg{[} {\rm erf} \left(\sqrt{\frac{\pi a_{12}}{j}}\left(t_0+t_2+\frac{b_{s_1,s_2}}{a_{12}}\right)\right)\\
\;-{\rm erf} \left(\sqrt{\frac{\pi a_{12}}{j}}\left(t_0+t_2+\frac{b_{s_1,s_2}}{a_{12}}\right)\right)\Bigg{]}, & {\rm if}\; a_{12}\neq 0, \\
e^{2j\pi c_{s_2}}e^{j\pi b_{s_1,s_2}(2t_0+t_1+t_2)}\\
\;\times\frac{ {\rm sinc}\left[\pi b_{s_1,s_2}(t_2-t_1)\right](t_2-t_1)}{\sqrt{T_{{\rm sym},1}T_{{\rm sym},2}}}, & {\rm if}\; a_{12}= 0. 
\end{cases}
\end{split}
\end{equation}
with $\mu_n=1/T_{{\rm sym},n}$ for $n=1,2$, $a_{12}=\mu_1 {\rm BW}_1-\mu_2 {\rm BW}_2$, $b_{s_1,s_2}=f_1+\mu_1 s_1-f_2-\mu_2 s_2 + \mu_2 \tau {\rm BW}_2$, $c_{s_2}=(f_2+\mu_2 s_2 - 0.5\mu_2\tau {\rm BW}_2)\tau$, $m_{s_1,s_2}(\tau)=\min(t_{s_1},\tau+t_{s_2})$, $M_{s_1,s_2}(\tau)=\max(t_{s_1},\tau+t_{s_2})$, and ${\rm BW}_c={\rm BW}_1 U(t_{s_2}+\tau-t_{s_1})-{\rm BW}_2 U(t_{s_1}-t_{s_2}-\tau)$. Also, note that $U(.)$ is the step function as $U(t)=1$ for $t\geq 0$ and $U(t)=0$ otherwise and $\mathds{1}(.)$ is the indicator function as $\mathds{1}(t\neq a)=1$ for $t\neq a$ and $\mathds{1}(t\neq a)=0$ otherwise. Moreover, we have ${\rm erf}(z)=(2/\sqrt{\pi})\int_0^z e^{-t^2}{\rm d}t$ and ${\rm sinc}(z)=\sin(z)/z$.

The cross-correlation between two continuous-time dechirped CSS waveforms $x_{s_1}\left(t\right)\times x^*_0(t)$ and $x_{s_2}\left(t\right)\times x_0^*(t)$ can be obtained using (\ref{orth_2}) with two substitutions as $\Tilde{c}_{s_1,s_2}=(2f_2+{\rm BW}_2)+\mu_2 s_2)\tau$ and $\Tilde{b}_{s_1,s_2}=\mu_1 s_1 - \mu_2 s_2 +2f_1-2f_2+{\rm BW}_1-{\rm BW}_2$ \cite{Benkhelifa_2022}. 

Let us assume that we use an identical sampling time $T_d$ for both continuous-time CSS waveforms $x_{s_1}(t)$ and $x_{s_2}(t)$. Based on this, we define parameters $\xi_n=\log_2 {\rm BW}_n T_d$ for $n=1,2$. Moreover, we define $m_0=t_0/T_d$ and $m_1=\tau/T_d$. Accordingly, the cross-correlation function between two discrete-time CSS waveforms $x_1[kT_d]$ and $x_2[kT_d]$ can be formulated as (\ref{orth_4}) at the above of next page \cite{Benkhelifa_2022}.
\begin{figure*}
\normalsize
\begin{equation}
\label{orth_4}
\begin{split}
&\mathcal{C}^{\rm (disc.)}_{s_1,s_2}(\tau,f_1,f_2)=\\
&\begin{cases}
\frac{e^{2j\pi c_{s_2}}e^{-j\pi a_{12}\left(\frac{b_{s_1,s_2}}{a_{12}}\right)^2}}{\sqrt{2^{{\rm SF}_1}2^{{\rm SF}_2}}}\sum^{2^{{\rm SF}_1}-\xi_1-1}_{n'=m_1} e^{j\pi a_{12}\left(n'T_d+\frac{b_{s_1,s_2}}{a_{12}}\right)^2}, & {\rm if}\; a_{12}\neq 0,\\
\frac{e^{2j\pi c_{s_2}}e^{2j\pi b_{s_1,s_2} T_d m_1}e^{j\pi (2^{{\rm SF}_1}-\xi_1-m_1-1)b_{s_1,s_2}T_d}}{\sqrt{2^{{\rm SF}_1}2^{{\rm SF}_2}}} {\rm cosec}(\pi b_{s_1,s_2}T_d)\sin\left[\pi(2^{{\rm SF}_1-\xi_1}-m_1)b_{s_1,s_2}T_d\right], & {\rm if}\; a_{12}\neq 0 \;\&\; b_{s_1,s_2}\neq 0,\\
\frac{e^{2j\pi c_{s_2}}}{\sqrt{2^{{\rm SF}_1}2^{{\rm SF}_2}}}\frac{T_{{\rm sym},1}-\tau}{T_d}, & {\rm if}\;a_{12}=0\;\&\;b_{s_1,s_2}=0.
\end{cases}
\end{split}
\end{equation}
\hrulefill
\vspace
*
{4pt}
\end{figure*}

Finally, the cross-correlation between two discrete-time dechirped CSS waveforms can be written as \cite{Benkhelifa_2022}:
\begin{equation}
\label{orth_5}
\begin{split}
&\mathcal{C}^{\rm (dechirped \;disc.)}_{s_1,s_2}(\tau,f_1,f_2)=\\
&\begin{cases}
\frac{e^{2j\pi \Tilde{c}_{s_2}}e^{2j\pi \Tilde{b}_{s_1,s_2}\tau}}{\sqrt{2^{{\rm SF}_1}2^{{\rm SF}_2}}}e^{j\pi(2^{{\rm SF}_1}-\xi_1-m_1-1)\Tilde{b}_{s_1,s_2}T_d}\\
\;\times \sin \left[\pi(2^{{\rm SF}_1}-\xi_1-m_1)\Tilde{b}_{s_1,s_2}T_d\right]\\
\;\times {\rm cosec}(\pi \Tilde{b}_{s_1,s_2}T_d), & {\rm if}\;\Tilde{b}_{s_1,s_2}\neq 0,\\
\frac{e^{2j\pi \Tilde{c}_{s_2}}}{\sqrt{2^{{\rm SF}_1}2^{{\rm SF}_2}}}\frac{T_{{\rm sym},1}-\tau}{T_d}, & {\rm if}\;\Tilde{b}_{s_1,s_2}=0.
\end{cases}
\end{split}
\end{equation}

\begin{figure*}[t!]
  \centering
\includegraphics[width=\textwidth]{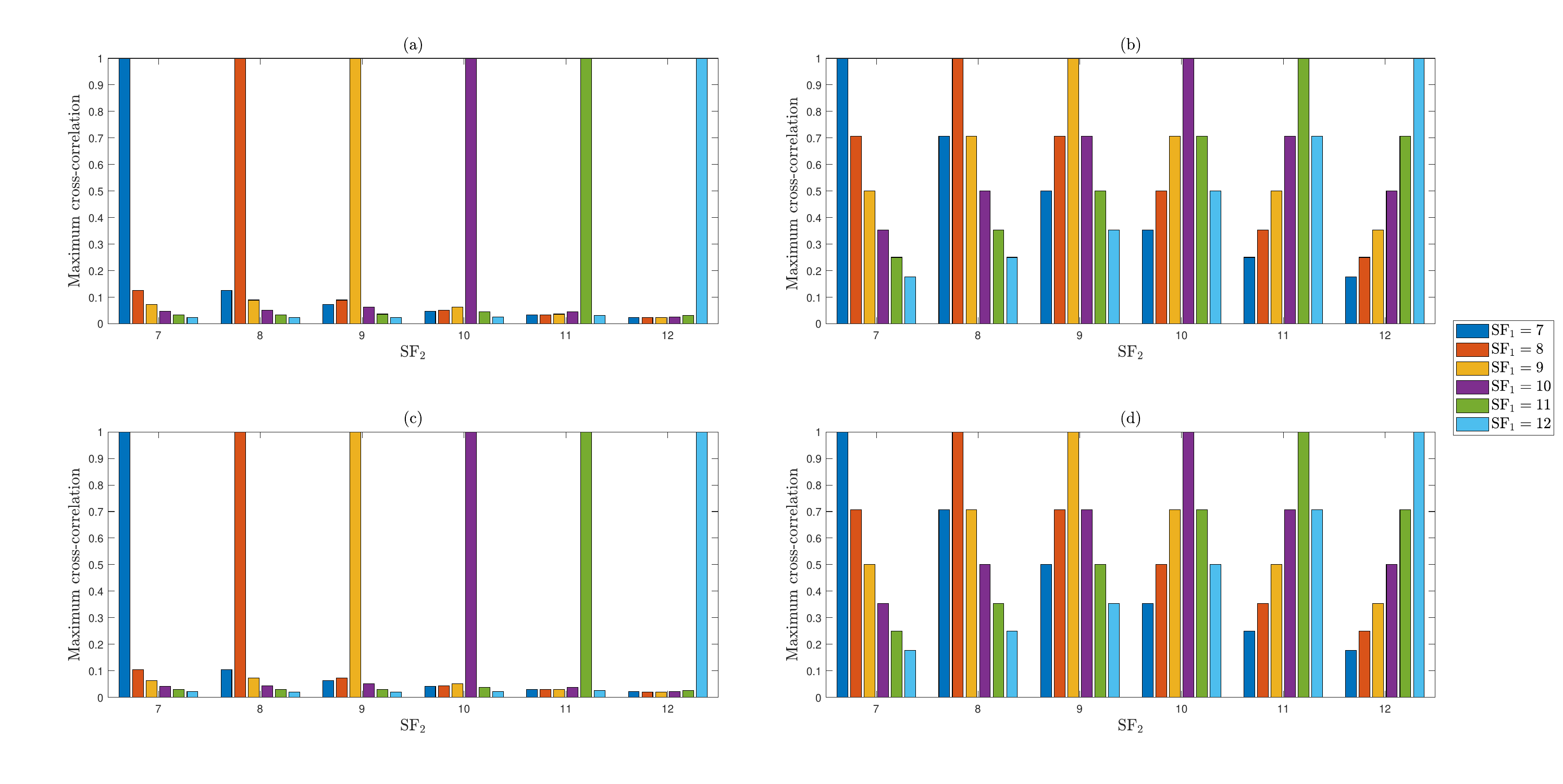}
  \caption{Maximum of cross-correlation with respect to the symbols $s_1$ and $s_2$ with $\tau=0$ and $f_1=f_2$ for two CSS signals in forms of (a) continuous-time, (b) dechirped continuous time, (c) discrete-time, and (d) dechirped discrete-time.}
  \label{max_cross}
\end{figure*}

Based on these equations and the data provided in \cite{Benkhelifa_2022}, we present Fig. \ref{max_cross} in which the maximum of the cross-correlation functions of (\ref{orth_2}), its dechirped version (\ref{orth_4}), and (\ref{orth_5}) are plotted with respect to the symbols $s_1$ and $s_2$ with $\tau=0$ and $f_1=f_2$. Based on these results, several important insights regarding the cross-correlation functions and orthogonality features of CSS modulation can be listed as follows:
\begin{itemize}
    \item From equations (\ref{orth_2}) and (\ref{orth_4}), we can interpret that the maximum cross-correlation between the same SFs in both continuous and discrete time domains is more significant compared to the case with different SFs. Moreover, this difference is particularly prominent prior to down-chirping.
    \item Before down-chirping, lower ${\rm SF}_2$ values than ${\rm SF}_1$ have a greater impact on the value of cross-correlation than higher ones, as their maximum cross-correlations are higher. Additionally, SFs that are closer to each other have a stronger influence on the maximum cross-correlation. This also can be shown numerically from (\ref{orth_2}) and (\ref{orth_4}).
    \item The maximum cross-correlation between different SFs after down-chirping is larger than the one before down-chirping. Moreover, this maximum cross-correlation does not depend on the time domain (discrete or continuous). 
\end{itemize}
Table \ref{table_orth} provides several useful orthogonality conditions for CSS modulation obtained from equations (\ref{orth_2}), (\ref{orth_4}), and (\ref{orth_5}).
\begin{table*}[t]
\caption{Orthogonality conditions of CSS modulated waveforms \cite{Benkhelifa_2022}.}
	\label{table_orth}
	\centering

      \begin{tabular}{| >{\centering\arraybackslash}m{0.7 cm} | >{\centering\arraybackslash}m{1.78 cm} |>{\centering\arraybackslash}m{1.9 cm}|>{\centering\arraybackslash}m{0.6 cm}|c|m{9 cm}|}
    \cline{2-6} 
    \multicolumn{1}{c|}{} & SF & BW & $\tau$  & $f_d=f_2-f_1$ & \vspace{7pt} Orthogonality conditions \vspace{5pt}\\  \hline
    
    \multirow{4}{*}{Cont.} & \multirow{4}{*}{${\rm SF}_1={\rm SF}_2$} & \multirow{4}{*}{${\rm BW}_1={\rm BW}_2$} & \multirow{2}{*}{$\tau=0$} & $f_d=0$ & SF even and $|s_1-s_2|\propto \sqrt{2^{\rm SF}}$ \\ \cline{5-6}
    
    &&&& $f_d\neq0$ & SF even, $f_d\propto\frac{{\rm BW}}{2^{\rm SF}}$, and $|s_1-s_2-\frac{f_d 2^{\rm SF}}{{\rm BW}}|\propto \sqrt{2^{\rm SF}}$ \\\cline{4-6}
    
    &&& \multirow{2}{*}{$\tau\neq0$} & $f_d=0$ & SF even, $\tau\propto \frac{1}{{\rm BW}}$, and $|s_1-s_2+{\rm BW}\tau|\propto \sqrt{2^{\rm SF}}$ \\ \cline{5-6}
    
    &&&& $f_d\neq0$ & SF even, $\tau\propto \frac{1}{{\rm BW}}$, $f_d\propto\frac{{\rm BW}}{2^{\rm SF}}$, and $|s_1-s_2+{\rm BW}\tau-\frac{f_d 2^{\rm SF}}{{\rm BW}}|\propto \sqrt{2^{\rm SF}}$ \\
    \hline
    
        \multirow{4}{*}{Disc.} & \multirow{4}{*}{${\rm SF}_1={\rm SF}_2$} & \multirow{4}{*}{${\rm BW}_1={\rm BW}_2$} & \multirow{2}{*}{$\tau=0$} & $f_d=0$ & $s_1-s_2\neq 0$ \\\cline{5-6}
        
    &&&& $f_d\neq0$ &  $f_d\propto \frac{{\rm BW}_1}{2^{{\rm SF}_1}}$ and $s_1-s_2-f_d\frac{2^{{\rm SF}_1}}{{\rm BW}_1}\neq 0$ \\\cline{4-6}
    
    &&& \multirow{2}{*}{$\tau\neq0$} & $f_d=0$ & $(2^{{\rm SF}_1}-{{\rm BW}_1}\tau)(s_1-s_2+{{\rm BW}_1}\tau)\propto2^{{\rm SF}_1}$ \\ \cline{5-6}
    
    &&&& $f_d\neq0$ & $f_d\propto \frac{{\rm BW}_1}{2^{{\rm SF}_1}}$, and $(2^{{\rm SF}_1}-{{\rm BW}_1}\tau)(s_1-s_2+{{\rm BW}_1}\tau-f_d\frac{2^{{\rm SF}_1}}{{\rm BW}_1})\propto2^{{\rm SF}_1}$  \\ 
    
    \hline
\multirow{9}{*}{Dechir.} & \multirow{4}{*}{${\rm SF}_1={\rm SF}_2$} & \multirow{4}{*}{${\rm BW}_1={\rm BW}_2$} & \multirow{2}{*}{$\tau=0$} & $f_d=0$ & $s_1-s_2\neq 0$ \\\cline{5-6}

    &&&& $f_d\neq0$ & $f_d\propto \frac{{\rm BW}_1}{2^{{\rm SF}_1}}$ and $s_1-s_2-2f_d\frac{2^{{\rm SF}_1}}{{\rm BW}_1}\neq 0$  \\\cline{4-6}
    
    &&& \multirow{2}{*}{$\tau\neq0$} & $f_d=0$ & $(2^{{\rm SF}_1}-{{\rm BW}_1}\tau)(s_1-s_2)\propto2^{{\rm SF}_1}$ \\ \cline{5-6}
    
    &&&& $f_d\neq0$ & $f_d\propto\frac{{\rm BW}_1}{2^{{\rm SF}_1}}$ and $(2^{{\rm SF}_1}-{{\rm BW}_1}\tau)(s_1-s_2-2f_d\frac{2^{{\rm SF}_1}}{{\rm BW}_1})\propto2^{{\rm SF}_1}$ \\\cline{2-6}
    
    & \multirow{4}{*}{${\rm SF}_2={\rm SF}_1+\xi$} & \multirow{4}{*}{${\rm BW}_2=2^\xi {\rm BW}_1$} & \multirow{2}{*}{$\tau=0$} & $f_d=0$ & $f_d\propto\frac{{\rm BW}_1}{2^{{\rm SF}_1}}$ and $(s_1-\frac{s_2}{2^\xi}+(1-2^\xi)2^{{\rm SF}_1})\neq0$ \\\cline{5-6}
    
    &&&& $f_d\neq0$ & $f_d\propto\frac{{\rm BW}_1}{2^{{\rm SF}_1}}$ and $(s_1-\frac{s_2}{2^\xi}-2f_d\frac{2^{{\rm SF}_1}}{{\rm BW}_1}+(1-2^\xi)2^{{\rm SF}_1})\neq0$  \\\cline{4-6}
    
    &&& \multirow{2}{*}{$\tau\neq0$} & $f_d=0$ & $(2^{{\rm SF}_1}-{{\rm BW}_1}\tau)(s_1-\frac{s_2}{2^\xi}+(1-2^\xi)2^{{\rm SF}_1})\propto2^{{\rm SF}_1}$ \\ \cline{5-6}
    
    &&&& $f_d\neq0$ & $f_d\propto\frac{{\rm BW}_1}{2^{{\rm SF}_1}}$ and $(2^{{\rm SF}_1}-{{\rm BW}_1}\tau)(s_1-\frac{s_2}{2^\xi}-2f_d\frac{2^{{\rm SF}_1}}{{\rm BW}_1}+(1-2^\xi)2^{{\rm SF}_1})\propto2^{{\rm SF}_1}$\\\cline{2-6}
    
    & ${\rm SF}_1\neq {\rm SF}_2$ & ${\rm BW}_1\neq {\rm BW}_2$ & $\forall{\tau}$ & $\forall{f_d}$ & $f_d\propto \frac{{\rm BW}_1}{2^{{\rm SF}_1}}$ and $\begin{aligned}[t] 
    &(2^{{\rm SF}_1}-{{\rm BW}_1}\tau)\Big{[}s_1-\frac{{{\rm BW}_2}2^{{\rm SF}_1}}{{{\rm BW}_1}2^{{\rm SF}_2}}s_2-2f_d\frac{2^{{\rm SF}_1}}{{\rm BW}_1} \\
    &+(1-\frac{{\rm BW}_2}{{\rm BW}_1})2^{{\rm SF}_1}\Big{]}\propto 2^{{\rm SF}_1}\end{aligned}$\\
    \hline
  \end{tabular}
\end{table*}
\subsection{CSS Waveform and Spectral Analysis}
Spectral analysis in CSS modulation is of profound significance as it provides insights into the frequency distribution of the transmitted signal. By examining the spectral characteristics, such as the spread of energy across different frequencies, one can assess the modulation's efficiency and bandwidth utilization. Based on the analytical derivation of the power spectrum of CSS modulation provided in \cite{Chiani_2019}, the power spectral density of the CSS signal $x_{s_n}\left(t\right)$ includes a continuous and a discrete part as follows:
\begin{equation}
\label{18}
S\left(f\right)=S_c\left(f\right)+S_d\left(f\right),
\end{equation}
where
\begin{equation}
\label{19}
\begin{split}
S_c\left(f\right)=&\frac{1}{T_{{\rm{sym}}}M}\times\\
&\left[\sum_{s_n=0}^{M-1}\left|X_{s_n}\left(f\right)\right|^2-\frac{1}{M}\left|\sum_{s_n=0}^{M-1}X_{s_n}\left(f\right)\right|^2\right],
\end{split}
\end{equation}
and
\begin{equation}
\label{20}
\begin{split}
S_d\left(f\right)=&\frac{1}{T_{{\rm{sym}}}^2M^2}\times\\
&\sum_{k=-\infty}^{\infty}{\left|\sum_{s_n=0}^{M-1}X_{s_n}\left(k\frac{{\rm{BW}}}{M}\right)\right|^2\delta\left(f-k\frac{{\rm{BW}}}{M}\right)},
\end{split}
\end{equation}
in which $\left\{X_{s_n}\left(f\right)\right\}_{s_n=0}^{M-1}$ are the CTFT of the continuous-time CSS signals $\left\{x_{s_n}\left(t\right)\right\}_{s_n=0}^{M-1}$. To find the CTFT $\left\{X_{s_n}\left(f\right)\right\}_{s_n=0}^{M-1}$, one can exploit Fresnel functions \cite{Int_2014} and obtain the following equation \cite{Chiani_2019}:
\begin{equation}
\label{21}
\begin{split}
X_{s_n}\left(f\right)&=W\left[{\rm{BW}}\left(\frac{{s_n}}{M}-\frac{1}{2}\right)-f;\frac{{\rm{BW}}^2}{2M};0;\frac{M-{s_n}}{\rm{BW}}\right]\\
&+W\left[{\rm{BW}}\left(\frac{{s_n}}{M}-\frac{3}{2}\right)-f;\frac{{\rm{BW}}}{2M};\frac{M-{s_n}}{{\rm{BW}}};\frac{M}{{\rm{BW}}}\right],
\end{split}
\end{equation}
where
\begin{equation}
\label{22}
\begin{split}
W\left(x;y;t_1;t_2\right)=&\frac{1}{\sqrt y}e^{-\frac{j2\pi x^2}{4y}}\Bigg{[}K\left(2\sqrt y\left(t_2+\frac{x}{2y}\right)\right) \\
&-K\left(2\sqrt y\left(t_1+\frac{x}{2y}\right)\right)\Bigg{]},
\end{split}
\end{equation}
in which $K\left(.\right)\triangleq C\left(.\right)+jS\left(.\right)$ and $C\left(.\right)$ and $S\left(.\right)$ are Fresnel integrals \cite{Int_2014}. 

The discrete spectrum in (\ref{18}) is due to the mean value of the signal. For CSS modulation, this mean value is not equal to zero. The total power of the discrete spectrum is exactly equal to $1/M$ of the total signal power \cite{Chiani_2019}. As an example for the total power of the discrete spectrum, for the case of ${\rm{SF}}=7$ which yields $M=128$, $0.78125\%$ of the signal power is carried in the discrete spectrum. 

\begin{figure}[t!]
  \centering
\includegraphics[width=0.5\textwidth]{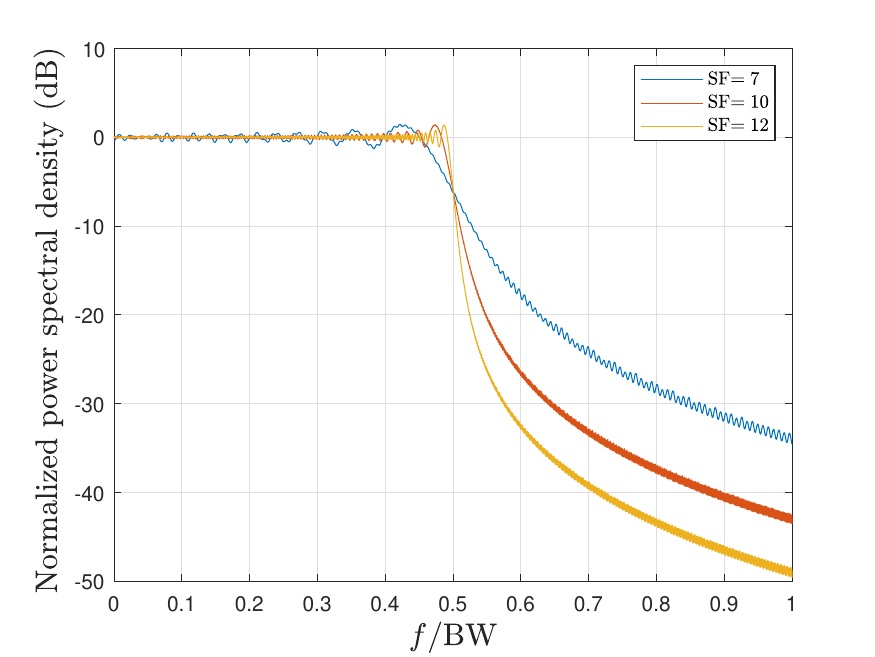}
  \caption{Continuous spectrum of the complex envelope for CSS modulation with different values of SF}
  \label{S_C}
\end{figure}

\begin{figure}[t!]
  \centering
\includegraphics[width=0.5\textwidth]{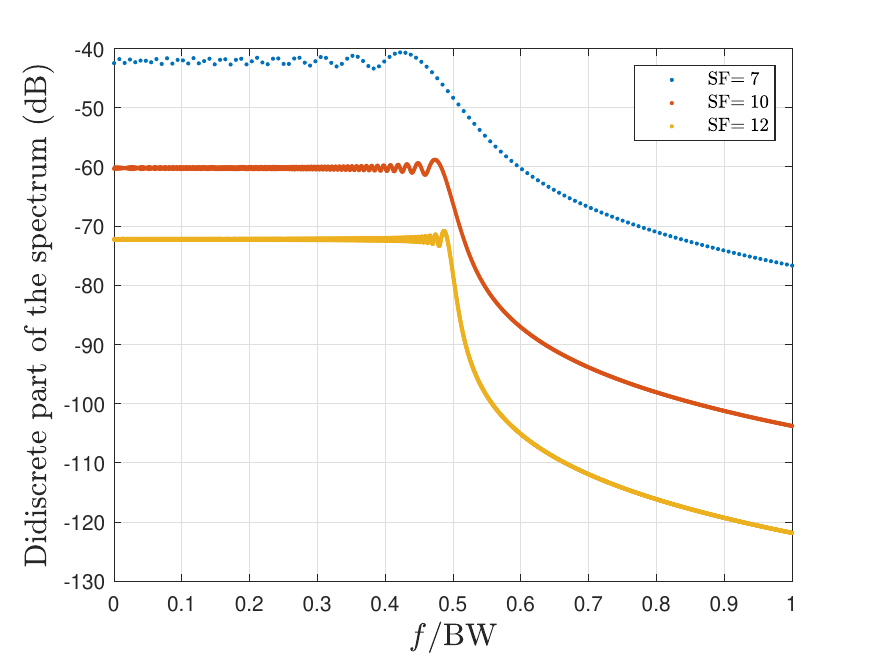}
  \caption{Discrete spectrum of the complex envelope for CSS modulation with different values of SF.}
  \label{S_D}
\end{figure}

In Fig. \ref{S_C} and Fig. \ref{S_D}, the normalized power spectral density, $10\log_{10}({S_c(f)}{\rm BW})$, and the discrete part of the spectrum of the CSS modulation, i.e., the power $10\log_{10}[|\sum_{s_n=0}^{M-1}X_{s_n}\left(k\frac{{\rm{BW}}}{M}\right)|^2/{T_{{\rm{sym}}}^2M^2}]$, are plotted, respectively \cite{Chiani_2019}, for different values of SFs. As can be seen, by increasing the value of $M$, the out-of-band radiation will be reduced\footnote{Please note that in Fig. \ref{S_D} as the SF increases, the distance between the samples decreases, causing the discrete power spectrum to not appear as a discrete-time curve, however, as can be seen, for ${\rm SF}=7$, these sample points are perfectly distinguishable.}. 

\subsection{CSS Modulation Error Performance}
\label{BER}
Here in this section, a thorough analysis of CSS modulation performance in terms of bit error rate (BER) is provided for additive white Gaussian noise (AWGN) and fading channel conditions. Generally, the discrete-time received signal can be formulated as follows:
\begin{equation}
\label{23}
r_{s_n}\left[k\right]=hx_{s_n}\left[k\right]+w\left[k\right],
\end{equation}
where $w\left[k\right]\sim\mathcal{CN}\left(0,\sigma^2\right)$ is a zero-mean AWGN sample and $h=\alpha\exp{\left(j\psi\right)}$ represents the complex channel coefficient (in which $\alpha$ denotes the channel attenuation and $\psi$ represents the phase shift). It is worth noting that the transmitter generates the chirp signal with a specific SF parameter known to the receiver. The dechirping process for demodulation of the received signal can be expressed as follows:
\begin{equation}
\label{24}
r_{s_n}\left[k\right]\times x^*_0[k]=hx_{{s_n}}\left[k\right]\times x^*_0[k]+\hat{w}\left[k\right],
\end{equation}
where $\hat{w}\left[k\right]$ is the dechirped version of AWGN signal ($\hat{w}\left[k\right]\sim\mathcal{CN}\left(0,\sigma^2\right)$).
After some mathematical operations and simplifications, we obtain:
\begin{equation}
\label{25}
r_{n}^{{\rm{(dech.)}}}\left[k\right]=hA_0^2\exp{\left[j2\pi\frac{k\left(s_n+M\right)}{M}\right]}+\hat{w}\left[k\right].
\end{equation}
Now, by taking an $M$-point discrete Fourier transform (DFT) of $r_{s_n}^{(\rm dech.)}$, the frequency domain signal is obtained as:
\begin{equation}
\label{26}
\begin{split}
V_{s_n}\left[u\right]=&\frac{1}{\sqrt M}\sum_{k=0}^{M-1}{r_{{s_n}}^{{\rm{(dech.)}}}\left[k\right]\exp{\left(-\frac{j2\pi uk}{M}\right)}}\\
=&\frac{1}{\sqrt M}\sum_{k=0}^{M-1}\left\{h A_0^2\exp{\left[j2\pi\frac{k\left(s_n+M\right)}{M}\right]}+{\hat{w}}\left[k\right]\right\}\\
&\times\exp{\left(-\frac{j2\pi uk}{M}\right)}\\
=&{\frac{1}{\sqrt M}}hA_0^2\sum_{k=0}^{M-1}\underbrace{\exp{\left\{j2\pi\frac{k\left(s_n+M-u\right)}{M}\right\}}}_{\text{is periodic with the period of $M$}}+W\left[u\right]\\
=&  \begin{cases}
\sqrt M \alpha e^{j\psi} A_0^2+W\left[u\right], & {\rm{if}}\ u=s_n, \\
                W\left[u\right], & {\rm{otherwise.}}
              \end{cases}
\end{split}
\end{equation}
From above results, one can obtain the detected symbol ${\hat{s}}_{n}$ using coherent detection as:
\begin{equation}
\label{27}
{\hat{s}}_{n}^{\rm{(coh.)}}= \argmax_{u = 0,1, \ldots ,\;M - 1} {V_{n}\left[u\right]\times \exp(-j\psi)}.
\end{equation}
Moreover, the non-coherent detection rule is formulated as follows:
\begin{equation}
\label{28}
{\hat{s}}_{n}^{\rm{(non-coh.)}}= \argmax_{u = 0,1, \ldots ,\;M - 1} {\left|V_{n}\left[u\right]\right|^2}.
\end{equation}

\begin{figure}[t!]
  \centering
\includegraphics[scale=0.6]{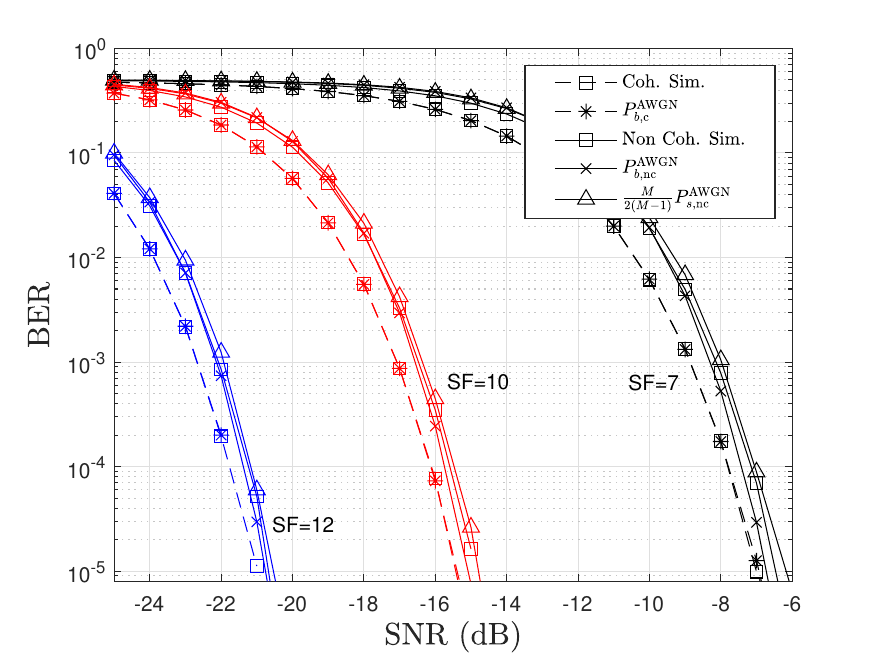}
  \caption{BER performance of the CSS modulation under AWGN condition for SF values of $7$, $10$, and $12$}
  \label{BER_AWGN}
\end{figure}

Assuming $h=1$ in (\ref{23}), our channel reduces to an AWGN channel. Several works exist in the literature that obtains closed-form expressions for CSS modulation performance in terms of BER and symbol error rate (SER) under AWGN condition \cite{TTNguyen_2019, Ferre_2018, Elshabrawy_2018, Vangelista_2017, Marquet_2019, Mroue_2018, Afisiadis_2020}. As a few examples, in \cite{TTNguyen_2019}, the theoretical
BER under coherent detection of CSS modulation in AWGN channels is identical to frequency-shift keying (FSK) modulation of the same cardinality (as both are $M$-ary orthogonal modulations) and is expressed as:  
\begin{equation}
\label{29}
\begin{split}
{P_{b,{\rm{c}}}^{\rm (AWGN)}} =& \frac{M}{{2\left( {M - 1} \right)}}\\
&\Bigg{[} 1 - \frac{1}{{\sqrt {2\pi } }}\mathop \int \nolimits_{ - \infty }^\infty  {{\left( {\frac{1}{{\sqrt {2\pi } }}\mathop \int \nolimits_{ - \infty }^y \exp \left\{ {\frac{{ - {x^2}}}{2}} \right\}{\rm{d}}x} \right)}^{M - 1}}\\
& \times \exp \left\{ { - \frac{1}{2}{{\left( {y - \sqrt {\frac{{2{{\log }_2}M{E_b}}}{{{N_0}}}} } \right)}^2}} \right\}{\rm d}y \Bigg{]},
\end{split}
\end{equation}
where ${E_b}/{N_0} = {\rm{SNR}}\times{M}/{\rm{SF}}$. Moreover, in \cite{Ferre_2018} an approximation of the bit error probability of CSS modulation under non-coherent detection is provided. In particular, the authors exploit Monte Carlo approximation for generating $N_{\rm MC}$ noise samples denoted by $W^{\left(i\right)}$ ($i=1,2,\dots,N_{\rm MC}$) following Gaussian distribution (because of the mathematical limitations in obtaining a closed-form expression for the error probability) and obtain the symbol error probability as:
\begin{equation}
\label{30}
\begin{split}
P_{s,{\rm{nc}}}^{\rm (AWGN)}\approx&\frac{1}{N_{MC}}\times \\
&\sum_{i=1}^{N_{MC}}{1-\left[F_{\chi^2}\left(\frac{\left|\sqrt N+W^{\left(i\right)}\right|^2}{{N_0}/{2}}\right)\right]^{N-1}},
\end{split}
\end{equation}
where $F_{\chi^2}\left(.\right)$ is the cumulative distribution function (CDF) of the Chi-square distribution. Another BER expression is also presented in \cite{Elshabrawy_2018} based on multiple approximations as follows:
\begin{equation}
\label{31}
P_{b,{\rm{nc}}}^{\rm (AWGN)}\approx0.5\mathcal{Q}\left(\sqrt{{\rm{SNR}}\times2^{{\rm{SF}}+1}}-\sqrt{1.386{\rm{SF}}+1.154}\right).
\end{equation}
Fig. \ref{BER_AWGN} is presented to show the CSS modulation performance under AWGN conditions for different values of SF and also to compare the above BER expressions with simulation results. As can be seen, the BER curve corresponding to the approximation (\ref{31}) is more accurate than (\ref{30}). Moreover, by decreasing the value of SF from $12$ to $7$, CSS modulation experiences performance degradation in terms of BER. 

\begin{figure}[t!]
  \centering
  \includegraphics[scale=0.63]{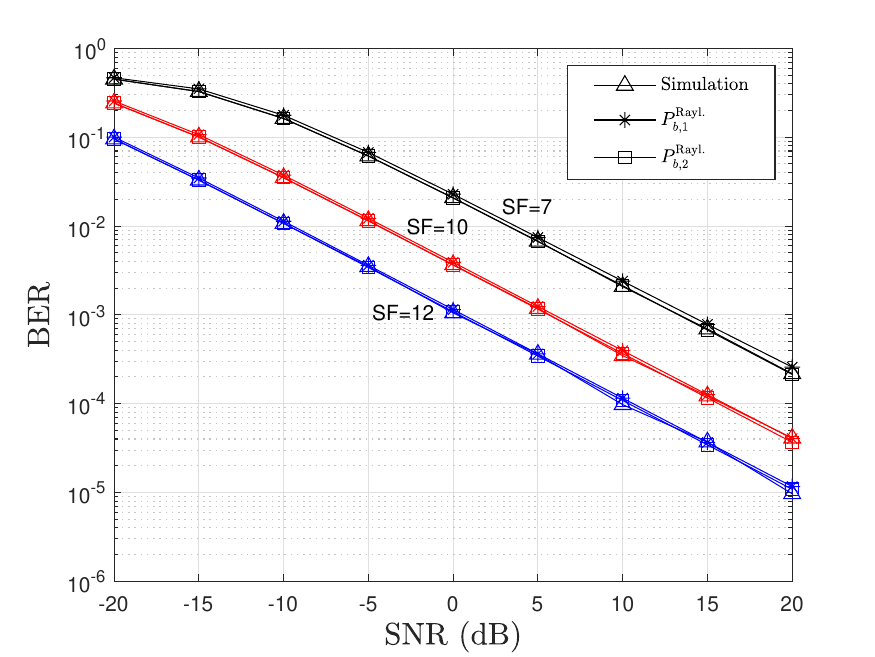}
  \caption{BER performance of the CSS modulation in Rayleigh channel for non-coherent detectors.}
  \label{BER_Rayleigh}
\end{figure}

Next, we consider the Rayleigh flat fading channel for evaluating the BER performance of CSS modulation. We assume $h=\alpha \exp(j\psi)$ in which $\alpha$ and $\psi$ have Rayleigh and Uniform (in $[0,2\pi]$) distributions, respectively. Since it is a challenging task to obtain perfect channel state information (CSI) in practical IoT networks, coherent detection can be achieved with extra training overhead which reduces the spectral efficiency (SE). Consequently, here we investigate non-coherent signal detection. As mathematically derived in \cite{Khai_2021}, the symbol error probability for the CSS modulation in the Rayleigh fading channel is formulated as:
\begin{equation}
\label{32}
\begin{split}
{P_{b,1}^{\rm{(Rayl.)}}} =& \frac{M}{2M-2}\\
&\times \Bigg{[}1 - \mathop \int \nolimits_0^\infty  \frac{1}{{1 + \overline {{\gamma _c}} }}\exp \left( { - \frac{{{y}}}{{1 + \overline {{\gamma _c}} }}} \right)\\
&\times {\left( {1 - \exp \left( {{y}} \right)} \right)^{M - 1}}{\rm d}{y}\Bigg{]},
\end{split}
\end{equation}
where $\overline {{\gamma _c}}=M\times\rm{SNR}$. In addition to this expression, a tight closed-form approximation of bit error probability for the Rayleigh fading scenario is derived in \cite{Elshabrawy_2018} as:
\begin{equation}
\label{33}
\begin{split}
{P_{b,2}^{\rm{(Rayl.)}}} \approx& 0.5 \times \Bigg{[} Q (-\sqrt{2H_{M-1}}) -\sqrt{\frac{\overline {{\gamma _c}}}{\overline {{\gamma _c}}+1}}\\
&\times \exp \left(-\frac{2H_{M-1}}{2(\overline {{\gamma _c}}+1)} \right) \times Q\Bigg{(} \sqrt{\frac{\overline {{\gamma _c}}}{\overline {{\gamma _c}}+1}}\\
&\left[-2\sqrt{2H_{M-1}}+\frac{\sqrt{2H_{M-1}}}{\overline {{\gamma _c}}+1}\right]\Bigg{)}\Bigg{]},
\end{split}
\end{equation}
where ${H_m} \approx \ln (m) + 1/(2m) + 0.57722$. BER curves for the Rayleigh fading scenario are presented in Fig. \ref{BER_Rayleigh}. As can be seen, the BER for all three cases of simulation, numerical based on the equation (\ref{32}) and approximation based on the equation (\ref{33}) are matching.
\section{Taxonomy of Research Topics}
\label{section:Tax}
\begin{figure*}[t!]
  \centering
  \includegraphics[width=\textwidth]{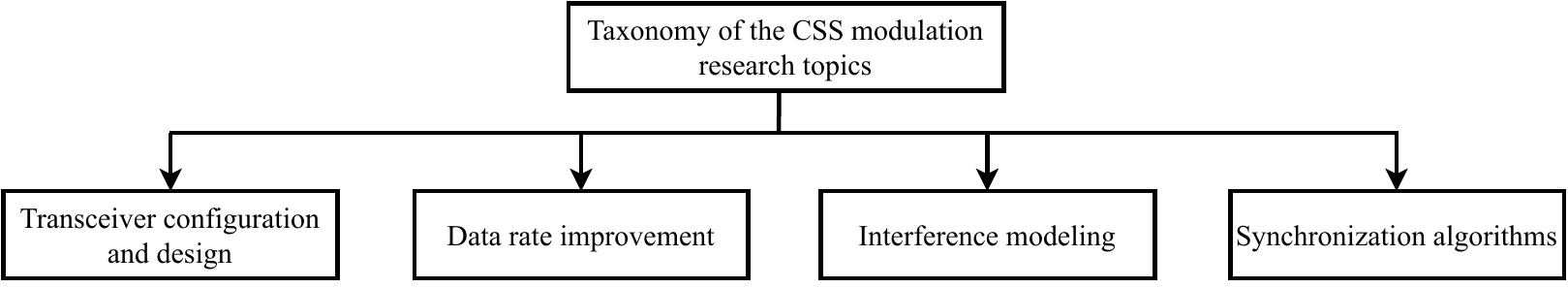}
  \caption{Research topics taxonomy of the CSS modulation with IoT application.}
  \label{Taxonomy}
\end{figure*}

So far, we have discussed the basic mathematical foundation related to CSS modulation and shed light on its SE characteristics, as well as its BER performance in different channel conditions. In the following section, we will investigate recent key advances existing in the literature related to the applications of CSS modulation in the IoT industry. A categorization has been made, as indicated in Fig. \ref{Taxonomy}, based on which these existing research studies are divided into four main groups, each of which will be briefly discussed in this section.  
\subsection{Transceiver Configuration and Design}
By using the equation (\ref{12}), it can be observed that CSS transmission is based on a look-up table (LUT) design that stores the phase samples of the reference up-chirp for $\rm{SF}=12$. However, despite its seemingly straightforward design, the flexibility requirement of CSS modulation would result in more expensive transceivers, as additional hardware resources are required to enable different SF and BW settings. Generally, the size of the LUT grows exponentially as SF increases. While this may not be a major issue for currently used $\rm{SF}=12$, which is supported by market LoRa transceivers, the LUT design will become prohibitively expensive if greater SF values are sought in the future. To tackle this issue, a novel receiver structure is presented which will be discussed in detail in Section ~\ref{section:TxRx}. 

Furthermore, due to the increasing need for deploying a vast number of IoT devices, traditional terrestrial LoRa networks may become highly dense, and in some regions (such as across oceans), they may not be deployable. Therefore, researching the CSS modulation performance for low Earth orbit (LEO) satellite IoT communication applications is imperative.

Moreover, the performance of CSS modulation in fading channels significantly degrades, and it remains a challenging task to exploit CSS modulation for these environments. To overcome this problem, a very new approach integrating CSS modulation with multiple-input multiple-output (MIMO) configuration has been recently proposed which will be discussed in more detail in Section \ref{section:TxRx}.     
\subsection{Data Rate Improvement}
The bit rate of the LoRa transmission scheme is calculated using the following equation:
\begin{equation}
    r_{b}={\rm SF}\times \frac{{\rm BW}}{M}.
\end{equation}
Due to the limited number of orthogonal chirps available in a given bandwidth and symbol duration, long-range communication provided by CSS modulation techniques comes at the cost of very low data rates in the order of a few hundred to a few thousand bits per second \cite{Elshabrawy_2019}. The values of SF and BW are the two main factors that influence the data rate of CSS modulation. For example, for $\rm{SF}=7$ and $\rm{BW}=500$ kHz, the highest data rate is $37.5$ kbps \cite{SemTech_Modem_Design}. Although this is suitable for most smart city applications in IoT networks, many other potential applications, such as smart home/building, image transmission, and indoor IoT, cannot be supported. Therefore, efforts have been made to improve the data rate of CSS-based transmission schemes. One approach to address the low data rate issue presented in \cite{SemTech_Modem_Design} is to exploit multiple transceivers in a device. However, this adds complexity, as several transceivers in a device require more hardware and software resources, particularly at the receiver end. Hence, the research trend in this area tends to modify CSS modulation to improve the data rate. Several works exist in the literature modifying traditional CSS modulation to embed more information bits in the transmitting symbol by combining it with other modulation schemes. These works are comparatively presented and discussed in Section ~\ref{section:DR}. 
\subsection{Interference Modeling}
Decoding the received CSS signals in the presence of interference has been addressed in the literature, as packet collisions are common due to the extended transmission times of LoRa packets and the ALOHA-based protocol. As the number of LoRa devices grows in the future, this problem will worsen, putting the scalability of LoRa networks at risk as they become interference-constrained. In existing works discussing the interference in LoRa networks, it is assumed that destructive collisions occur when two or more signals are received in the same frequency band and with the same SF because CSS signals with different spreading factors are practically orthogonal since the rejection gain ranges from $16$ to $36$ dB \cite{Georgiou_2017}. On the other hand, as discussed in Section \ref{section:Base}, the authors in \cite{Benkhelifa_2022} have thoroughly discussed the orthogonality of CSS modulation based on mathematically derived cross-correlation functions and it has been confirmed that CSS modulation orthogonality relies on many factors such as the value of the SF, bandwidth, and time delay. We will review several works discussing the interference in LoRa networks and investigate their proposed solutions in Section \ref{section:Int}.
\subsection{Synchronization Algorithms}
Considering that a LoRa frame has a specific pattern of CSS symbols before the payload part, after receiving a LoRa frame at the receiver, the first step is to detect the preamble part and synchronization. In most practical cases, a CSS receiver must compensate for carrier frequency offsets (CFO), symbol time offsets (STO), and sampling frequency offsets (SFO). The following definitions are presented to provide a clear understanding of these parameters:
\begin{itemize}
    \item CFO: Inexpensive crystal oscillators inherently exhibit a discrepancy from their intended frequency value. As a result, the process of down-conversion takes place using a frequency different from that used for up-conversion. In the context of CSS signals, this results in CFO and is revealed as a shift in the swept BW of the CSS signal regenerated at the receiver and can be introduced as a multiply of a CSS signal frequency bin, e.g., a shift of $4.1$ frequency bins \cite{Bernier_2020}.
    \item STO: When the receiver's clock for sampling the received signal is not perfectly synchronized with the transmitter's clock, it leads to misalignment in the sampling instants. So the regenerated CSS signal at the receiver experiences a shift in the time domain compared to the transmitted version. The STO can be measured as a shift equal to the multiply of a CSS sample, e.g., a shift of $4.1$ samples \cite{Bernier_2020}. 
    \item SFO: The mismatch between the oscillators of the transmitter and receiver leads to different sampling rates. As a consequence, the CSS signal acquired through sampling at the receiver encounters a sampling frequency offset. For example, if the receiver uses a higher sampling frequency than $f_s$, an accumulating sampling discrepancy gradually shifts the samples away from the intended frequency bins. As an example \cite{Bernier_2020}, consider low-cost quartz crystals with $\pm20$ parts per million (ppm) precision at the transmitter and the receiver. In the worst case, this can result in a drift of $0.16$ sample after a CSS symbol regeneration at the receiver.
\end{itemize}
Large synchronization errors can cause significant interference between chirps, resulting in unreliable detection performance. Currently, there is limited research on synchronization in the context of CSS modulation. In Section ~\ref{section:Sync}, we review the effects of each of the mentioned factors on a CSS signal and review the current synchronization procedure during a LoRa packet transmission. Moreover, existing key works on various proposed synchronization algorithms are discussed.
\section{Key Advances in Transceiver Configuration and Design}
\label{section:TxRx}
In this section, we review key works that aim to improve the implementation and signal detection of CSS-based communication systems \cite{TTNguyen_2019, Khai_2021, Kang_2022_Pre}, redesign CSS signal transmission for adoption in satellite-based IoT scenarios \cite{Doroshkin_2019, Qian_2018, Qian_2019_Per, Roy_2019, Yang_2019}, and propose a new MIMO-CSS configuration \cite{Ma_2021, Kang_2022}.
\subsection{Implementation and Signal Detection}
Nguyen \textit{et al.} in \cite{TTNguyen_2019} present the orthogonal chirp generator (OCG), a low-cost CSS transmitter architecture that supports various SF and BW settings. Because of the flexibility of the OCG, it may be used to not only generate complex baseband or real pass-band signals at the transmitter but also to demodulate the chirp signal at the receiver. The expression for the phase function allowing the OCG to generate the CSS signal is:
\begin{equation}
\label{34}
\begin{split}
\phi^{\rm (OCG)}_{s}[k]=&\frac{1}{2ML^2}\sum\limits_{n=0}^{k}(2n+1+2sL\mod 2ML)-ML,\\
&k=0,1,\dots,ML-1
\end{split}
\end{equation}
where $L$ is the up-sampling factor, i.e., $f_s^{\rm (up)}=L\times\rm{BW}$. The OCG block diagram is shown in Fig. \ref{OCG}. As can be seen, the frequency accumulator, frequency manipulator, phase accumulator, and vector rotator are the four primary components. The first three components generate the phase function as in (\ref{34}). The frequency accumulator generates a ramp, i.e., a frequency that increases linearly over time. According to the input symbol $s$, the selected BW, and SF, the frequency manipulator introduces jumps into the frequency ramp at the symbol boundaries. To provide phase samples for the created chirp symbol, the phase accumulator conducts discrete integration of the frequency samples. Eventually, the vector rotator rotates an input vector $x_{\rm in}[k]$ by an amount of $\phi^{\rm (OCG)}_{s}[k]$ as follows:
\begin{figure}[t!]
  \centering
  \includegraphics[scale=0.44]{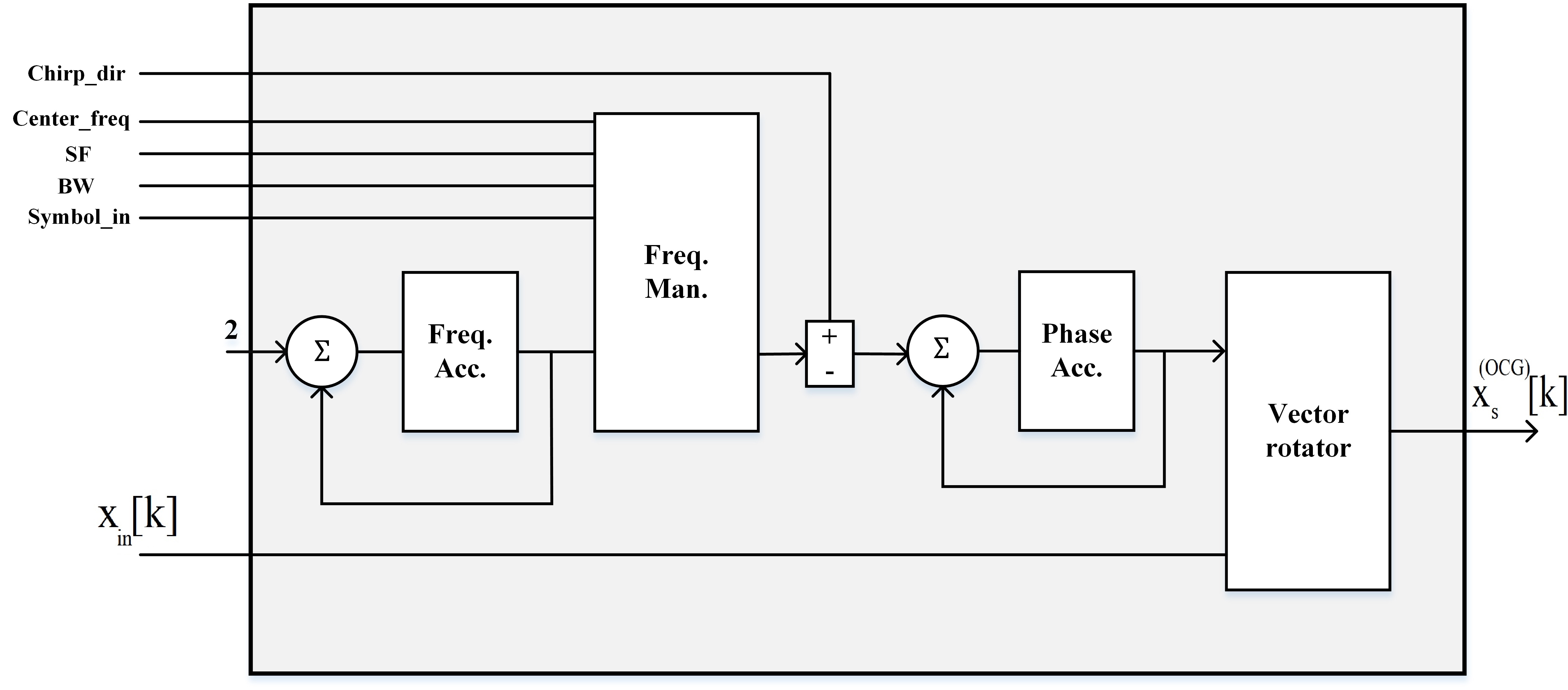}
  \caption{Orthogonal chirp generator of \cite{TTNguyen_2019}.}
  \label{OCG}
\end{figure}
\begin{equation}
\label{35}
x^{\rm (OCG)}_{s}[k]=x_{\rm in}[k]\exp\left({j2\pi \phi^{\rm (OCG)}_{s}[k]}\right).
\end{equation}
To obtain a continuous phase CSS signal, $x_{\rm in}[k]$ is set to a scalar determined by the CSS signal magnitude. Otherwise, in a case that the continuous phase is not required between two adjacent CSS symbols, $x_{\rm in}[k]$ can carry extra information as its phase. To upconvert the CSS signal to the passband, the OCG can be used in conjunction with a quadrature modulator. Also, direct digital synthesis (DDS) technology can be utilized to lower the cost of the up-converter. Moreover, an OCG-based receiver structure is presented which by means of simulation, is shown to have a much better performance than Semtech's receiver. The SNR enhancement ranges from $0.9$ dB at $\rm{SF} = 6$ to a remarkable $2.5$ dB at $\rm{SF} = 12$.

\begin{figure}[t!]
  \centering
  \includegraphics[scale=0.5]{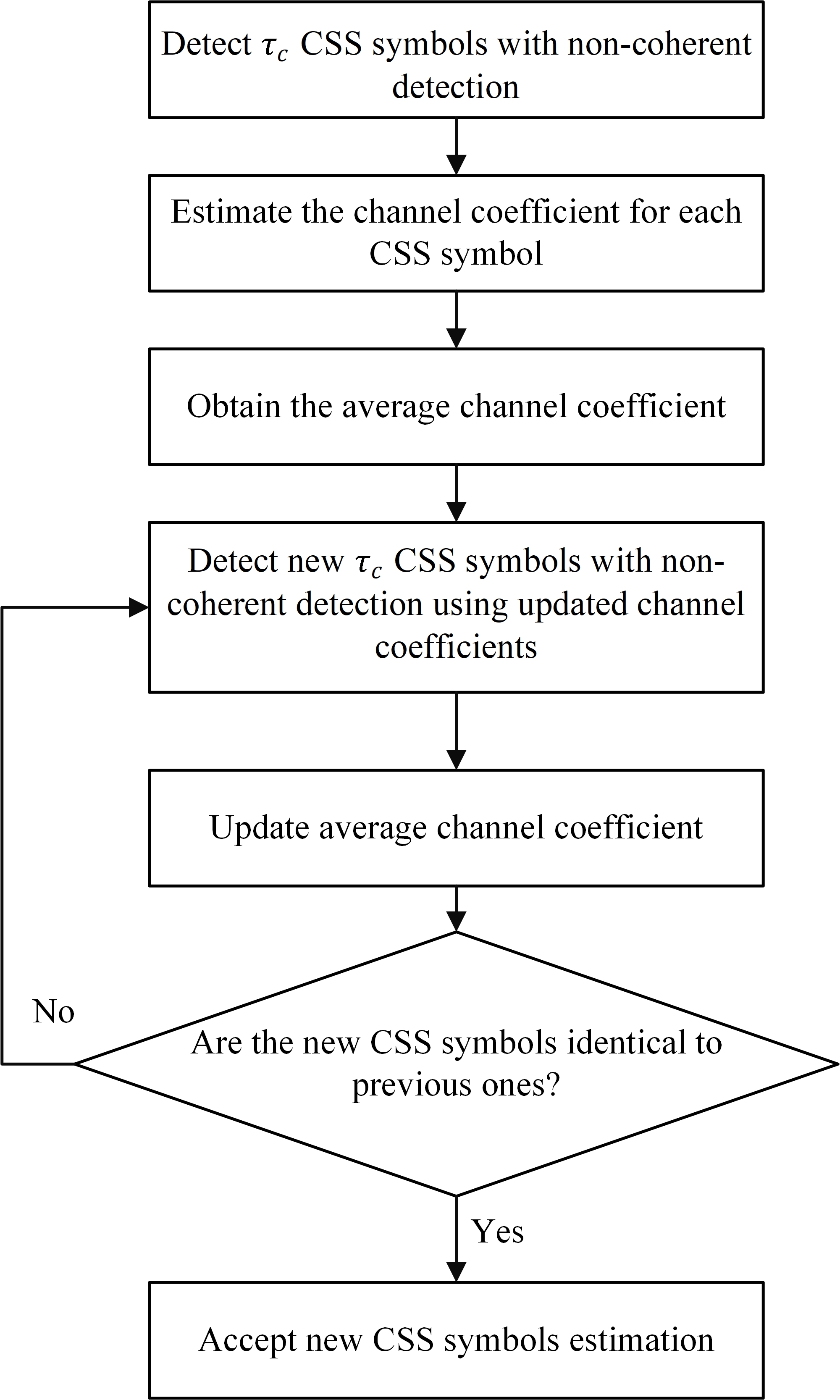}
  \caption{Semi-coherent detection algorithm flowchart of \cite{Khai_2021}.}
  \label{Semi}
\end{figure}

Using signal combining and semi-coherent detection, Nguyen \textit{et al.} in \cite{Khai_2021} offer a performance improvement strategy for CSS modulation. In their system model, the gateway is assumed to have multiple antennas rather than a single antenna. According to BER results of signal combining methods at LoRa gateways, appropriately integrating diverse incoming signals at the gateway can considerably improve CSS modulation performance, resulting in reduced transmit power and/or improved coverage for EDs. The authors also present a novel iterative semi-coherent detection strategy that outperforms non-coherent detection, especially as the number of antennas at the gateway increases. The term ``semi-coherent'' is used due to the fact that there is no extra overhead used for channel estimation, i.e., the channel is estimated using different versions of the same CSS signal all carrying information. The flowchart of the semi-coherent detection algorithm is indicated in Fig. \ref{Semi}. It is assumed that the fading channel has a coherence time of $\tau_c$ CSS symbols. When the system operates over AWGN (coherent) and Rayleigh fading (non-coherent) channels, simulation results show that increasing the number of antennas at the gateway from $1$ to $4$ results in approximately $8$ and $29$ dB savings in transmit power of the IoT ED (which translates to $2.5$ times and $27$ times larger distances for the same transmit power of the device). Moreover, the results show that the semi-coherent detection BER performance is close to the coherent case (the difference is smaller than $1$ dB).  
\begin{figure*}[t]
  \centering
  \includegraphics[scale=0.5]{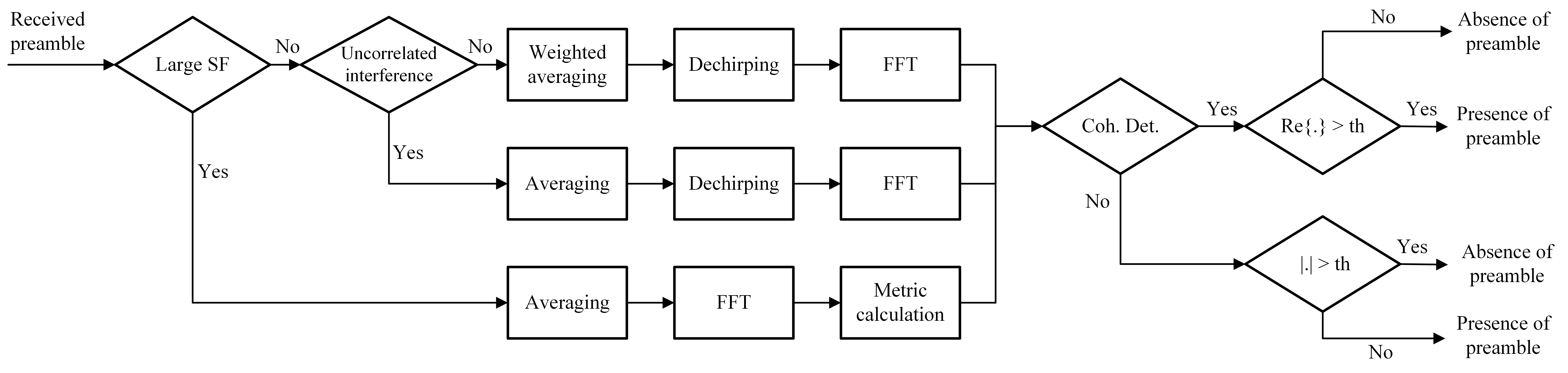}
  \caption{Preamble detection algorithm of \cite{Kang_2022_Pre}.}
  \label{Kang}
\end{figure*}

Kang \textit{et al.} \cite{Kang_2022_Pre} propose a novel optimal detection algorithm for the preamble part of CSS modulation. Every frame is started with a preamble in many CSS modulation implementations, which is known a priori at the transceiver. The authors assume that $N$ repetition of a specific CSS signal is adopted as the preamble. The purpose of this work is to come up with the best preamble detection strategy for CSS modulation in terms of maximizing detection probability with the false alarm rate constraint. The optimization problem presented in this work is non-convex in general. As can be seen in Fig. \ref{Kang}, the first branch corresponds to the general case when SF is not large and also the interference from other devices is correlated (the interference covariance matrix is not diagonal). The weighted average refers to calculating the mean sample of the received signal and left-multiplying it by the inverse of the interference covariance matrix. The rest of the operations are similar to the common CSS signal detection procedure explained in previous sections. For a special scenario of uncorrelated interference, the weighted averaging step reduces to only computing the sample mean of the received signal. However, since SF values in practice are large, the $M\times M$ interference covariance matrix has a large size making its inverse calculation very complex. To handle this issue, the authors propose a simplified version of their detection algorithm (third branch of Fig. \ref{Kang}) which does not include the weighted averaging and dechirping steps. The proposed method is resulted from wide-sense stationary (WSS) random process assumption for the interference and can be accomplished using non-heavy and basic operations like FFT, addition, and multiplication, which reduces the computing complexity of preamble detection dramatically. This outcome is desirable in many real-world CSS-modulated IoT applications. Finally, by means of simulation, it has been shown that the proposed method outperforms the state-of-the-art techniques in terms of detection probability and complexity.     

\subsection{Satellite-based IoT Scenarios}
For the application of the CSS modulation in satellite-based IoT, there are several challenges such as the effect of the Doppler shift, and the interference rejection if the network becomes dense (due to the high coverage of the satellites). Doroshkin \textit{et al.} present an experimental study of CSS modulation in \cite{Doroshkin_2019} to evaluate its robustness against the Doppler effect in LEO satellite IoT scenarios. The laboratory experiments conducted in this paper showed that the CSS modulation has high resistance to static Doppler shift (the value of the shift does not vary with time) for all SF values, while its robustness against dynamic Doppler shift (the value of the shift varies with time) highly depends on the SF value. As the SF increases, the maximum tolerable Doppler rate decreases. The outdoor experiments, where the CSS receiver was installed in a moving car, fully confirmed the results of the laboratory experiments. Finally, the experiments showed that the CSS modulation can be used in radio communication between the ground station and a satellite in a circular orbit of more than $550$ km in height without any restrictions associated with the Doppler effect.

\begin{figure}[t!]
  \centering
  \includegraphics[scale=0.6]{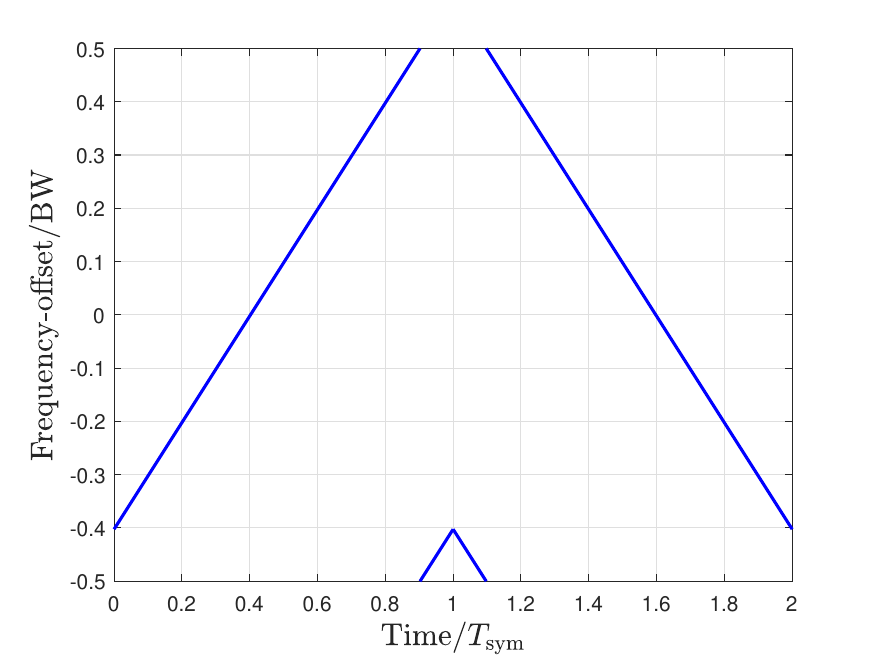}
  \caption{SCS instantaneous frequency-offset for $s=50$, $\rm{SF}=9$, $\rm{BW}=250$ kHz, and $T_{\rm sym}=2.048$ ms.}
  \label{SCSS}
\end{figure}

Qian \textit{et al.} in \cite{Qian_2018} modify the CSS modulation and propose symmetry chirp signal (SCS) to be more compatible with dense satellite-base IoT deployment in which the interference at the CSS receiver may be much larger than terrestrial scenarios. It is shown that when the difference between the starting frequency offsets of two chirps tends to zero or BW, the cross-correlation of the two chirps is large with a value close to CSS signal energy and its maximum value cannot be lower than half of CSS signal energy. Based on the fact that the cross-correlation between chirps with negative and positive rates is relatively small, the authors introduce SCS in which $T_{\rm{sym}}$ is divided into two equal periods consisting of up- and down-chirp signals respectively. Based on this modification, the resulting instantaneous frequency offset of Fig. \ref{Chir} changes to the curve in Fig. \ref{SCSS}. After evaluation of the SCS cross-correlation features, it can be seen that the SCS can obtain a maximum cross-correlation of less than half of the signal energy. Also, the authors investigate the effect of the Doppler frequency shift (DFS) on the SCS modulation scheme in \cite{Qian_2019_Per} which results in a high ambiguity function (AF) value. Therefore, an asymmetry chirp signal (ACS) is obtained as a solution for the optimization of AF. Using simulation, it has been shown that ACS can achieve better BER and acquisition performance compared to SCS in the presence of DFS.   

Roy and Nemade in \cite{Roy_2019} design a new set of chirps namely symmetry chirp with multiple chirp rates (SC-MCR) to reduce the overall cross-correlation level caused by possible delays in LEO satellite communication. The main difference between SC-MCR and SCS is to use different chirp rates for each half of the $T_{\rm{sym}}/2$ period. It is shown that the cross-correlation of SC-MCR is lower than the SCS method. Thus, for a multiuser communication scenario, SC-MCR provides less multiple access interference (MAI) compared to SCS. Further, the combination of up-SC-MCR and down-SC-MCR through time domain multiplexed SC-MCR (TDM SC-MCR) is presented which allows an increase in the overall system data rate without much degradation in performance. Thus, SC-MCR waveforms are able to provide a more reliable LEO satellite IoT communication and maintain a higher data rate compared to that of SCS waveforms.

\begin{figure*}[t!]
  \centering
  \includegraphics[scale=0.55]{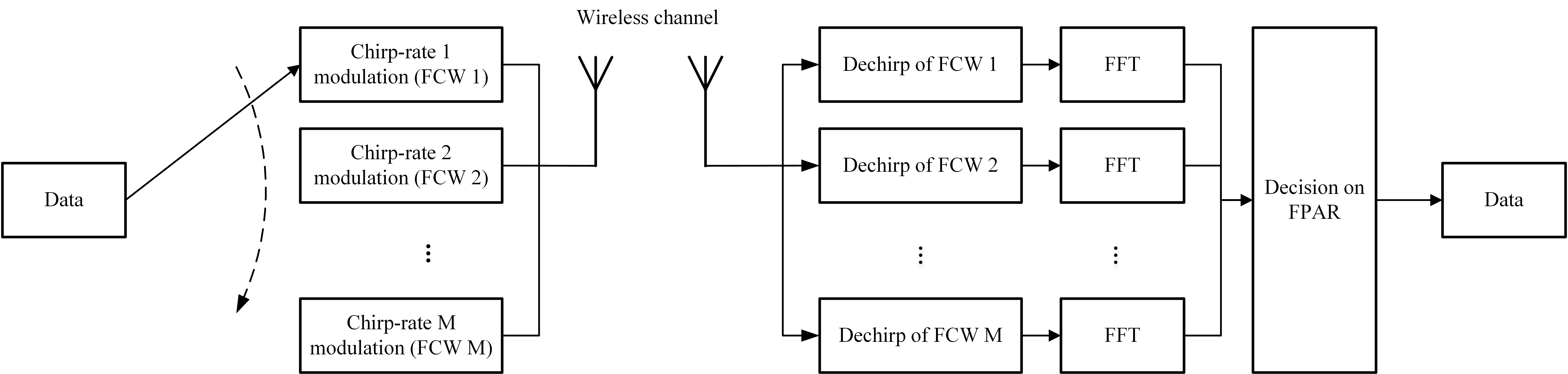}
  \caption{FCrSK transceiver block diagram \cite{Yang_2019}.}
  \label{FCW}
\end{figure*}

Yang \textit{et al.} in \cite{Yang_2019} introduce folded chirp-rate shift keying (FCrSK) modulation for the application in satellite-based IoT networks. The authors discuss two common problems existing in chirp-rate shift keying systems. First, for chirp signals with identical $T_{\rm{sym}}$, different chirp rates can be achieved using different BWs which results in different spreading gains. To avoid sampling distortion, the sampling rate must be set to handle the largest rate existing among chirp signals. This translates to an increase in the burden on analog-to-digital converter (ADC). Second, identical signals with chirp rates resulted in different $T_{\rm{sym}}$s making the symbol synchronization a challenging task at the receiver. To deal with these issues, folded chirp waveform (FCW) is proposed as:
\begin{equation}
\label{36}
\begin{split}
x^{\rm{(FCW)}}(t)=&A_0\sum_{n=0}^{n_c-1} \exp{(j\pi{\frac{\rm{BW}}{T_c}}t^2-2j\pi n \rm{BW}t)}\\
&\times g_{T_c}\left(t-nT_c\right),
\end{split}
\end{equation}
where the complete FCW is combined of $n_c$ chirp-chips each has a duration of $T_c$ (${n_c}{T_c}=T_{\rm{sym}}$). The block diagram of the complete FCrSK system is indicated in Fig. \ref{FCW}. After converting the binary data into the decimal value of $s\in\mathcal{S}$ similar to CSS modulation, the results are mapped into one FCW out of $M$ possible waveforms using the following equation:
\begin{equation}
\label{37}
\begin{split}
x^{\rm (FCW)}(t,s)=&A_0\sum_{n=0}^{n_c-1} \exp{(j\pi s{\frac{\rm{BW}}{T_{\rm{sym}}}}t^2-2j\pi n \rm{BW}t)}\\
&\times g_{T_c}\left(t-nT_c\right).
\end{split}
\end{equation}
Followed by dechirping and $M$-point DFT procedures, the decision is made based on frequency peak average rate (FPAR) which is insensitive to the Doppler effect. Due to the interference between distinct chirp rates, FCrSK performs somewhat worse than the CSS modulation without Doppler shift, however, FCrSK outperforms CSS modulation against large Doppler shift or large Doppler rate, as demonstrated by simulation and experimental data. Moreover, it can maintain a sufficient BER without frequency synchronization even when Doppler is very large and Doppler variation is very fast making it appropriate for satellite-based IoT applications.  
\subsection{MIMO-CSS}
To tackle the issue of poor performance of CSS modulation in the presence of fading, Ma \textit{et al.} in \cite{Ma_2021} propose a novel MIMO configuration exploiting space-time block coding (STBC). The authors consider a $N_t\times N_r$ MIMO system in which the CSS symbols are coded with a complex orthogonal STBC transmission $U_c\times N_t$ matrix. Note that this matrix is utilized to encode $J_c$ input symbols ($J_c\geq 2$) into an $N_t$-dimensional vector sequence of $U_c$ time slots, with the transmission rate of $J_c/U_c$. The encoding algorithm for STBC-MIMO CSS is summarized in the following steps:
\begin{enumerate}
	\item Initializing $N_t$ and SF
	\item Performing CSS modulation on information data bits
	\item Selecting the STBC transmission matrix ($U_c\times N_t$) and generating the matrix $\mathfrak{I}_{U_c\times {N_t}{M}}$ which its entries are linear combinations of transmitted CSS symbols and their conjugates.
	\item Transmitting CSS signals in the $u_c$th row, $[(n_t-1)M+1]$th column to ${n_t}{M}$th column of $\mathfrak{I}$ via $n_t$th antenna ($1\leq n_t \leq N_t$), for each time-slot $1\leq u_c \leq U_c$
\end{enumerate}
At the receiver end, the decoding algorithm consists of the following steps:
\begin{enumerate}
	\item Initializing $N_r$ and SF
	\item Performing channel estimation exploiting CSS preamble (the authors assume both cases of perfect and imperfect CSI)
	\item Canceling the channel effect from the received signal in each time slot for each antenna in a symbol-by-symbol manner
	\item Summing up the resulted signals over all time-slots and antennas for each symbol
	\item Performing CSS demodulation
\end{enumerate}
From the simulation results provided in \cite{Ma_2021}, it can be seen that the use of the STBC-MIMO scheme can considerably improve the diversity gain of the CSS-based communication system. Moreover, as an example, the STBC-MIMO CSS with $2$ transmit antennas and one receiver antenna outperforms conventional CSS modulation by $16$ dB at the BER of $10^{-4}$ in the perfect CSI scenario. 

\begin{figure}[t!]
  \centering
  \includegraphics[scale=0.5]{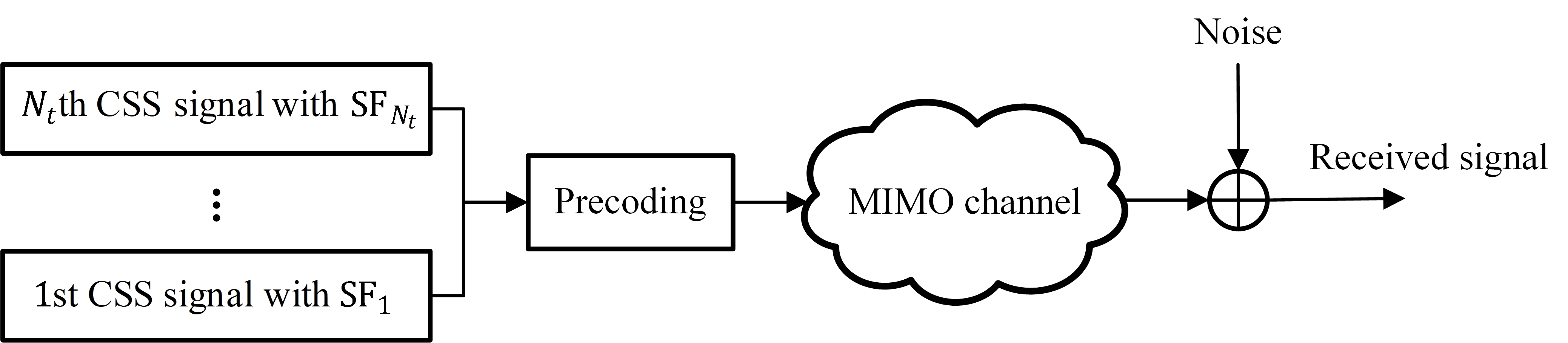}
  \caption{MIMO-CSS system model of \cite{Kang_2022}.}
  \label{MIMO}
\end{figure}

Recently, Kang in a short paper \cite{Kang_2022} present the concepts and precoding design for MIMO-based CSS modulation with the application in high data rate IoT scenarios. Fig. \ref{MIMO} depicts the proposed configuration. A CSS signal is carried by each transmit antenna with different SF to prevent interference among the CSS signals. Considering that conventional CSS modulation can carry SF bits in each symbol, the MIMO-CSS configuration conveys $\sum_{i=1}^{N_t}{\rm{SF}}_i$ bits. In the second part of the paper, the authors also present a precoding design by maximizing the received SNR with total and peak transmit power constraints. Based on this scheme, the number of information bits slightly reduces precoding. The suggested MIMO-CSS system greatly outperforms the standard CSS modulation in terms of BER and effective data rate, according to simulation results. 
\subsection{Discussion}
So far in this section, we investigated the recent key works regarding the advances in transceiver configurations for CSS modulation. The followings are the key insights inferred from the existing works:
\begin{itemize}
	\item LUT-based design of CSS transmitter can result in high memory consumption in the near future. Therefore, the OCG design presented in \cite{TTNguyen_2019} seems like an interesting low-cost replacement at the industrial level.
	\item Exploiting multiple antennas at the gateway improves the reception quality of CSS signals as expected \cite{Khai_2021}.
	\item Application of CSS modulation in its current for satellite-based IoT networks may not result in favorable performance due to the massive number of IoT end-devices (EDs) served by a single satellite \cite{Qian_2018}.
	\item Compared to CSS and SCS modulations, ACS modulation can deliver better BER performance in the presence of DFS which is the case in satellite-based IoT scenarios \cite{Qian_2019_Per}.
	\item SC-MCR modulation provides more reliable LEO satellite IoT communication and achieves higher data rates compared to SCS \cite{Roy_2019}.
	\item Modified chirp-rate modulation of FCrSK outperforms CSS modulation in the presence of large Doppler shifts or large Doppler rate \cite{Yang_2019}.
	\item The Concept of MIMO communication can be integrated with CSS modulation to overcome challenges such as poor fading performance and low data rate transmission of conventional CSS modulation \cite{Ma_2021,Kang_2022}.  
\end{itemize}
\section{Key Advances in Data Rate Improvement}
\label{section:DR} 
As mentioned, long-range communication provided by CSS modulation techniques comes with the cost of very low data rates in the order of a few hundred to a few thousand bits per second. Although this is suitable for most of the smart city applications in IoT networks, efforts have been made to improve the data rate of CSS-based transmission schemes. In the following, recent key advances \cite{Elshabrawy_2019_DR,Bomfin_2019,Hanif_2021_SSK,Almeida_2020,Hanif_2021_IM,Almeida_2021,Baruffa_2021,Ma_2021_FBI,Azim_2022,Yu_2022,An_2022,Mondal_2022,Shi_2022} on this topic are discussed in detail. 

\begin{figure}[t!]
  \centering
  \includegraphics[width=0.5\textwidth]{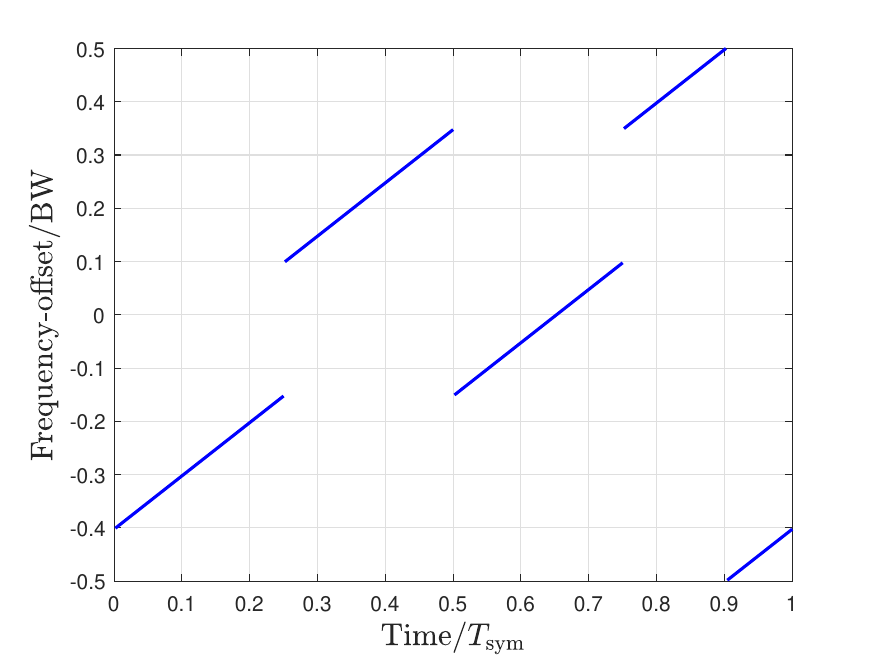}
  \caption{ICS-LoRa instantaneous frequency offset of CSS signal for $s=50$, $\rm{SF}=9$, $\rm{BW}=250$ kHz, and $T_{\rm sym}=2.048$ ms.}
  \label{ICS_CSS}
\end{figure}

Elshabrawy and Robert \cite{Elshabrawy_2019_DR} propose a technique called interleaved chirp spreading LoRa (ICS-LoRa) with the aim of improving data rates in LoRa networks. This approach results in doubling the size of the signal set, i.e., adding $1$ additional bit per symbol compared to the conventional encodable information data with the same BW. The interleaved version of the CSS signal discrete-time frequency-offset $\Delta f_s[k]$ can be formulated as follows:
\begin{equation}
\label{ICS}
\begin{split}
\Delta f^{({\rm I})}_s[k]=\begin{cases}
 \Delta f_s[k], & 0\leq k<2^{{\rm SF}-2},  \\
 \Delta f_s[k+2^{{\rm SF}-2}], & 2^{{\rm SF}-2}\leq k <2^{{\rm SF}-1},\\
 \Delta f_s[k-2^{{\rm SF}-2}], & 2^{{\rm SF}-1}\leq k <3\times2^{{\rm SF}-2}, \\
 \Delta f_s[k], & 3\times 2^{{\rm SF}-2}\leq k <2^{{\rm SF}}.
              \end{cases}
\end{split}
\end{equation}
Fig. \ref{ICS_CSS} illustrates these interleaving performed on the discrete-time version of the same CSS signal frequency offset of Fig. \ref{Chir}. Based on this technique, the transmitter input is a bit vector of the length ${\rm SF}+1$. That is because instead of $M$ possible CSS waveforms in conventional LoRa, there are also $M$ interleaved versions of CSS waveforms and $\log 2M=\log M + \log 2={\rm SF}+1$. This extra bit (considered a less significant bit of the input vector \cite{Elshabrawy_2019_DR}) is used at the transmitter to decide whether a conventional CSS signal or the interleaved version should be used to transmit the remaining SF bits. By applying the SE formula as $\eta=r_b/{\rm BW}$ ($r_b$ is the bit rate), the SE of ICS-LoRa is $\eta^{({\rm ICS-LoRa})}=({\rm SF}+1)/M$ bit/s/Hz. At the receiver end, the received samples and the interleaved samples undergo signal dechirping. Subsequently, both dechirped signals are fed into DFT blocks to produce two resulting vectors. The ICS-LoRa detector then generates an estimated bit indicating whether the normal CSS or interleaved version is sent, along with an estimated non-binary symbol. The results of the BER performance indicate that for an SF of $7$, the increased capacity of interleaved CSS comes with a slight degradation in BER performance, equivalent to a $0.8$ dB gap. However, for higher SFs, the capacity advantages can be achieved with minimal impact on BER performance compared to the standard LoRa modulation.    

Hanif and Nguyen in \cite{Hanif_2021_SSK} propose slope-shift keying LoRa (SSK-LoRa) in which down-chirps are used instead of interleaved ones. It is indicated in \cite{Hanif_2021_SSK} that because of the non-orthogonality between the interleaved chirps and the conventional ones, the ICS-LoRa scheme hurts the performance in terms of BER. The SSK-LoRa is performed by using $M$ up-chirps and $M$ down-chirps for a symbol. Therefore, one more bit can be carried by this method similar to the ICS-LoRa approach. Similar to the ICS-LoRa, the SE of SSK-LoRa is $\eta^{({\rm SSK-LoRa})}=({\rm SF}+1)/M$ bit/s/Hz. It should be noted that the sets of chirps used in SSK-LoRa are not also perfectly orthogonal and therefore, BER is slightly degraded compared to conventional CSS. But, as provided in the simulation part of this paper, the achievable rate is high enough to be able to neglect this degradation.

Almeida \textit{et al.} propose an in-phase and quadrature (IQ) CSS technique in \cite{Almeida_2020}. IQ-CSS exploits both IQ axes to convey two superposed LoRa signals. This enables encoding $2{\rm SF}$ total bits, ${\rm SF}$ bits in the in-phase component, and ${\rm SF}$ bits in the quadrature component. The IQ-CSS symbol waveform is formed as
\begin{equation}
\label{IQ_CSS}
x^{\rm (IQ)}[k]=\frac{1}{\sqrt{2}}\left(x^{\rm (NC))}_{s_{\rm I}}[k]+jx^{\rm (NC)}_{s_{\rm Q}}[k]\right),
\end{equation}
in which $x^{\rm (NC))}_{s_{\rm I}}[k]$ (in-phase component) and $x^{\rm (NC)}_{s_{\rm Q}}[k]$ (quadrature component) are non-continuous phase CSS signals encoding two information symbols $s_{\rm I}$ and $s_{\rm Q}$, respectively. The $M^2$ resulting waveforms from (\ref{IQ_CSS}) are not orthogonal, but it is shown that the two symbols $s_{\rm I}$ and $s_{\rm Q}$ can be independently decoded using coherent detection, each one similar to an orthogonal LoRa system. Clearly, the bit rate of IQ-CSS is doubled with respect to LoRa, at the same bandwidth. Thus, the SE of IQ-CSS is $\eta^{({\rm IQ-CSS})}=2{\rm SF}/M$ bits/s/Hz.

Discrete chirp rate keying CSS (DCRK-CSS), proposed by Almeida \textit{et al.} \cite{Almeida_2021} further extends the waveform set of IQ-CSS using multiple up-chirps and down-chirps via IQ structure with different chirp rates. This scheme uses $N_c$ additional bits to encode $2^{N_c}$ different chirp rates as $R^{'}\in\{\pm 1,\pm 2,\dots,\pm 2^{N_c-1}\}$. Hence, the DCRK-CSS waveform can be written as:
\begin{equation}
\label{DCRK}
x^{\rm (DCRK)}[k]=A_0\exp\left[2\pi j\frac{R^{'}(k+s)^2-M(k+s)}{2M}\right].
\end{equation}
Therfore, the SE of DCRK-CSS is $\eta^{({\rm DCRK-CSS})}=({\rm SF}+N_c)/M$ bit/s/Hz. It should be noted that the received DCRK-CSS signal goes through a bank of dechirping modules with different chirp rates in order to estimate $R^{'}$ used in the transmitter.

Phase shift keying-LoRa (PSK-LoRa), as proposed by Bomfin \textit{et al.} \cite{Bomfin_2019}, aims to improve the SE of LoRa by modifying its waveform assignment. In PSK-LoRa, instead of using only SF bits for each waveform, ${\rm SF}+N_p$ bits are assigned, where $N_p$ represents the number of phase-shift bits. The concept involves extending the existing set of $M$ LoRa waveforms by introducing an additional set of $(2^{N_p}-1)M$ waveforms that share the same chirps but have a phase shift that is a multiple of $\pi/2^{N_p-1}$. The bit rate in PSK-LoRa is then equal to $r_b =({\rm SF}+N_p)f_s/M$, while the bandwidth remains unchanged compared to LoRa. The SE of PSK-LoRa can be calculated as:
\begin{equation}
\label{42}
\eta^{\rm (PSK-LoRa)}=\frac{{\rm SF}+N_p}{M}\;\;\;{\rm bits/s/Hz}.
\end{equation}

By choosing $N_p=1$, PSK-LoRa is reduced to BPSK-LoRa which has a waveform as follows:
\begin{equation}
\label{43}
\begin{split}
x^{\rm (BP)}[k]=&A_0 A_{\rm BP}\exp\left[2\pi j\frac{(k+\lfloor\frac{s}{2}\rfloor)^2-M(k+\lfloor\frac{s}{2}\rfloor)}{2M}\right],
\end{split}
\end{equation}
where $A_{\rm BP}=\left[1-2(s \mod 2)\right]$ and $(s \mod 2)$ is the bit that determines the sign of the up-chirp. Similarly, for $N_p=2$, we will achieve the QPSK-LoRa waveform as:
\begin{equation}
\label{44}
\begin{split}
x^{\rm (QP)}[k]=&A_0 A_{\rm QP}\exp\left[2\pi j\frac{(k+\lfloor\frac{s}{4}\rfloor)^2-M(k+\lfloor\frac{s}{4}\rfloor)}{2M}\right],
\end{split}
\end{equation}
where $A_{\rm QP}=A_{\rm BP}+j[1-2(\lfloor{s/2}\rfloor\mod{2})]$. The detection of PSK-LoRa is performed in two steps, i.e., first, using a conventional CSS detector, the information embedded in CSS part is decoded, and in the second step, the phase estimate is obtained by applying the maximum likelihood (ML) PSK estimator. After derivation of the approximate BER for PSK-LoRa in AWGN and Rayleigh channels, the results in \cite{Bomfin_2019} show that in certain scenarios, PSK-LoRa offers the capability to transmit an equivalent amount of data as LoRa, but in a more efficient manner that saves time resources. Consequently, implementing PSK-LoRa modulation in LPWAN can be advantageous in reducing energy consumption, which is a significant issue for these systems. 

The index modulation (IM) technique is considered as one of the most recent approaches deployed in various modulation techniques for data rate improvement purposes \cite{Ishikawa_2018}. A similar approach to use IM in CSS modulation is presented in Zhang and Liu’s work \cite{Zhang_2006} and Yoon \textit{et al.} work \cite{Yoon_2008}. In these works, basic time-delayed up-chirps are transmitted in multiple manners with the basic up-chirp, and the information bits are embedded in the phase and/or the amplitude. The difference between these schemes and IM-based methods is that in IM, information bits are encoded in the signal indices, which, in the case of CSS modulation, correspond to the initial phase and frequency of up-chirps.  In \cite{Hanif_2021_IM}, Hanif and Nguyen propose integrating IM with CSS resulting in a technique called frequency shift chirp spread spectrum with index modulation (FSCSS-IM). In FSCSS-IM, IM is integrated with CSS modulation similar to the case of IM application in the frequency bins of orthogonal frequency-division multiplexing (OFDM) \cite{Başar_2013}. However, in OFDM-IM, a fraction of BW is allocated to each symbol, but in FSCSS-IM, each symbol occupies the whole available BW. The FSCSS-IM waveform is formed as the summation of $W$ different CSS waveforms, each encoding the information symbols $s_{n_w}$ ($w=0,1,\dots W$). Therefore, the waveform of FSCSS-IM can be formulated as:  
\begin{equation}
\label{FSCSS_IM}
x^{\rm (IM)}[k]=\frac{1}{\sqrt{W}}\sum_{w=0}^{W-1}x_{w}[k].
\end{equation}
Based on this, the number of bits that can be carried by a single FSCSS-IM signal is $N_b^{({\rm IM})}=\lfloor\log_2{\binom{{M}}{W}}\rfloor$. It is worth noting that two FSCSS-IM symbols are orthogonal if the CSS orthogonality conditions hold for any one-by-one combination of their individual CSS signals. Moreover, the SE of FSCSS-IM is $\eta^{({\rm IM})}=N_b^{({\rm IM})}/M$ bits/s/Hz. When multiple chirps are combined and utilized for signal transmission to enhance the data rate, the optimal receiver would necessitate searching through all the patterns employed by the transmitter. There is also a suboptimal receiver proposed in \cite{Hanif_2021_IM} to overcome this issue. From the simulation results provided in \cite{Hanif_2021_IM}, it can be observed FSCSS-IM has an acceptable performance compared to CSS in terms of BER.  

As an extended approach resulting from the combination of FSCSS-IM and IQ-CSS methods, Baruffa and Rugini in \cite{Baruffa_2021} present an approach to enhance CSS waveforms called in-phase and quadrature chirp index modulation (IQCIM). In fact, the IQCIM waveform is formed as the sum of two FSCSS-IM signals as in-phase and quadrature parts. Hence, the waveform of IQCIM can be written as:
\begin{equation}
\label{IQCIM}
x^{\rm (IQCIM)}[k]=\frac{1}{\sqrt{2}}\left(x_{\rm NC}^{({\rm IM,I})}[k]+jx_{\rm NC}^{({\rm IM,Q})}[k]\right).
\end{equation}
the bit rate of IQCIM is double the bit rate of FSCSS with the same value of $W$ and for the same bandwidth. Thus, the SE in IQCIM is $2\eta^{({\rm IM})}$ bits/s/Hz. Based on the performance analysis, it has been observed that the IQCIM offers a notable improvement in SE compared to CSS while maintaining a similar level of complexity. This enhanced efficiency comes at a marginal energy cost of approximately $0.9$ dB.

Ma \textit{et al.} in \cite{Ma_2021_FBI} propose frequency-bin-index LoRa (FBI-LoRa) via two schemes for data rate improvement of LoRa modulation. In scheme I, the CSS symbol set $\mathcal{S}$ is divided into $g_{\rm num}$ groups where the number of CSS symbols in each group is $N_g=M/g_{\rm num}$. Then, the summation of $N_{\rm sum}$ symbols, which are equally taken from each group, is selected to shape the CSS waveform. Therefore, the number of selected symbols from each group is $f_{\rm num}=N_{\rm sum}/g_{\rm num}$. Based on these, the waveform of the FBI-LoRa scheme I can be written as:   
\begin{equation}
\label{FBI}
\begin{split}
x^{\rm (FBI_I)}[k]=&\sum_{i=0}^{g_{\rm num}-1}\sum_{i'=0}^{f_{\rm num}-1}A0\sqrt{\frac{M}{f_{\rm num}g_{\rm num}}}\\
&\times\exp\left[2\pi j\frac{(k+s_{i,i'})^2-M(k+s_{n,i,i'})}{2M}\right],
\end{split}
\end{equation}
where $s_{i,i'}$ represent the $i'$th CSS symbol in $i$th group. The SE of FBI-LoRa scheme I can be obtained as $\eta^{\rm FBI_I}=g_{\rm num}\lfloor \log_2{\binom{N_g}{f_{\rm num}}}\rfloor/M$ bit/s/Hz. In scheme II, the groups are indexed, and $g'_{\rm num}$ groups out of total $g_{\rm num}$ groups are selected ($1\leq g'_{\rm num}<g_{\rm num}$). Therefore, the number of symbols in each selected group is $f'_{\rm num}=N_{\rm sum}/g'_{\rm num}$ and the vector of selected groups is denoted as $\{{\rm Grp}(0),{\rm Grp}(1),\dots,{\rm Grp}(g'_{\rm num}-1)\}$. Therefore, the waveform of FBI-LoRa scheme II can be formulated as:
\begin{equation}
\label{FBI_II}
\begin{split}
x^{\rm (FBI_{II})}[k]=&\sum_{i=0}^{g'_{\rm num}-1}\sum_{i'=0}^{f'_{\rm num}-1}A0\sqrt{\frac{M}{f'_{\rm num}g'_{\rm num}}}\\
&\times\exp\left[2\pi j\frac{(k+s'_{i,i'})^2-M(k+s'_{n,i,i'})}{2M}\right],
\end{split}
\end{equation}
where $s'_{i,i'}$ is the $i'$th CSS symbol in group ${\rm Grp}(i)$. The SE of the FBI-LoRa scheme II is then 
\begin{equation}
\label{SE_FBI_II}
\begin{split}
\eta^{\rm FBI_{II}}=&\frac{1}{M}\Bigg{(}g'_{\rm num}\Bigg{\lfloor} \log_2{\binom{N_g}{f'_{\rm num}}}\Bigg{\rfloor}\\
&+\Bigg{\lfloor}\log_2\binom{g_{\rm num}}{g'_{\rm num}}\Bigg{\rfloor}\Bigg{)}\;\;{\rm bit/s/Hz}.
\end{split}
\end{equation}
Additionally, the theoretical analyses and simulations show that the FBI-LoRa system surpasses conventional CSS, ICS-LoRa, SSK-LoRa, and PSK-LoRa systems, albeit with a minor compromise in the BER performance. 

To improve the PSK-LoRa scheme, Azim \textit{et al.} in \cite{Azim_2022} propose enhanced PSK-LoRa (ePSK-LoRa) to increase the CSS modulation data rate and make it more energy efficient. Based on this approach, the BW of CSS modulation is divided into $N_{\rm sub}$ subbands, where, each one contains $M/N_{\rm sub}$ CSS symbols (it is equivalent to dividing the set $\mathcal{S}$ into $N_{\rm sub}$ subsets). Calling the starting frequencies of CSS modulation in the first subband as fundamental frequencies (FF), one can encode $\log_2 (M/N_{\rm sub})$ bits by selecting an FF. Then, the redundancy is generated by selecting the harmonics of the selected FF ($f_{\rm sel}$) in all $N_{\rm sub}$ subbands as $f_{\rm sel}+M/N_{\rm sub}$, $f_{\rm sel}+2M/N_{\rm sub}$, $\dots$, and $f_{\rm sel}+(N_{\rm sum}-1)M/N_{\rm sum}$. After this, FF and its harmonics are multiplied by a phase shift coefficient as $\exp(2\pi j pl/2^{N_{\rm PS}})$ where $N_{\rm PS}$ is the number of bits encoded into the phase shifts and $l\in[\![0,N_{\rm sum}-1]\!]$ is the index of the subband. Based on these, the ePSK-LoRa waveform can be formulated as:
\begin{equation}
\label{ePSK}
\begin{split}
x^{({\rm ePSK})}[k]=&\sum_{l=0}^{N_{\rm sub}-1} \exp\left(\frac{2\pi j[f_{\rm sel}+l(M/N_{\rm sub})]k}{M}\right)\\
&\times \exp\left(\frac{2\pi j pl}{2^{N_{\rm PS}}}\right) x_{0}[k].
\end{split}
\end{equation}
Therefore, the SE of ePSK-LoRa is $\eta^{\rm (ePSK)}=(1/M)[\log_2 (M/N_{\rm sub})+N_{\rm sub}N_{\rm PS}]$ bit/s/Hz. Based on the provided results, ePSK-LoRa offers appealing characteristics compared to traditional alternatives, including satisfactory resilience to both phase offset and frequency offset.

Yu \textit{et al.} present a group-based CSS (GCSS) modulation in \cite{Yu_2022} in order to improve the SE of CSS modulation. Based on this approach, the set of $\mathcal{S}$ is divided into $GN$ groups resulting in $K=\log_2(M/GN)$ transmittable information bits per group. Therefore, the total number of $N_s=GN\times K$ bits can be modulated for all $GN$ groups. Based on this, every $N_s$ information bits are divided into groups of $b^{(m)}=[b_0^{(m)},b_1^{(m)},\dots,b_{K-1}^{(m)}]$ where $m\in\{1,2,\dots,GN\}$ and the mapping symbol $s^{(m)}$ is generated as:
\begin{equation}
\label{s_n_m}
s^{(m)}=(m-1)\frac{M}{GN}+\sum_{i=0}^{K-1}b_i^{(m)}2^i.
\end{equation}
Finally, the GCSS waveform can be formulated as:
\begin{equation}
\label{GCSS}
\begin{split}
x^{\rm (GCSS)}[k]=&A_0\sqrt{\frac{1}{GN}}\\
&\times\sum_{m=1}^{GN}\exp\left[2\pi j\frac{(k+s^{(m)})^2-(k+s^{(m))}}{2M}\right].
\end{split}
\end{equation}
The SE of GCSS is then obtained as $\eta^{\rm (GCSS)}=GN\times K/M$ bit/s/Hz.

Time domain multiplexed LoRa (TDM-LoRa) is another data rate improvement technique proposed by An \textit{et al.} in \cite{An_2022}. The basic idea of this approach is to form the TDM-LoRa waveform based on a CSS up-chirp and a CSS down-chirp as:
\begin{equation}
\label{TDM}
x^{\rm (TDM)}[k]=\sqrt{\frac{1}{2}}\left(x^{\rm (NC)}_{s_1}[k]+x^{{\rm (NC)}}_{s_1,{\rm D}}[k]\right).
\end{equation}
Hence, the SE of TDM-LoRa is $\eta^{\rm (TDM)}=2{\rm SF}/M$ bit/s/Hz. Moreover, an IQTDM-LoRa is also proposed in which in-phase and quadrature components are TDM-LoRa waveforms. The SE of IQTDM LoRa can be written as $\eta^{\rm (IQTDM)}=4{\rm SF}/M$.

Inspired by ICS-LoRa and SSK-LoRa, Mondal \textit{et al.} propose SSK-ICS-LoRa in which the signal set includes CSS up-chirps, down-chirps, interleaved up-chirps, and interleaved down-chirps. Building upon this, the SSK-ICS-LoRa waveform has the following form:
\begin{equation}
\label{IS}
\begin{split}
x_{s_n}^{({\rm SSK-ICS})}[k]=\begin{cases}
 x^{\rm (NC)}_{s_n}[k], & 0\leq s_n<M,  \\
 x^{\rm (NC)}_{s_n,{\rm D}}[k], & M\leq s_n<2M,\\
 x_{s_n}^{\rm (I)}[k], & 2M\leq s_n<3M, \\
 x_{s_n,{\rm D}}^{{\rm (I)}}[k], & 3M\leq s_n<4M,
              \end{cases}
\end{split}
\end{equation}
where $x_{s_n}^{\rm (I)}[k]$ and $x_{s_n,{\rm D}}^{{\rm (I)}}[k]$ are ICS-LoRa and down-chirp ICS-LoRa waveforms generated according to (\ref{ICS}). Consequently, the SE of SSK-ICS-LoRa is $\eta^{\rm (SSK-ICS)}=({\rm SF}+2)/M$ bit/s/Hz.

In another work based on ICS-LoRa, Shi \textit{et al.} \cite{Shi_2022} introduce the enhanced ICS-LoRa (EICS-LoRa) by generating random interleavers for ICS-LoRa. As mentioned, the interleaved version of the CSS in ICS-LoRa consists of four segments. Based on EICS-LoRa, the transmission order of these four segments can be used to carry more information bits. Therefore, in this case, $4!=24$ possible interleavers can be used. However, because of the increasing complexity and degradation of BER, the authors assumed a random $32$ possible interleavers for an eight-segment CSS signal. Hence, the SE of this method at the best scenario is $\eta^{\rm EICS}=({\rm SF}+5)/M$.
\subsection{Discussion}
Table \ref{DR} includes the SE formula for all discussed data rate improvement methods. It should be noted that the parameters for each method are set in a way that an acceptable receiver complexity and BER performance can be achieved. Based on this, the FBI-LoRa scheme I provides the largest SE improvement compared to the conventional CSS modulation.   
\begin{table*}[t]
\caption{SE comparison of various CSS data improvement schemes}
\label{DR}
\centering
\begin{tabularx}{\textwidth}{ccX}
\toprule[0.5pt]
\textbf{Method} & \textbf{SE (bit/s/Hz)} & \textbf{Improvement percentage compare to CSS (for ${\rm SF}=8$)} \\ 
\midrule[0.5pt] 
\vspace{5pt}
\textbf{Conventional CSS} & $\frac{{{\rm{SF}}}}{M}$ & -- \\
\vspace{5pt}
\textbf{ICS-LoRa \cite{Elshabrawy_2019_DR}} & $\frac{{{\rm{SF}} + 1}}{M}$ & $112.5\%$ \\
\vspace{5pt}
\textbf{SSK-LoRa \cite{Hanif_2021_SSK}} & $\frac{{{\rm{SF}} + 1}}{M}$ & $112.5\%$ \\
\vspace{5pt}
\textbf{IQ-CSS \cite{Almeida_2020}} & $\frac{2{\rm SF}}{M}$ & $200\%$ \\
\vspace{5pt}
\textbf{DCRK-CSS \cite{Almeida_2021}} & $\frac{{\rm SF}+N_c}{M}$ & $125\%$ \\
\vspace{5pt}
\textbf{PSK-LoRa \cite{Bomfin_2019}} & $\frac{{\rm SF}+N_p}{M}$  & $112.5\%$ (for $N_p=1$) and $125\%$ (for $N_p=2$) \\
\vspace{5pt}
\textbf{FSCSS-IM \cite{Hanif_2021_IM}} & $\frac{1}{M}\left\lfloor\log_2\binom{M}{W}\right\rfloor$ & $262.5\%$ (for $W=3$) \\
\vspace{5pt}
\textbf{IQCIM \cite{Baruffa_2021}} & $\frac{2}{M}\left\lfloor\log_2\binom{M}{W}\right\rfloor$ & $525\%$ (for $W=3$) \\
\vspace{5pt}
\textbf{FBI-LoRa (I) \cite{Ma_2021_FBI}} & $\frac{g_{\rm num}}{M}\left\lfloor\log_2\binom{N_g}{f_{\rm num}}\right\rfloor$ & $800\%$ (for $g_{\rm num}=8$ and $f_{\rm num}=2$) \\
\vspace{5pt}
\textbf{FBI-LoRa (II) \cite{Ma_2021_FBI}} & $\frac{1}{M}\Bigg{(}g'_{\rm num}\Bigg{\lfloor} \log_2{\binom{N_g}{f'_{\rm num}}}\Bigg{\rfloor}+\Bigg{\lfloor}\log_2\binom{g_{\rm num}}{g'_{\rm num}}\Bigg{\rfloor}\Bigg{)}$ & $250\%$ (for $g_{\rm num}=8$, $g'_{\rm num}=2$, and $f'_{\rm num}=2$) \\
\vspace{5pt}
\textbf{ePSK-LoRa \cite{Azim_2022}} & $\frac{1}{M}\left\lfloor\log_2\left(\frac{M}{N_{\rm sub}}\right)+N_{\rm sub}N_{\rm PS}\right\rfloor$ & $112.5\%$ (for $N_{\rm sub}=2$ and $N_{\rm PS}=1$) and $137.5\%$ (for $N_{\rm sub}=2$ and $N_{\rm PS}=2$)  \\
\vspace{5pt}
\textbf{GCSS \cite{Yu_2022}} & $\frac{GN}{M}\log_2\left(\frac{M}{GN}\right)$ & $500\%$ (for $GN=8$) \\
\vspace{5pt}
\textbf{TDM-LoRa \cite{An_2022}} & $\frac{2{\rm SF}}{M}$ & $200\%$ \\
\vspace{5pt}
\textbf{IQTDM-LoRa \cite{An_2022}} & $\frac{4{\rm SF}}{M}$ & $400\%$ \\
\vspace{5pt}
\textbf{SSK-ICS-LoRa \cite{Mondal_2022}} & $\frac{{\rm SF}+2}{M}$ & $125\%$ \\
\vspace{5pt}
\textbf{EICS-LoRa \cite{Shi_2022}} & $\frac{{\rm SF}+5}{M}$ & $162.5\%$ \\
\bottomrule[0.5pt]
\end{tabularx}
\end{table*}

\section{Key Advances in Interference Modeling}
\label{section:Int}
In this section, we review the existing key works in the literature corresponding to the performance of CSS modulation in the presence of interference \cite{Elshabrawy_2018_IntBER, Edward_2019,Temim_2020_Enh,Garlisi_2021,Afisiadis_2020,Afisiadis_2021,Demeslay_2022_BER,Demeslay_2022_Receiver,Benkhelifa_2022} and investigate the proposed solutions. 

Before proceeding with the literature review of CSS signal interference, we briefly present the interference model and discuss the BER performance of CSS modulation in the presence of the same SF interference. According to \cite{Croce_2018}, whereas the signal-to-interference ratio (SIR) threshold for same-SF interference is $0$ dB, interferers with different SF have an average rejection SIR threshold of $-16$ dB. Therefore, we assume that the CSS signal transmission is being performed in the presence of the same SF interference from another LoRa ED. Hence, equation (\ref{23}) can be modified as:
\begin{equation}
\label{r_int}
r_s[k]=hx_s[k]+h_{\rm int}x_{{\rm int}}[k]+w[k],
\end{equation}
where $h_{\rm int}$ is the channel gain between the interferers and the LoRa gateway and $x_{{\rm int}}[k]$ is the interfering signal. Also, due to the lack of complete synchronization between LoRa transmitters, we can assume that the interfering part is due to two different CSS symbols in such a way that:
\begin{equation}
\label{x_int}
x_{\rm int}[k]=\begin{cases}
    x_{s_1}[k] & 0\leq k < \tau_{\rm int} \\
   x_{s_2}[k] & \tau_{\rm int}\leq k < M, 
\end{cases}
\end{equation}
where $\tau_{\rm int}$ is the number of sample shift between two EDs. Based on this model, an approximate expression for the BER performance of CSS modulation under the same SF interference and AWGN channel condition is obtained in \cite{Elshabrawy_2018_IntBER} as:
\begin{equation}
\label{P_bint}
P_b^{\rm (Int)}=0.5\times \left[P_{\rm N}+(1-P_{\rm N})P_{\rm I}\right],
\end{equation}
where we have:
\begin{equation}
\label{P_N}
\begin{split}
P_{\rm N}=&\int_0^\infty \left\{1-\left[1-\exp\left(-\frac{\beta^2}{2\sigma^2}\right)\right]^{M-2}\right\}\\
&\times \frac{1}{\sqrt{2\pi\sigma^2}}\exp\left[-\frac{(\beta-1)^2}{2\sigma^2}\right]{\rm d}\beta,
\end{split}
\end{equation}
and 
\begin{equation}
\label{P_I}
P_{\rm I}\approx\frac{1}{M/2+1}\sum_{\tau_{\rm int}=0}^{M/2}\left[\frac{1}{M}\sum_{z_1=0}^{M-1}\mathcal{Q}\left(\frac{1-\mathcal{U}_0(\tau_{\rm int},z_1)}{\sqrt{2\sigma^2}}\right)\right],
\end{equation}
in which 
\begin{equation}
\label{U}
\mathcal{U}_0(\tau_{\rm int},z_1)=\frac{1}{M\sqrt{\gamma_{\rm int}}}\left[M-\tau_{\rm int}+\left|\frac{\sin(-\pi z_1 \tau_{\rm int}/M)}{\sin(-\pi z_1/M)}\right|\right].
\end{equation}
where $\gamma_{\rm int}$ is the SIR value.

\begin{figure}[t!]
  \centering
\includegraphics[scale=0.6]{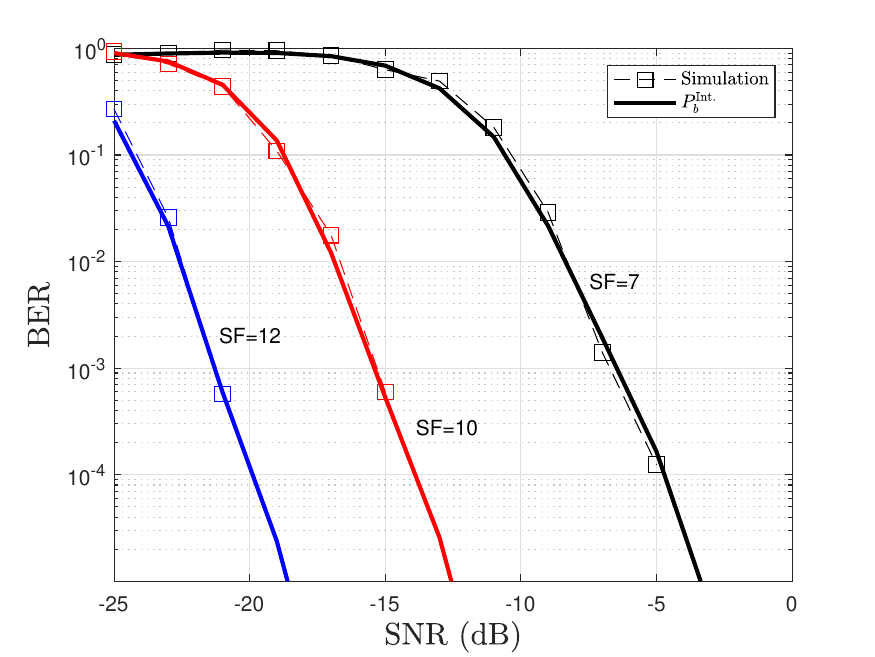}
  \caption{BER performance of the CSS modulation under same SF interference and AWGN channel for SF values of $7$, $10$, and $12$ and SIR of $\gamma_{\rm int}=6$ dB.}
  \label{BER_Int}
\end{figure}

Based on the numerical evaluation of equation (\ref{P_bint}) and computer simulations of the same SF interference model for CSS modulation with non-coherent detection, Fig. \ref{BER_Int} indicates the BER performance of the CSS modulation in such scenarios. As can be seen, compared to Fig. \ref{BER_AWGN}, a nearly $1$ to $2$ dB degradation in performance can be observed.

The discussed interference model is based on Elshabrawy \textit{et al.} work in \cite{Elshabrawy_2018_IntBER}. By means of the results provided in this work, it can be observed that in the presence of shadowing, the EDs experiencing the same SF interference can benefit from diversity resulting from shadowing to achieve a better performance compared to the ideal case of no shadowing in terms of coverage probability. Moreover, Aifisiadis \textit{et al.} in \cite{Afisiadis_2020}, extend the same SF interference model by assuming not only integer values of $\tau_{\rm int}$, but also the fractional values which can be the case in practical situations. The SER and frame error rate (FER) expressions are also derived for this case. 

In \cite{Edward_2019}, Edward \textit{et al.} present another interference model similar to \cite{Elshabrawy_2018_IntBER}, with this difference that instead of CSS signals, the desired CSS signal is being interfered from two ICS-LoRa signals. The authors propose a simple receiver structure to retrieve both CSS and ICS-LoRa signals. Based on their proposed structure, at the receiver gateway, the intended CSS signal is demodulated and detected using dechirping and DFT methods. To recover the interfering ICS-LoRa signals, the detected CSS signal is re-modulated and then subtracted from the combined received signal. However, to accurately eliminate the re-modulated CSS signal, a delay compensation process is necessary. The subtracted outcome is demodulated using the ICS-LoRa demodulation technique, which begins with the de-interleaver to convert the ICS-LoRa chirp signal back into the corresponding CSS chirp signal. Simulation results indicate that the BER performance is not almost affected in the presence of ICS-LoRa interference.

Ben Temim \textit{et al.} propose an enhanced receiver to detect superposed CSS signals \cite{Temim_2020_Enh}. This work deals with the case of receiving multiple CSS signals with the same SF in a gateway simultaneously with the aim of decoding these unsynchronized signals in the presence of frequency offset. In the proposed interference model, it is assumed that multiple LoRa frames each consisting of different CSS signals (generated according to LoRa specifications \cite{Ghanaatian_2019}) arrive at the GW taking into account the time desynchronization, and the frequency offset. There are three main steps toward decoding these superposed LoRa frames:
\begin{enumerate}
    \item Detection of the strongest CSS signal.
    \item Decoding the information embedded in the strongest CSS signal.
    \item Regenerating the complex envelope of the strongest CSS signal and subtracting it from the received signal exploiting successive interference cancellation (SIC).
\end{enumerate}
In the simulation section, the immunity of the proposed detection algorithm against the same SF interference has been proved. As a matter of fact, from the findings, it can be inferred that this receiver has the ability to successfully decode three overlapping signals with a ${\rm SF}=12$, as long as a minimum power ratio of $6$ dB between the interfering CSS signals is maintained. Moreover, the authors provide experimental results to further validate the performance of their proposed receiver structure.

In \cite{Afisiadis_2021}, Affisiadis \textit{et al.} investigate the effect of the same SF interference in CSS-modulated LoRa frames and derive error rate expressions for SER and FER. The interference modeling is similar to \cite{Elshabrawy_2018_IntBER}, but the results are explored for coherent and non-coherent CSS detection. Based on the findings, it can be deduced that the performance difference between coherent and non-coherent detection in AWGN and no interference conditions is nearly $0.7$ dB. On the other hand, this difference is shown to be around $10$ dB when the SIR is $0$ dB. According to this observation, dense LoRa networks with a high likelihood of same-SF packet collisions are a good fit for coherent CSS signal detection.

Demeslay \textit{et al.} present an analysis for CSS modulation error performance in the multi-path channel (MPC) along with the same SF interference \cite{Demeslay_2022_BER}. Based on the proposed model for MPC, and assuming the receiver's perfect synchronization to the desired CSS signal, the effect of MPC at the receiver can be translated to the detection of a single CSS waveform in the presence of time-shifted versions of the same CSS signal and its previous one. This is also the case in the same SF interference model with some modification of the SIR value. The authors provide SER expressions for the MPC effect as well as two colliding CSS signals for both coherent and non-coherent detection schemes. Based on the provided results, the CSS signal appears to be less susceptible to the other lower path gains and just sensitive to the first path of the exponential decay channel. This makes it possible to estimate basic LoRa equalization schemes that just take into account a small number of channel pathways (or the strongest path gain). Moreover, it is shown that in the case of the same SF interference, the error performance is at its best when $\tau_{\rm in}$ is around half of the CSS symbol interval.

A CSS receiver scheme is presented in \cite{Demeslay_2022_Receiver} by Demeslay \textit{et al.}. In this work, to eliminate the effect of MPC, authors first investigate the performance of an ideal matched filter (MF) based on the MPC which depends on the transmitted CSS symbol. However, it is not practical because the CSS symbol is not known at the receiver. One way is to search over all possible CSS symbols and select the one that maximizes the DFT output of the MF which results in very high complexity. Therefore, the authors present fixed candidate MF in which they select $N_{\rm peak}$ highest peak values of $M$-point DFT of the received signal. But this requires a sorting algorithm which again results in high complexity. To deal with this issue, a variable candidate MF in which a threshold is used to select the candidates is proposed in which the search domain is generated from a list of the most probable candidate symbols. This threshold can be designed as a fraction of the maximum value of the $M$-point DFT output. Finally, by presenting the equivalent RAKE receiver model of the MF receiver, it is shown that the RAKE approach is by far preferred to MF in practice as its complexity outperforms MF in all cases.

\subsection{Discussion}
In this section, we investigated the recent key works regarding interference modeling and its effect in CSS-modulated LoRa networks. We can infer the followings from the reviewed works:
\begin{itemize}
    \item In the presence of the same SF CSS interference, shadowing can be helpful and it improves the network performance in terms of coverage probability \cite{Elshabrawy_2018_IntBER}.
    \item SCI can be utilized for joint detection of CSS and ICS-LoRa signals without almost any degradation in BER performance \cite{Edward_2019}.
    \item An SCI approach can also be used to detect three overlapping CSS signals with a minimum of $6$ dB power difference between interfering parts \cite{Temim_2020_Enh}.
    \item The gap between coherent and non-coherent detection schemes in the presence of interference is notably larger than this gap for the case of having AWGN and no interference. \cite{Afisiadis_2021}. 
    \item In the presence of interference due to MPC, CSS modulation is almost only affected by the first path of the channel \cite{Demeslay_2022_BER}.
\end{itemize}
  
\section{Key Advances in Synchronization Algorithms}
\label{section:Sync}
\begin{figure}[t!]
  \centering
\includegraphics[scale=0.6]{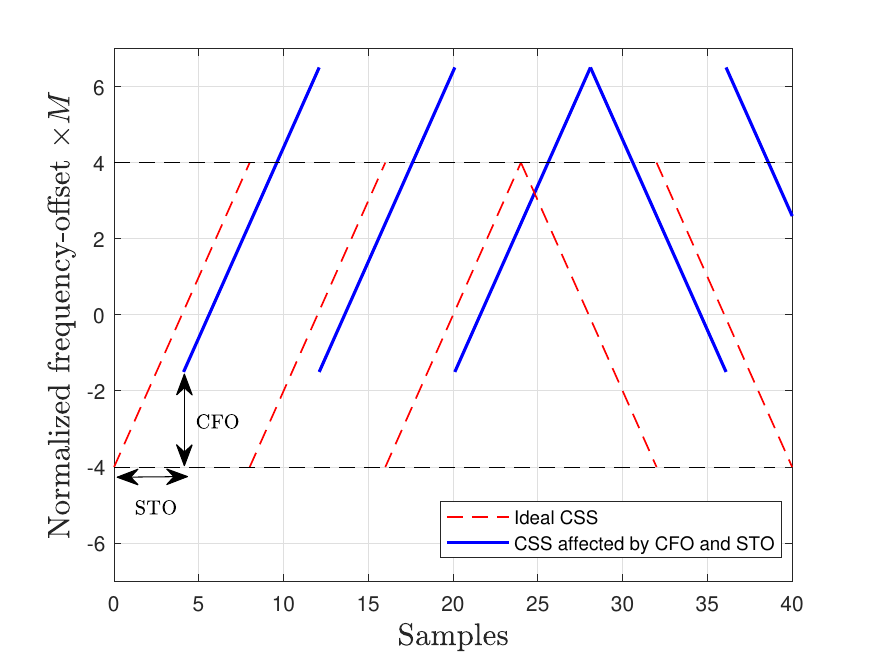}
  \caption{Effect of CFO and STO on CSS signals at the receiver side}
  \label{Off}
\end{figure}

\begin{figure}[t!]
  \centering
\includegraphics[scale=0.6]{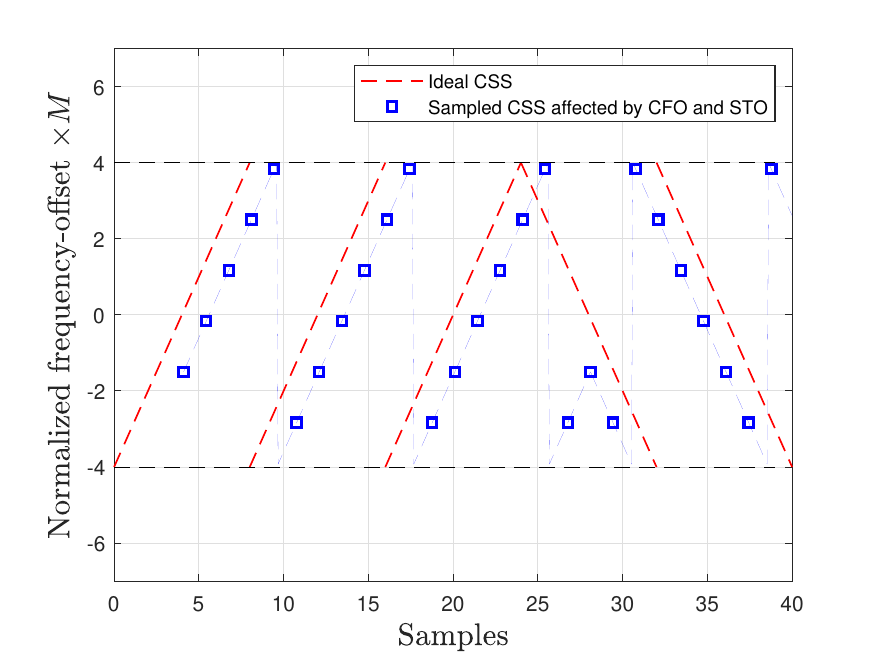}
  \caption{Effect of CFO and STO on sampled CSS signals at the receiver side}
  \label{Off_samp}
\end{figure}

It is a known fact that in practice, any transmission technique experiences time and frequency offsets and CSS modulation is no exception. It is necessary to provide an accurate estimation scheme in the synchronization procedure in order to have a reliable signal recovery at the receiver end. Considering that a LoRa frame is made of a sequence of CSS signals in a specific format, i.e. the payload preceded by multiple basic up-chiprs as the preamble followed by two specific CSS waveforms as sync word and $2.25$ basic down-chirps as start frame delimiter (SFD), with the help of \cite{LoRa_pat,Bernier_2020}, we present the foundation for LoRa frame synchronization including the offset cancellation procedure. The time and frequency offsets in CSS transmission result in having CFO, STO, and SFO. It should be noted that both CFO and STO can be separated into integer (denoted by subscript ``int'' in this paper) and fractional (denoted by subscript ``frac'' in this paper) parts. It is worth noting that the fractional part of CFO can result in sensitivity degradation at the receiver. To indicate the effect of these offsets on a CSS waveform, Fig. \ref{Off} and Fig. \ref{Off_samp} are provided for three consecutive up-chirps and two consecutive down-chirps in the presence of CFO of $2.5$ frequency bins and STO of $4.1$ samples. For illustration purposes, we select $M=8$. As can be seen, in the presence of CFO and STO, the sampled CSS at the receiver can result in erroneous detection in some cases.

To simply model the offsets at the receiver side, we assume that the CFO and STO only contain an integer part and the channel is in ideal condition. Therefore, the received CSS signal in the presence of offset can be formulated as follows:
\begin{equation}
\label{r_off}
\begin{split}
r^{\rm off}_{s_n}[k]=&A_0\exp\left[2\pi j\frac{k^2+2k(s_n-{\tau}_{\rm int})-kM}{2M}\right]\\
&\times \exp\left(2\pi j\frac{k{\mu_{\rm int}}}{\rm BW}+j\psi\right),
\end{split}
\end{equation}
where $\tau_{\rm int}$ and $\mu_{\rm int}$ denote the integer parts of STO and CFO, respectively.  Based on this, the synchronization algorithm can be divided into $6$ main steps as follows:
\begin{enumerate}
    \item \textit{Preamble synchronization:} The initial task of the receiver involves identifying the preamble and aligning it with the frame boundary. This is achieved by utilizing the repetitive basic up-chirps present in the preamble. Firstly, the receiver gathers a set of samples and performs CSS symbol detection. If the magnitude of at least one frequency bin surpasses a predetermined threshold, the index of the bin with the highest magnitude is recorded, and the process continues. If these consecutive CSS blocks belong to a LoRa preamble, this highest magnitude remains unchanged throughout subsequent DFT blocks. Moreover, since the intended CSS symbol for the LoRa preamble is $s=0$, and the detected value of $\hat{s}_{\rm pre}$ in the preamble blocks may not be equal to zero, the receiver synchronizes to the start of the header by skipping $M-\hat{s}_{\rm pre}$ samples. 
    \item \textit{Frame synchronization:} After aligning the CSS blocks, the search for the sync word begins. As mentioned, sync word contains two identical CSS signals encoding a specific CSS symbol $s_{\rm sync}$. However, the fractional part of CFO, i.e., $\mu_{\rm frac}$ can shift this CSS symbol to $s_{\rm sync}+1$ or $s_{\rm sync}-1$. Consequently, the search is performed for two consecutive CSS symbols each carrying $s_{\rm sync}$, $s_{\rm sync}+1$, or $s_{\rm sync}-1$. To ensure efficient frame filtering, the receiver can be programmed to anticipate a specific count of preamble symbols before successfully identifying the sync word symbols. This approach enables the automatic rejection of transmitted frames that have longer preambles than anticipated. 
    \item \textit{Coarse time and frequency synchronization:} Using the first step, it is shown in \cite{Bernier_2020} that $\hat{s}_{\rm pre}=\mu_{\rm int}/{\rm BW}+(M-\tau_{\rm int})/M$. On the other hand, by multiplying one of the CSS signals in the aligned version of the sync word to the basic up-chirp (not the complex conjugate one) and exploiting FFT, one can obtain $s_{\rm sync}=\mu_{\rm int}/{\rm BW}-(M-\tau_{\rm int})/M$ \cite{Bernier_2020}. Using these equations, the receiver can estimate $\tau_{\rm int}$ and $\mu_{\rm int}$ simultaneously. 
    \item \textit{Fine frequency synchronization:} The $\mu_{\rm frac}$ can also be simply estimated using the phase difference ($\Delta\phi_{\rm sync}$) between two identical samples from each of the CSS signals existing in the sync word as $\mu_{\rm frac}=\Delta \phi_{\rm sync}\times{\rm BW}/M$.
    \item \textit{Fine time synchronization:} This step corresponds to estimating the fractional part of STO, i.e., $\tau_{\rm frac}$ which can result in inter-symbol interference. For every CSS block being processed at the receiver, there is an amount of energy shifted into the next nearest frequency bins due to fractional timing and frequency errors. Based on this, in \cite{LoRa_pat}, a ``raw timing error'' is estimated that can be converted into a ``timing error'' using a conversion function \cite{Bernier_2020}. This timing error can be considered as an approximation of $\tau_{\rm frac}$, if the fractional frequency error is ignored.   
    \item \textit{Sampling frequency offset (SFO) compensation:} The frequency mismatch between crystal oscillators can lead to different sampling frequencies at the transmitter and receiver ends. This is known as SFO. In other words, while the sampling frequency at the transmitter is equal to the $f_s={\rm BW}$, at the receiver, we have a different sampling frequency of $f'_s\neq{\rm BW}$. Therefore, a received CSS signal under this condition can be written as \cite{Ghanaatian_2019}: 
\begin{equation}
\label{x_SFO}
\begin{split}
x_{\rm SFO}[k,s]=&A_0\exp\Bigg{[}2\pi j\frac{\left(\frac{\rm BW}{f'_s}k\right)^2+2\left(\frac{\rm BW}{f'_s}k\right)s}{2M}\\
&-\frac{\left(\frac{\rm BW}{f'_s}k\right)M}{2M}\Bigg{]}.
\end{split}
\end{equation}
As can be seen, there is a re-scaling by a factor of $BW/f'_s$ which can lead to a side-lobe in frequency bins other than the desired one which ultimately, increases noise sensitivity. The fractional part of this SFO can be removed using the same ``timing error'' introduced in the previous step. However, the integer part can be compensated using a method proposed in \cite{Ghanaatian_2019}.
\end{enumerate}

After becoming familiar with the practical synchronization procedure for LoRa communication, we investigate the literature corresponding to the improvement of LoRa synchronization \cite{Bernier_2020,Temim_2020_Syn,Vangelista_2021,Xhonneux_2022,Nguyen_2022_sync,Savaux_2022,Ameloot_2022} and discuss the provided solutions. 

Bernier \textit{et al.} in \cite{Bernier_2020} state that the $\mu_{\rm frac}$ can be estimated using the preamble part of the frame via a successive estimate averaging approach. Moreover, a low-complexity algorithm is also proposed for resolving the start of data block ambiguity. The authors suggest that after step 4, it becomes straightforward to calculate the expected number of down-chirp samples in a sequence of blocks suspected to contain SFD samples. These expected values are stored in a vector (vector 1). Additionally, another vector (vector 2) is created by extracting the FFT magnitude outputs at the index $s_{\rm sync}$ for each block. In high SNR scenarios, it is expected that this vector reflects the proportion of down-chirp samples in each block. Finally, by convolving Vector 1 with Vector 2, the maximum value of the convolution result can be used to determine the index of the block that contains the start of the data symbols.     

In the work of Ben Temim \textit{et al.} \cite{Temim_2020_Syn}, a novel approach is proposed to improve the performance of LoRa synchronization. Based on this scheme, called differential CSS (DCSS), another version of the CSS signal is transmitted which does not encode directly the value of the CSS symbols but rather their cumulative sum so that, at the receiver, they can be retrieved by differentiation. Using this feature, the SFD part is no longer needed. Simulation results show that DCSS is performing well in the presence of large CFO values while having a negligible impact on the receiver sensitivity. 

Vangelista \textit{et al.} \cite{Vangelista_2021} propose a novel algorithm to compensate for the synchronization errors in CSS-modulated signals. Based on this approach, the time offset can be estimated using the difference of the received CSS signal cross-correlations with up-chirp and down-chirp, which are generated locally at the receiver side. This is achieved based on the fact that the cross-correlation peak can move to the left or right in the time axis in the presence of positive or negative CFO, respectively. Based on simulation results, it is shown that the proposed algorithm can recover time offset in SNRs lower than $-6.35$ dB while the signal experiences a CFO of $30$ kHz.  

Xhonneux \textit{et al.} suggest that $\mu_{\rm frac}$ and $\tau_{\rm frac}$ have different and separable effects on the DFT output of the received CSS signal based on which $\mu_{\rm frac}$ results in a linear continues phase term across CSS symbols, while $\tau_{\rm frac}$ introduces a phase term which is periodic with the period of $M$. Accordingly, the authors propose a synchronization algorithm in which the $\tau_{\rm frac}$ is estimated without the timing error parameter in step 5. In fact, in the proposed algorithm, the preamble part is exploited to initially estimate and compensate $\tau_{\rm frac}$. Results show that the synchronization algorithm suggested in this work allows a receiver to achieve a target packet error rate of $10^{-3}$ with only a $1$ or $2$ dB increase in SNR compared to a receiver that is perfectly synchronized while introducing a negligible additional complexity.

In \cite{Nguyen_2022_sync}, Nguyen \textit{et al.} propose a CSS receiver design to cope with timing, frequency, and phase synchronization taking into account the transmitter pulse shaping and receiver matched-filter. Based on this design, the authors have developed simple first-order and second-order phase-locked loop (PLL) circuits with dynamic gain control. The outcome of this design is a practical non-coherent receiver that can be utilized in conventional LoRa/CSS systems. This receiver surpasses existing designs, either by providing enhanced performance or by decreasing computational complexity. Moreover, the proposed coherent version of the designed receiver enables PSK-LoRa efficient signal detection in which the phase of the DFT output containing the peak can reveal the PSK modulated bits after the phase loop is locked.

Savaux \textit{et al.} in \cite{Savaux_2022} present a low complexity synchronization algorithm for CSS signal receivers. It is demonstrated that the synchronization problem of time and frequency can be transformed into an optimization problem. However, the cost function associated with it is not concave. The proposed approach involves a thorough analysis of the cost function and relies on a rapid search for a nearly optimal estimate of the time delay, avoiding local optima. Subsequently, a search for the optimum within the vicinity of the initial estimate is performed. This approach significantly reduces the complexity compared to an exhaustive search. Also, the results indicate that the maximum likelihood (ML) estimation performance is achieved, accompanied by reduced computation time, enabling real-time software implementation of synchronization on standard hardware.

Finally, Ameloot \textit{et al.} in \cite{Ameloot_2022} propose a LoRa packet synchronization algorithm with a focus on BER and packet reception rate (PRR) performance. The synchronization process of \cite{Ameloot_2022} can be summarized in four steps. First, a coarse symbol synchronization is performed to align the start of the packet with the detected symbols. Then, the first symbol of the preamble is identified by analyzing the number of up-chirps and down-chirps before and after the detected symbol. Afterward, an additional synchronization step corrects any constant frequency mismatch between the detector and the received packet by considering the symbol levels of the up-chirps and down-chirps. Finally, sub-sample synchronization methods are applied to further improve synchronization by making timing corrections smaller than a certain threshold, ensuring accurate data detection without energy distribution between adjacent frequency bins.
\subsection{Discussion}
In this section, we discussed the synchronization procedure for a LoRa frame and reviewed the existing works in the literature. The following are the key insights regarding the problem of LoRa synchronization and the potential solutions:
\begin{itemize}
    \item The complexity of recovering CSS signaling at the receiver, considering the current synchronization procedure performed in practice, is relatively high \cite{Bernier_2020,Ghanaatian_2019}.
    \item The start of data symbols can be found using a low-complexity algorithm based on FFT operations \cite{Bernier_2020}.
    \item Transmission of DCSS instead of CSS can be helpful to improve synchronization performance despite having a negligible added complexity \cite{Temim_2020_Syn}.
    \item Instead of using frequency domain operation for recovering the frequency offset in CSS signaling as in \cite{Bernier_2020}, time domain cross-correlation can be used more effectively to recover time offset based on the algorithm provided in \cite{Vangelista_2021}.
    \item Timing error parameter which introduces high complexity to CSS signaling synchronization can be removed from the procedure by replacing the algorithm introduced in \cite{Xhonneux_2022} for estimation and compensation of $\tau_{\rm frac}$.
    \item Exploiting PLL circuits at the receiver results in better synchronization performance for CSS and PSK-LoRa signaling \cite{Nguyen_2022_sync}.
\end{itemize}

\section{Future Research Directions}
\label{section:Fut}
Here, we discuss some potential open research areas in each of the four introduced categories.
\begin{itemize}
    \item \textit{Transceiver configuration and design:} The immunity of CSS modulation to the Doppler effect is shown in \cite{Doroshkin_2019}. However, there is a lack of work in the literature that analytically incorporates the effect of static and dynamic Doppler shifts into CSS signal transmission considering different channel conditions. All of this analysis can also be applied to SCS and FCrSK to verify their compatibility for use in satellite-based IoT networks.
    \item \textit{Data rate:} We have investigated several existing key works regarding the data rate improvement of LoRa modulation. One main challenge regarding this part is that by increasing the data rate, the interference will degrade the performance of non-coherent detection significantly. Therefore, low-complexity non-coherent detectors can be developed to tackle this issue.
    \item \textit{Interference modeling:} LoRa communication can be extended to networks including other types of LoRaWAN protocols such as long-range frequency hopping spread spectrum (LR-FHSS) \cite{10,Maleki_2023}. The physical layer modulation used in LR-FHSS is Gaussian minimum shift keying (GMSK). Hence, an interference model including both CSS and GMSK signaling can be developed to investigate the performance of LoRa GWs in the joint detection of LoRa and LR-FHSS signals. Moreover, novel low-complexity detection techniques can be proposed for such scenarios. 
    \item \textit{Synchronization:} CSS signal detection on a software-defined radio (SDR) platform can enable many potentials for the implementation of the discussed synchronization algorithms. One approach can be a try to integrate parallel computing available in SDR for effective implementation of CSS synchronization algorithms. Moreover, the synchronization of the LoRa frame while the modulated signal is a variant of CSS modulation, e.g., SSK-LoRa, IQCIM, SSK-ICS-LoRa, etc., also needs further investigation in order to design more simple and efficient synchronization algorithms.
\end{itemize}

\section{Conclusion}
\label{section:Con}
Considering the lack of a foundation work discussing all the aspects of the CSS modulation application in LoRaWAN and investigating the recent key advances in this field, this paper presents the basic mathematical foundation of CSS signaling in LoRaWAN networks, exploring various aspects such as signal generation, detection, and error performance. Moreover, an overview of the existing key recent works in the literature is presented to extract some key insights and possible research potentials that can help to further improve this newly introduced technology, i.e., LoRa modulation.

\bibliography{IoT_Transmission_Techs_PhD_V13} 

\begin{thebibliography}{10}

\bibitem{Vaezi_2022}
{M. Vaezi, A. Azari, S. R. Khosravirad, M. Shirvanimoghaddam, M. M. Azari, D.
  Chasaki and P. Popovski}, ``Cellular, wide-area, and non-terrestrial {IoT}: A
  survey on {5G} advances and the road toward {6G},'' {\em IEEE Commun. Surv.
  Tutor.}, vol.~24, no.~2, pp.~1117--1174, second quarter, 2022.

\bibitem{Jouhari_2023}
{M. Jouhari, N. Saeed, M. -S. Alouini and E. M. Amhoud}, ``A survey on scalable
  {LoRaWAN} for massive {IoT}: Recent advances, potentials, and challenges,''
  {\em IEEE Commun. Surv. Tutor.}, pp.~1--1, third quarter, 2023.

\bibitem{Milarokostas_2023}
{C. Milarokostas, D. Tsolkas, N. Passas and L. Merakos}, ``A comprehensive
  study on {LPWANs} with a focus on the potential of {LoRa/LoRaWAN} systems,''
  {\em IEEE Commun. Surv. Tutor.}, vol.~25, no.~1, pp.~825--867, first quarter,
  2023.

\bibitem{Lueth_2020}
{K. L. Lueth}, ``State of the {IoT} 2020: 12 billion {IoT} connections,
  surpassing {Non-IoT} for the first time,'' 2020.
\newblock [Online]. Available:
  \url{https://iot-analytics.com/state-of-the-iot-2020-12-billion-iotconnections-surpassing-non_iot-for-the-first-time/}~(Accessed
  Nov. 2022).

\bibitem{5G}
``{5G}: The future of {IoT},'' 2019.
\newblock
  [Online].~Available:~\url{https://www.5gamericas.org/wp-content/uploads/2019/07/5G_Americas_White_Paper_on_5G_IOT_FINAL_7.16.pdf}~(Accessed
  Jan. 2022).

\bibitem{market}
``Market research report.''
\newblock
  [Online].~Available:~\url{https://www.fortunebusinessinsights.com/industry-reports/internetof-things-iot-market-100307}~(Accessed
  Nov. 2022).

\bibitem{Vangelista_2017}
L.~Vangelista, ``Frequency shift chirp modulation: {The LoRa} modulation,''
  {\em IEEE Signal Process. Lett.}, vol.~24, pp.~1818--1821, Dec. 2017.

\bibitem{Pasolini_2022}
G.~Pasolini, ``On the {LoRa} chirp spread spectrum modulation: {Signal}
  properties and their impact on transmitter and receiver architectures,'' {\em
  IEEE Trans. on Wireless Commun.}, vol.~21, pp.~357--369, Jan. 2022.

\bibitem{Chiani_2019}
M.~Chiani and A.~Elzanaty, ``On the {LoRa} modulation for {IoT}: {Waveform}
  properties and spectral analysis,'' {\em IEEE Internet Things J.}, vol.~6,
  pp.~8463--8470, Oct. 2019.

\bibitem{Elshabrawy_2018_IntBER}
{T. Elshabrawy and J. Robert}, ``Analysis of {BER} and coverage performance of
  {LoRa} modulation under same spreading factor interference,'' in {\em Proc.
  IEEE Annu. Int. Symp. Pers., Indoor, Mobile Radio Commun. (PIMRC)}, pp.~1--6,
  2018.

\bibitem{Benkhelifa_2022}
{F. Benkhelifa, Y. Bouazizi and J. A. McCann}, ``How orthogonal is {LoRa}
  modulation?,'' {\em IEEE Internet Things J.}, vol.~9, pp.~19928--19944, Oct.
  2022.

\bibitem{TTNguyen_2019}
{T. T. Nguyen, H. H. Nguyen, R. Barton and P. Grossetete}, ``Efficient design
  of chirp spread spectrum modulation for low-power wide-area networks,'' {\em
  IEEE Internet Things J.}, vol.~6, pp.~9503--9515, Dec. 2019.

\bibitem{LoRa_White}
{LoRa Alliance}, ``{A technical overview of LoRa and LoRaWAN},'' 2015.
\newblock {San Ramon, CA, USA}.

\bibitem{LoRa_base}
{SemTech}, ``{AN1200.22: LoRa modulation basics},'' May 2015.

\bibitem{LoRa_Tx_Patent}
``{EP2763321A1},'' 2020.
\newblock [Online]. Available:
  \url{https://patents.google.com/patent/EP2763321A1}~(Accessed Sep. 2022).

\bibitem{Almeida_2020}
{I. B. F. de Almeida, M. Chafii, A. Nimr and G. Fettweis}, ``In-phase and
  quadrature chirp spread spectrum for {IoT} communications,'' in {\em Proc.
  IEEE Global Telecommun. Conf.}, pp.~1--6, 2020.

\bibitem{Baruffa_2021}
{G. Baruffa and L. Rugini}, ``Performance of {LoRa}-based schemes and
  quadrature chirp index modulation,'' {\em IEEE Internet Things J.}, vol.~9,
  pp.~7759--7772, May 2022.

\bibitem{An_2022}
{S. An, H. Wang, Y. Sun, Z. Lu and Q. Yu}, ``Time domain multiplexed {LoRa}
  modulation waveform design for {IoT} communication,'' {\em IEEE Commun.
  Lett.}, vol.~26, pp.~838--842, Apr. 2022.

\bibitem{CSS_Patent}
``{EP2449690B1},'' 2020.
\newblock [Online]. Available:
  \url{https://patents.google.com/patent/EP2449690B1/en}~(Accessed Jun. 2021).

\bibitem{Proakis_2007}
Proakis, {\em Digital Communications 5th Edition}.
\newblock McGraw Hill, 2007.

\bibitem{Int_2014}
{I. S. Gradshteyn and I. M. Ryzhik}, {\em Table of integrals, series, and
  products; 8th ed.}
\newblock Amsterdam: Academic Press, 2014.

\bibitem{Ferre_2018}
{G. Ferre and A. Giremus}, ``{LoRa} physical layer principle and performance
  analysis,'' in {\em Proc. IEEE Int. Conf. Electron. Circuits Syst. (ICECS)},
  pp.~65--68, 2018.

\bibitem{Elshabrawy_2018}
T.~Elshabrawy and J.~Robert, ``Closed-form approximation of {LoRa} modulation
  {BER} performance,'' {\em IEEE Commun. Lett}, vol.~22, pp.~1778--1781, Sep.
  2018.

\bibitem{Marquet_2019}
{A. Marquet, N. Montavont and G. Z. Papadopoulos}, ``Investigating theoretical
  performance and demodulation techniques for {LoRa},'' in {\em Proc. Int.
  Symp. "A World of Wireless, Mobile and Multimedia Networks" (WoWMoM)},
  pp.~1--6, 2019.

\bibitem{Mroue_2018}
{H. Mroue, A. Nasser, B. Parrein, S. Hamrioui, E. Mona-Cruz and G. Rouyer},
  ``Analytical and simulation study for {LoRa} modulation,'' in {\em Proc. Int.
  Conf. Telecommun. (ICT)}, pp.~655--659, 2018.

\bibitem{Afisiadis_2020}
{O. Afisiadis, M. Cotting, A. Burg and A. Balatsoukas-Stimming}, ``On the error
  rate of the {LoRa} modulation with interference,'' {\em IEEE Trans. on
  Wireless Commun.}, vol.~19, pp.~1292--1304, Feb. 2020.

\bibitem{Khai_2021}
{T. K. Nguyen, H. H. Nguyen and E. Bedeer}, ``Performance improvement of {LoRa}
  modulation with signal combining and semi-coherent detection,'' {\em IEEE
  Commun. Letters}, vol.~25, pp.~2889--2893, Sep. 2021.

\bibitem{Elshabrawy_2019}
T.~Elshabrawy and J.~Robert, ``Evaluation of the {BER} performance of {LoRa}
  communication using {BICM} decoding,'' in {\em Proc. IEEE 9th Int. Conf.
  Consumer Electron. (ICCE-Berlin)}, pp.~162--167, 2019.

\bibitem{SemTech_Modem_Design}
{SemTech}, ``{AN1200.13: SX1272/3/6/7/8: LoRa modem designer’s guide},'' July
  2013.

\bibitem{Georgiou_2017}
{O. Georgiou and U. Raza}, ``Low power wide area network analysis: Can {LoRa}
  scale?,'' {\em IEEE Wireless Commun. Lett.}, vol.~6, pp.~162--165, Apr. 2017.

\bibitem{Bernier_2020}
{C. Bernier, F. Dehmas and N. Deparis}, ``Low complexity {LoRa} frame
  synchronization for ultra-low power software-defined radios,'' {\em IEEE
  Trans. Commun.}, vol.~68, pp.~3140--3152, May 2020.

\bibitem{Kang_2022_Pre}
{J. -M. Kang, D. -W. Lim and K. -M. Kang}, ``On the {LoRa} modulation for
  {IoT}: Optimal preamble detection and its performance analysis,'' {\em IEEE
  Internet Things J.}, vol.~9, pp.~4973--4986, Apr. 2022.

\bibitem{Doroshkin_2019}
{A. A. Doroshkin, A. M. Zadorozhny, O. N. Kus, V. Y. Prokopyev and Y. M.
  Prokopyev}, ``Experimental study of {LoRa} modulation immunity to {Doppler}
  effect in {CubeSat} radio communications,'' {\em IEEE Access}, vol.~7,
  pp.~75721--75731, 2019.

\bibitem{Qian_2018}
{Y. Qian, L. Ma and X. Liang}, ``Symmetry chirp spread spectrum modulation used
  in {LEO} satellite internet of things,'' {\em IEEE Commun. Lett.}, vol.~22,
  pp.~2230--2233, Nov. 2018.

\bibitem{Qian_2019_Per}
{Y. Qian, L. Ma and X. Liang}, ``The performance of chirp signal used in {LEO}
  satellite internet of things,'' {\em IEEE Commun. Lett.}, vol.~23,
  pp.~1319--1322, Aug. 2019.

\bibitem{Roy_2019}
{A. Roy, H. B. Nemade and R. Bhattacharjee}, ``Symmetry chirp modulation
  waveform design for {LEO} satellite {IoT} communication,'' {\em IEEE Commun.
  Lett.}, vol.~23, pp.~1836--1839, Oct. 2019.

\bibitem{Yang_2019}
{C. Yang, M. Wang, L. Zheng and G. Zhou}, ``Folded chirp-rate shift keying
  modulation for {LEO} satellite iot,'' {\em IEEE Access}, vol.~7,
  pp.~99451--99461, 2019.

\bibitem{Ma_2021}
{H. Ma, G. Cai, Y. Fang, P. Chen and G. Han}, ``Design and performance analysis
  of a new {STBC-MIMO LoRa} system,'' {\em IEEE Trans. Commun.}, vol.~69,
  pp.~5744--5757, Sep. 2021.

\bibitem{Kang_2022}
{J. -M. Kang}, ``{MIMO-LoRa} for high-data-rate {IoT}: Concept and precoding
  design,'' {\em IEEE Internet Things J.}, vol.~9, pp.~10368--10369, Jan. 2022.

\bibitem{Elshabrawy_2019_DR}
{T. Elshabrawy and J. Robert}, ``Interleaved chirp spreading lora-based
  modulation,'' {\em IEEE Internet Things J.}, vol.~6, pp.~3855--3863, Apr.
  2019.

\bibitem{Bomfin_2019}
{R. Bomfin, M. Chafii and G. Fettweis}, ``A novel modulation for {IoT}:
  {PSK-LoRa},'' in {\em Proc. IEEE Veh. Technol. Conf.}, pp.~1--5, 2019.

\bibitem{Hanif_2021_SSK}
{M. Hanif and H. H. Nguyen}, ``Slope-shift keying {LoRa}-based modulation,''
  {\em IEEE Internet Things J.}, vol.~8, pp.~211--221, Jan. 2021.

\bibitem{Hanif_2021_IM}
{M. Hanif and H. H. Nguyen}, ``Frequency-shift chirp spread spectrum
  communications with index modulation,'' {\em IEEE Internet Things J.},
  vol.~8, pp.~17611--17621, Dec. 2021.

\bibitem{Almeida_2021}
{I. Bizon Franco de Almeida, M. Chafii, A. Nimr and G. Fettweis}, ``Alternative
  chirp spread spectrum techniques for {LPWANs},'' {\em IEEE Trans. Green
  Commun. Netw.}, vol.~5, pp.~1846--1855, Dec. 2021.

\bibitem{Ma_2021_FBI}
{H. Ma, Y. Fang, G. Cai, G. Han and Y. Li}, ``A new frequency-bin-index {LoRa}
  system for high-data-rate transmission: Design and performance analysis,''
  {\em IEEE Internet Things J.}, vol.~9, pp.~12515--12528, July 2022.

\bibitem{Azim_2022}
{A. W. Azim, J. L. G. Monsalve and M. Chafii}, ``Enhanced {PSK-LoRa},'' {\em
  IEEE Wireless Commun. Lett.}, vol.~11, pp.~612--616, Mar. 2022.

\bibitem{Yu_2022}
{Q. Yu, H. Wang, Z. Lu and S. An}, ``Group-based {CSS} modulation: A novel
  enhancement to {LoRa} physical layer,'' {\em IEEE Wireless Commun. Lett.},
  vol.~11, pp.~660--664, Mar. 2022.

\bibitem{Mondal_2022}
{A. Mondal, M. Hanif and H. H. Nguyen}, ``{SSK-ICS LoRa}: A {LoRa}-based
  modulation scheme with constant envelope and enhanced data rate,'' {\em IEEE
  Commun. Lett.}, vol.~26, pp.~1185--1189, May 2022.

\bibitem{Shi_2022}
{Y. Shi, W. Xu and L. Wang}, ``An enhanced interleaved chirp spreading {LoRa}
  modulation scheme for high data transmission,'' in {\em Proc. Wireless
  Telecommun. Symp.}, pp.~1--6, 2022.

\bibitem{Ishikawa_2018}
{N. Ishikawa, S. Sugiura and L. Hanzo,}, ``50 years of permutation, spatial,
  and index modulation: From classic {RF} to visible light communications and
  data storage,'' {\em IEEE Commun. Surv. Tutor.}, vol.~20, no.~3,
  pp.~1905--1938, third quarter, 2018.

\bibitem{Zhang_2006}
{P. Zhang and H. Liu}, ``An ultra-wide band system with chirp spread spectrum
  transmission technique,'' in {\em Proc. Int. Conf. ITS Telecommun.},
  pp.~294--297, 2006.

\bibitem{Yoon_2008}
{T. Yoon, S. Ahn, S. Y. Kim and S. Yoon}, ``Performance analysis of an
  overlap-based {CSS} system,'' in {\em Proc. Int. Conf. Adv. Commun.
  Technol.}, vol.~1, pp.~424--426, 2008.

\bibitem{Başar_2013}
{E. Başar, Ü. Aygölü, E. Panayırcı and H. V. Poor}, ``Orthogonal
  frequency division multiplexing with index modulation,'' {\em IEEE Trans.
  Signal Process.}, vol.~61, no.~22, pp.~5536--5549, 2013.

\bibitem{Edward_2019}
{P. Edward, S. Elzeiny, M. Ashour and T. Elshabrawy}, ``On the coexistence of
  {LoRa}- and interleaved chirp spreading {LoRa}-based modulations,'' in {\em
  Proc. Int. Conf. Wireless Mobile Comput. Netw. Commun. (WiMob)}, pp.~1--6,
  2019.

\bibitem{Temim_2020_Enh}
{M. A. Ben Temim, G. Ferré, B. Laporte-Fauret, D. Dallet, B. Minger and L.
  Fuché}, ``An enhanced receiver to decode superposed {LoRa}-like signals,''
  {\em IEEE Internet Things J.}, vol.~7, pp.~7419--7431, Aug. 2020.

\bibitem{Garlisi_2021}
{D. Garlisi, S. Mangione, F. Giuliano, D. Croce, G. Garbo and I. Tinnirello},
  ``Interference cancellation for {LoRa} gateways and impact on network
  capacity,'' {\em IEEE Access}, vol.~9, pp.~128133--128146, 2021.

\bibitem{Afisiadis_2021}
{O. Afisiadis, S. Li, J. Tapparel, A. Burg and A. Balatsoukas-Stimming}, ``On
  the advantage of coherent {LoRa} detection in the presence of interference,''
  {\em IEEE Internet Things J.}, vol.~8, pp.~11581--11593, July 2021.

\bibitem{Demeslay_2022_BER}
{C. Demeslay, P. Rostaing and R. Gautier}, ``Theoretical performance of {LoRa}
  system in multi-path and interference channels,'' {\em IEEE Internet Things
  J.}, pp.~1--1, May 2021.

\bibitem{Demeslay_2022_Receiver}
{C. Demeslay, P. Rostaing and R. Gautier}, ``Simple and efficient {LoRa}
  receiver scheme for multi-path channel,'' {\em IEEE Internet Things J.},
  pp.~1--1, Sept. 2022.

\bibitem{Croce_2018}
{D. Croce, M. Gucciardo, S. Mangione, G. Santaromita and I. Tinnirello},
  ``Impact of {LoRa} imperfect orthogonality: Analysis of link-level
  performance,'' {\em IEEE Commun. Lett.}, vol.~22, pp.~796--799, Apr. 2018.

\bibitem{Ghanaatian_2019}
{R. Ghanaatian, O. Afisiadis, M. Cotting and A. Burg}, ``{LoRa} digital
  receiver analysis and implementation,'' in {\em Proc. IEEE Int. Conf.
  Acoust., Speech, Signal Process.}, pp.~1498--1502, 2019.

\bibitem{LoRa_pat}
{O. Seller and N. Sornin}, ``Low complexity, low power and long range radio
  receiver.''
\newblock {Eur. Patent 3 264 622, Jul. 1, 2016.}

\bibitem{Temim_2020_Syn}
{M. A. Ben Temim, G. Ferré and R. Tajan}, ``A novel approach to enhance the
  robustness of {LoRa}-like {PHY} layer to synchronization errors,'' in {\em
  Proc. IEEE Global Telecommun. Conf.}, pp.~1--6, 2020.

\bibitem{Vangelista_2021}
{L. Vangelista and A. Cattapan}, ``Start of packet detection and
  synchronization for {LoRaWAN} modulated signals,'' {\em IEEE Trans. Wirel.
  Commun.}, vol.~21, pp.~4608--4621, June 2022.

\bibitem{Xhonneux_2022}
{M. Xhonneux, O. Afisiadis, D. Bol and J. Louveaux}, ``A low-complexity {LoRa}
  synchronization algorithm robust to sampling time offsets,'' {\em IEEE
  Internet Things J.}, vol.~9, pp.~3756--3769, Mar. 2022.

\bibitem{Nguyen_2022_sync}
{T. T. Nguyen and H. H. Nguyen}, ``Design of noncoherent and coherent receivers
  for chirp spread spectrum systems,'' {\em IEEE Internet Things J.}, vol.~9,
  pp.~19988--20002, Oct. 2022.

\bibitem{Savaux_2022}
{V. Savaux, C. Delacourt and P. Savelli}, ``On time-frequency synchronization
  in {LoRa} system: From analysis to near-optimal algorithm,'' {\em IEEE
  Internet Things J.}, vol.~9, no.~12, pp.~10200--10211, 2022.

\bibitem{Ameloot_2022}
{T. Ameloot, H. Rogier, M. Moeneclaey and P. Van Torre}, ``{LoRa} signal
  synchronization and detection at extremely low signal-to-noise ratios,'' {\em
  IEEE Internet Things J.}, vol.~9, no.~11, pp.~8869--8882, 2022.

\bibitem{10}
{G. Boquet, P. Tuset-Peiro, F. Adelantado, T. Watteyne and X. Vilajosana},
  ``{LR-FHSS}: Overview and performance analysis,'' {\em IEEE Commun. Mag.},
  vol.~59, pp.~30--36, Mar. 2021.

\bibitem{Maleki_2023}
{A. Maleki, H. H. Nguyen and R. Barton}, ``Outage probability analysis of
  {LR-FHSS} in satellite iot networks,'' {\em IEEE Commun. Lett.}, vol.~27,
  no.~3, pp.~946--950, 2023.

\end{thebibliography}
\bibliographystyle{ieeetr}

\end{document}